\documentclass[amsmath,amssymb,twocolumn,superscriptaddress,prb,longbibliography]{revtex4-2}
\usepackage{physics}
\usepackage{bm}
\usepackage{graphicx}
\usepackage{xcolor}
\usepackage{float}
\usepackage{verbatim}
\usepackage{MnSymbol}
\usepackage{diagbox}
\usepackage{tikz}
\usepackage{amsmath}
\usepackage{tikz-feynman} 
\usetikzlibrary{decorations.pathmorphing} 
\usetikzlibrary{arrows.meta}
\usepackage[colorlinks=true, breaklinks=true, linkcolor=red, citecolor=blue, urlcolor=blue]{hyperref}

\usetikzlibrary{decorations.markings}
\tikzset{
    arrowmid/.style={
        postaction={
            decorate,
            decoration={
                markings,
                mark=at position 0.5 with {\arrow{latex}}
            }
        }
    }
}

\begin{document}
\title{Third-order strong-coupling impurity solver for real-frequency DMFT: 
Accurate spectral functions for antiferromagnetic and photo-doped states
}
\author{Lei Geng}
\affiliation{Department of Physics, University of Fribourg, 1700 Fribourg, Switzerland}
\email{lei.geng@unifr.ch}
\author{Aaram J. Kim}
\affiliation{Department of Physics and Chemistry, DGIST, Daegu 42988, Korea}
\email{aaram@dgist.ac.kr}
\author{Philipp Werner}
\affiliation{Department of Physics, University of Fribourg, 1700 Fribourg, Switzerland}

\begin{abstract}
We present a real-frequency third-order strong-coupling impurity solver which employs quantics tensor cross interpolation (QTCI) for an efficient evaluation of the diagram weights. Applying the method to dynamical mean-field theory (DMFT) calculations of the single-band Hubbard model on the Bethe lattice, we clarify the interaction and temperature range in which the third-order approach yields accurate results. Since the calculations are implemented on the real-time/frequency axis, the detailed structure of spectral functions can be obtained without analytical continuation, as we demonstrate with examples for paramagnetic, antiferromagnetic and photo-doped states. Our work establishes a viable path toward high-order, real-frequency impurity solvers for both equilibrium and non-equilibrium DMFT studies.
\end{abstract}
\maketitle

\section{Introduction}

Quantum impurity models are minimal models of many-body physics which provide fundamental insights into local correlation effects in strongly correlated electron systems~\cite{anderson1961localized}. These models are also a key ingredient in the dynamical mean-field theory (DMFT) approximation to correlated lattice models~\cite{georges1996dynamical}. By neglecting spatial correlations, DMFT maps the original lattice problem onto an effective quantum impurity model, i.~e., a single correlated site embedded in a self-consistently determined bath that captures the influence of the surrounding lattice. This mapping becomes exact in the limit of infinite dimensions~\cite{metzner1989correlated} and offers a non-perturbative, yet computationally accessible approach to study a wide range of strongly correlated electron systems, including Mott insulators~\cite{rozenberg1992mott,Brito2016}, heavy fermion materials~\cite{shim2007modeling,Nagai2020}, and unconventional superconductors~\cite{werner2020nickelate,Karp2022}.

The central component of a DMFT calculation is the impurity solver—an algorithm that computes the local Green's function and self-energy of the effective impurity model. The accuracy, efficiency, and applicability of the DMFT method are thus in many situations limited by the choice of the impurity solver. Over the past two decades, several families of impurity solvers have been developed, including numerically exact quantum Monte Carlo (QMC) methods~\cite{rubtsov2005,werner2006continuous}, exact diagonalization (ED)~\cite{amaricci2022edipack}, numerical renormalization group (NRG) solvers~\cite{bulla2008numerical,stadler2016interleaved}, tensor-network approaches~\cite{ganahl2015efficient,grundner2025tensor,cao2021tree}, influence functional methods~\cite{Chen2024,Chu2024,ng2023real,nayak2025steady} and various diagrammatic perturbation schemes~\cite{eckstein2010nonequilibrium,kirchner2004dynamical}. Among these, continuous-time QMC solvers have become the method of choice for many equilibrium applications due to their high accuracy and efficiency~\cite{gull2011continuous}. However, they are primarily used in equilibrium settings, where calculations can be performed on the Matsubara axis. To obtain real-time or real-frequency information, one then typically performs an analytic continuation from imaginary to real frequencies, for example using the Maximum Entropy method \cite{Jarrell1996}. This however introduces errors which are difficult to control. Alternatively, the QMC solvers can be implemented directly on a real-time contour \cite{muhlbacher2008real,werner2009diagrammatic,profuno2015}, but such calculations suffer from a severe dynamical sign problem and rapidly increasing computational cost, making them of limited use for non-equilibrium DMFT \cite{eckstein2009thermalization,tsuji2011dynamical}, the study of transport problems \cite{muhlbacher2008real,werner2010weak,profuno2015} and real-frequency spectroscopy. Recently, variants of the hybridization-expansion QMC method, in particular the inchworm algorithm~\cite{yu2025inchworm,eidelstein2020multiorbital,antipov2017currents}, have employed a resummed real-time diagrammatic expansion to significantly alleviate the sign problem, offering a promising route toward real-time simulations of strongly correlated systems.

In contrast, diagrammatic impurity solvers based on low-order perturbation theory (both weak-coupling \cite{tsuji2013nonequilibrium} and strong-coupling expansions \cite{keiter1971diagrammatic,pruschke1989anderson}) provide versatile and cost-effective methods for real-time and real-frequency DMFT. For nonequilibrium DMFT simulations of Mott insulators, pseudo-particle strong-coupling expansions have proven particularly effective~\cite{eckstein2010nonequilibrium}. These methods, which include the first-order non-crossing approximation (NCA) \cite{keiter1971diagrammatic} and second-order one-crossing approximation (OCA) \cite{pruschke1989anderson}, formulate the impurity problem in terms of pseudo-particles and systematically sum selected classes of skeleton diagrams up to a given order in the hybridization function. As a lowest order approximation, NCA is computationally cheap and captures qualitative features of the Mott insulating phase, but it fails to provide quantitative accuracy near the metal-insulator transition or in the presence of low-temperature coherence effects~\cite{costi1996spectral,kroha2005conserving,sposetti2016qualitative}. OCA offers significant improvements by including the one-crossing vertex corrections~\cite{sposetti2016qualitative,eckstein2010nonequilibrium}, but accurate solutions in the low-temperature metal regime, or in magnetically ordered insulators requires going beyond the first crossing level to incorporate higher-order processes.

However, the inclusion of higher-order skeleton diagrams leads to a rapid growth in computational complexity, since high-dimensional time integrals need to be evaluated. This computational bottleneck has long limited the practical applicability of higher-order strong-coupling impurity solvers. One possible strategy is to evaluate these integrals using Monte Carlo sampling, as is done in the inchworm approach~\cite{antipov2017currents}. Another related idea has been explored in the context of diagrammatic decompositions using separable basis functions~\cite{Kaye2024}. Recently, advances in low-rank tensor decomposition techniques -- particularly tensor cross interpolation (TCI) and quantics tensor cross interpolation (QTCI) ~\cite{oseledets2011tensor,oseledets2010tt,nunez2022learning,ritter2024quantics,shinaoka2023multiscale,fernandez2024learning} -- have opened up new possibilities for addressing this challenge. These techniques enable an efficient representation, compression, and manipulation of high-dimensional tensor objects, significantly reducing the memory and computational cost associated with evaluating high-order Feynman diagrams~\cite{sroda2024high,murray2024nonequilibrium}. It is thus appealing to use these techniques for the calculation of the high-order integrals for the self-energies and Green's functions that appear in strong-coupling impurity solvers.

Encouraging progress has been made in applying TCI techniques to Matsubara-axis impurity solvers~\cite{erpenbeck2023tensor,yu2025inchworm,matsuura2025tensor}, and, in parallel, TCI and QTCI have also proven effective in the context of real-time impurity solvers~\cite{kim2025strong,eckstein2024solving,paprotzki2025high}. These recent implementations achieve high accuracy at a manageable computational cost. Building upon these developments, we present here a third-order QTCI-based strong-coupling impurity solver for real-frequency DMFT. Our method extends the established NCA and OCA frameworks to include the complete set of third-order skeleton diagrams, using QTCI as implemented in the xfac library~\cite{fernandez2024learning} for the efficient evaluation of high-dimensional integrals. We benchmark our method on the single-band Hubbard model and demonstrate that it yields fully converged results in the strongly correlated regime, as well as well-resolved correlation-induced structures in the spectral functions of equilibrium and photo-doped systems.

The paper is organized as follows. Section~\ref{sec2} outlines the real-time formalism and strong-coupling pseudo-particle methods for DMFT. It also introduces the numerical techniques and implementation strategies based on the QTCI framework. Section~\ref{sec3} presents numerical benchmarks: first for the single-impurity Anderson model in Section~\ref{sec3a}, then for DMFT in the paramagnetic state (Section~\ref{sec3b}), antiferromagnetic state (Section~\ref{sec3c}), and photodoped state (Section~\ref{sec3d}). Finally, Section~\ref{sec4} provides conclusions and a brief outlook.

\section{Theory and Methods}\label{sec2}
\subsection{Model and Real-Time Formalism for DMFT}\label{sec2a}
We consider the single-band Hubbard model defined on a lattice, 
\begin{equation}
	H = -\sum_{\langle ij \rangle,\sigma} t_{ij}  c_{i\sigma}^\dagger c_{j\sigma}^{} + U \sum_i n_{i\uparrow} n_{i\downarrow} - \mu \sum_{i,\sigma} n_{i\sigma},
    \label{eq1}
\end{equation}
with $c^\dagger_{i\sigma}$ the creation operator for an electron with spin $\sigma$ on site $i$ and $n_{i\sigma}=c^\dagger_{i\sigma}c^{}_{i\sigma}$ denoting the corresponding spin-resolved density. This model describes electrons hopping between nearest-neighbor lattice sites with amplitude $t_{ij}$, and the effects of the on-site Coulomb interaction $U$ and chemical potential $\mu$. Within DMFT, this lattice problem is mapped to an effective quantum impurity model where a single interacting site is coupled to a noninteracting bath. Introducing an index $p$ to label the bath sites, the impurity Hamiltonian reads
\begin{equation}
	H_{\text{imp}} = \sum_{p\sigma} \epsilon_p a_{p\sigma}^\dagger a_{p\sigma}^{} + \sum_{p\sigma} \left( V_p a_{p\sigma}^\dagger c_\sigma^{} + \text{h.c.} \right) + U n_\uparrow n_\downarrow - \mu \sum_\sigma n_\sigma,
    \label{eq2}
\end{equation}
where $a_{p\sigma}^\dagger$ creates a bath electron with spin $\sigma$ and energy $\epsilon_p$, and $V_p$ denotes the hybridization with the impurity orbital.

\begin{figure}[t]
  \centering
    \begin{tikzpicture}[thick, ->, >=stealth]
    \draw[->] (0,0) -- (6,0) node[anchor=west] {$\text{Re}~ t$};
    \draw[->] (0,0) -- (0,-2.8) node[anchor=north] {$\text{Im}~ t$};

    \draw[->,blue] (0,0.1) -- (5,0.1) node[pos=0.5, above] {forward branch};
    \draw[->,blue] (5,-0.1) -- (0,-0.1) node[pos=0.5, below] {backward branch};
    \draw[->,red] (0,-0.1) -- (0,-2.3) node[pos=0.5, right] {imaginary branch};

    \filldraw (0,0.0) circle (1pt) node[anchor=south] {$t=0$};
    \filldraw (5,0.0) circle (1pt) node[anchor=south] {$t=t_\text{max}$};
    \filldraw (0,-2.3) circle (1pt) node[anchor=west] {$t = -i\beta$};
    \end{tikzpicture}
    \caption{Kadanoff-Baym contour with forward, backward, and imaginary-time branches.}
  \label{fig1}
\end{figure}
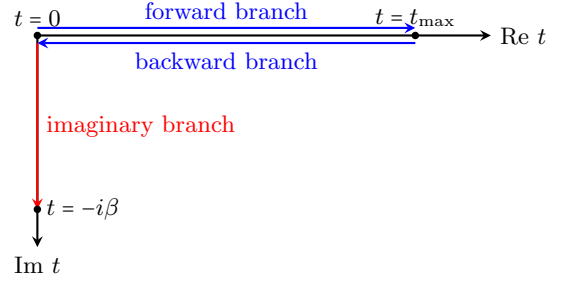

For the description of a general nonequilibrium state, the impurity model must be solved on the Kadanoff-Baym contour $\mathcal{C}_{K}$, which consists of forward, backward, and imaginary-time branches~\cite{keldysh2024diagram,schuler2020nessi}. A sketch of the contour is shown in Fig.~\ref{fig1}. The contour-ordered Green's function is defined as
\begin{equation}
    G(t,t') = -\mathrm{i} \langle \mathcal{T}_\mathcal{C} c(t) c^\dagger(t') \rangle,
    \label{eq3}
\end{equation}
where $\mathcal{T}_\mathcal{C}$ indicates time-ordering along the contour $\mathcal{C}_{K}$. Under certain physical conditions simplifications are possible. The most common case is the thermal equilibrium state. If the expectation values satisfy the Kubo-Martin-Schwinger (KMS) condition 
\begin{equation}
    \langle A(t) B(t') \rangle = \langle B(t') A(t + \mathrm{i}\beta) \rangle
    \label{eq4}
\end{equation}
for any pair of operators $A$ and $B$, 
it has been demonstrated that the real-time Green's functions can be reconstructed from the imaginary-time branch using the known thermal distributions~\cite{kubo1957statistical,martin1959theory}. Consequently, many equilibrium studies restrict the calculations to the imaginary branch, which significantly simplifies the numerical treatment.

Another important case is the non-equilibrium steady state (NESS). A typical example is a system with applied voltage bias and dissipation, or (in an approximate description) the long-lived photo-doped state induced in a Mott insulator by a laser excitation~\cite{murakami2023photo,li2021nonequilibrium}. Although such systems no longer satisfy the KMS condition and are therefore not in thermal equilibrium, they exhibit time-translation invariance in the steady-state regime. This implies that Green's functions and other two-point correlation functions depend only on the time difference, rather than on two separate arguments. As a result, the full three-branch Kadanoff-Baym contour can be reduced to the Keldysh contour with two real-time branches (forward and backward), which now extend from $t=-\infty$ to $t=\infty$. It is then sufficient to work with the univariate lesser, greater, and retarded Green's functions defined on the real axis, 
\begin{align}
G^{<}(t) &= \mathrm{i} \langle c^\dagger(0) c(t) \rangle, \label{eq5a} \\
G^{>}(t) &= -\mathrm{i} \langle c(t) c^\dagger(0) \rangle, \label{eq5b} \\
G^{R}(t) &=  \theta(t) \left[G^{>}(t) - G^{<}(t) \right]. \label{eq5c}
\end{align}

In a NESS, it is possible to define an effective distribution function $f_{\mathrm{eff}}(\omega)$ that characterizes the relation between the lesser and retarded Green's functions in frequency space~\cite{picano2021quantum}, analogous to the Fermi-Dirac distribution in thermal equilibrium. In practice, we define $f_{\mathrm{eff}}(\omega)$ by the relation
\begin{equation}
G^<(\omega) = -2\mathrm{i}f_{\mathrm{eff}}(\omega) \text{Im} G^R(\omega).
\label{eq6}
\end{equation}
In the equilibrium case, $f_{\mathrm{eff}}(\omega)$ becomes the Fermi-Dirac distribution, which imposes the fluctuation-dissipation relation between the Green's functions. In nonequilibrium calculations, we may employ a physically motivated form of $f_{\mathrm{eff}}(\omega)$, e.~g. to model photo-doped steady states, and use it to obtain a self-consistent solution of the DMFT iterations on the real-frequency axis \cite{kunzel2024numerically,kim2025strong}. In the present work, we adopt this approach and determine $G^{<}(\omega)$ from $G^{R}(\omega)$ through Eq.~\eqref{eq6} with given $f_{\mathrm{eff}}(\omega)$.

Having established the formalism of real-frequency Green's functions under steady-state conditions, we now turn to the DMFT formulation. In the impurity model shown in Eq.~\eqref{eq2}, the effect of the lattice environment on the impurity is encoded in the hybridization function $\Delta^R(\omega)=\sum_p \frac{|V_p|^2}{\omega-\epsilon_p+\mathrm{i}0_+}$, which is the function that is determined self-consistently. On the Bethe lattice, the DMFT self-consistency condition simplifies considerably and relates the hybridization function $\Delta(\omega)$ directly to the local Green's function $G_{\text{loc}}(\omega)$ as~\cite{georges1996dynamical}
\begin{equation}
\Delta(\omega) = v^2 G_{\text{loc}}(\omega),
\label{eq7}
\end{equation}
where $v$ is the hopping amplitude (properly rescaled in the limit of infinite connectivity \cite{metzner1989correlated}). We will focus on the infinitely-connected Bethe lattice in this study.

The impurity problem is then solved with the given hybridization function $\Delta(\omega)$ to obtain the local self-energy $\Sigma(\omega)$, which encodes the effects of the Hubbard interaction $U$ on the propagation of the electrons. The local Green's function is updated using the Dyson equation:
\begin{equation}
G^R_{\text{loc}}(\omega) = \left[ \omega + \mu - \Sigma^R(\omega) - \Delta^R(\omega) \right]^{-1}.
\label{eq8}
\end{equation}
This closed set of equations defines the DMFT self-consistency loop on the Bethe lattice within the real-frequency steady-state formalism.

\subsection{Strong-Coupling Pseudo-Particle Expansion}\label{sec2b}

\subsubsection{Pseudo-particle formalism}

The pseudo-particle approach provides a powerful framework to describe strongly interacting quantum impurity systems by reformulating the local Hilbert space in terms of auxiliary degrees of freedom~\cite{barnes1976new,coleman1984new}. In the context of DMFT, it is particularly suited for studying Mott insulating systems and the strong-coupling regime.

In this formulation, the local impurity states $| m \rangle$ (for the model in Eq.~\eqref{eq2}, $| m \rangle$=$| 0 \rangle$,$|\!\uparrow\rangle$,$|\!\downarrow\rangle$,$|\!\uparrow\downarrow\rangle$) are assigned auxiliary 
pseudo-particles $d_m^\dagger$~.
The physical electron creation operator $c^\dagger_\sigma $ is then expressed as a combination of these pseudo-particle operators via matrix elements~\cite{aoki2014nonequilibrium,eckstein2010nonequilibrium,murakami2018high},
\begin{equation}
c^\dagger_\sigma = \sum_{m,n} \bar{F}^\sigma_{mn} \, d^\dagger_m d_n,
\label{eq11}
\end{equation}
where $ \bar{F}^{\sigma}_{mn} = \langle m | c^\dagger_\sigma | n \rangle $ are the matrix elements of the electron creation operator between impurity eigenstates. In this way, the original impurity Hamiltonian can be mapped onto a model of free pseudo-particles interacting with an electron bath, subject to a constraint that ensures physical states:
\begin{equation}
Q = \sum_m d^\dagger_m d_m = 1.
\label{eq:Qconstraint}
\end{equation}
This constraint enforces that only one pseudo-particle is present at any given time, corresponding to a single physical local many-body state. 

The action of the impurity model on the Keldysh contour $\mathcal{C}_K$ can be written as
\begin{equation}
  \begin{split}
S =& -\mathrm{i} \int_{\mathcal{C}_K} \mathrm{d}t \sum_m E_m(t) d_m^\dagger(t) d_m(t) \\
& - \mathrm{i} \int_{\mathcal{C}_K} \mathrm{d}t \, \mathrm{d}t' \sum_{mn, m'n', \sigma} d_m^\dagger(t) d_n(t) \bar{F}^{\sigma}_{mn} \\
& \qquad\quad \times \Delta_{\sigma}(t,t') F^{\sigma}_{n'm'} d_{n'}^\dagger(t') d_{m'}(t'),
  \end{split}
\label{eq:action}
\end{equation}
where $E_m(t)$ are the eigenenergies of the local many-body states, and the partition function and expectation values of the system become
\begin{align}
Z &= \mathrm{Tr} \left[\mathcal{T}_{\mathcal{C}} e^{S[d^\dagger, d]} \Big|_{Q=1} \right],\label{eq:pp_partition}\\
\langle \cdots \rangle &= \frac{1}{Z} \mathrm{Tr} \left[\mathcal{T}_{\mathcal{C}} (\ldots) e^{S[d^\dagger, d]} \Big|_{Q=1}\right].\label{eq:expectation}
\end{align}

To evaluate the expectation values \eqref{eq:expectation}, one needs to compute the pseudo-particle Green's functions. These are defined analogously to physical Green's functions in Eq.~\eqref{eq3}, but with the electron operators $c^\dagger$, $c$ replaced by pseudo-particle operators $d^\dagger$, $d$. However, a crucial difference is that the pseudo-particles reside in an enlarged (unphysical) Hilbert space, and physical observables are only recovered after projection onto the physical subspace via Eq.~\eqref{eq:Qconstraint}. We need to define the projected Green's function $\mathcal{G}$ based on the original grand canonical Green's function. This projection leads to the constraint that time arguments appear in cyclic order along the contour $\mathcal{C}_K$~\cite{eckstein2010nonequilibrium,coleman1984new}. In the case of a NESS, one commonly defines the following projected pseudo-particle Green's function components~\cite{murakami2018high}:
\begin{align}
\mathcal{G}_{mn}^>(t) &= -\mathrm{i} \langle \mathcal{T}_\mathcal{C} d_m(t) d_n^\dagger(0) \rangle_{Q=0}, \label{eq13a} \\
\mathcal{G}_{mn}^<(t) &= -\mathrm{i} \langle \mathcal{T}_\mathcal{C} d_m(t) d_n^\dagger(0) \rangle_{Q=1}, \label{eq13b} \\
\mathcal{G}_{mn}^R(t) &= \theta(t) \mathcal{G}_{mn}^>(t), \label{eq13c}\\
\mathcal{G}_{mn}^A(t) &= -\theta(-t) \mathcal{G}_{mn}^>(t), \label{eq13d}
\end{align}
Here, the subscripts  $Q=0$ and  $Q=1$ indicate the pseudo-particle number for the initial state of the forward branch. 

\begin{figure}[t]
  \centering
  \includegraphics[width=0.45\textwidth]{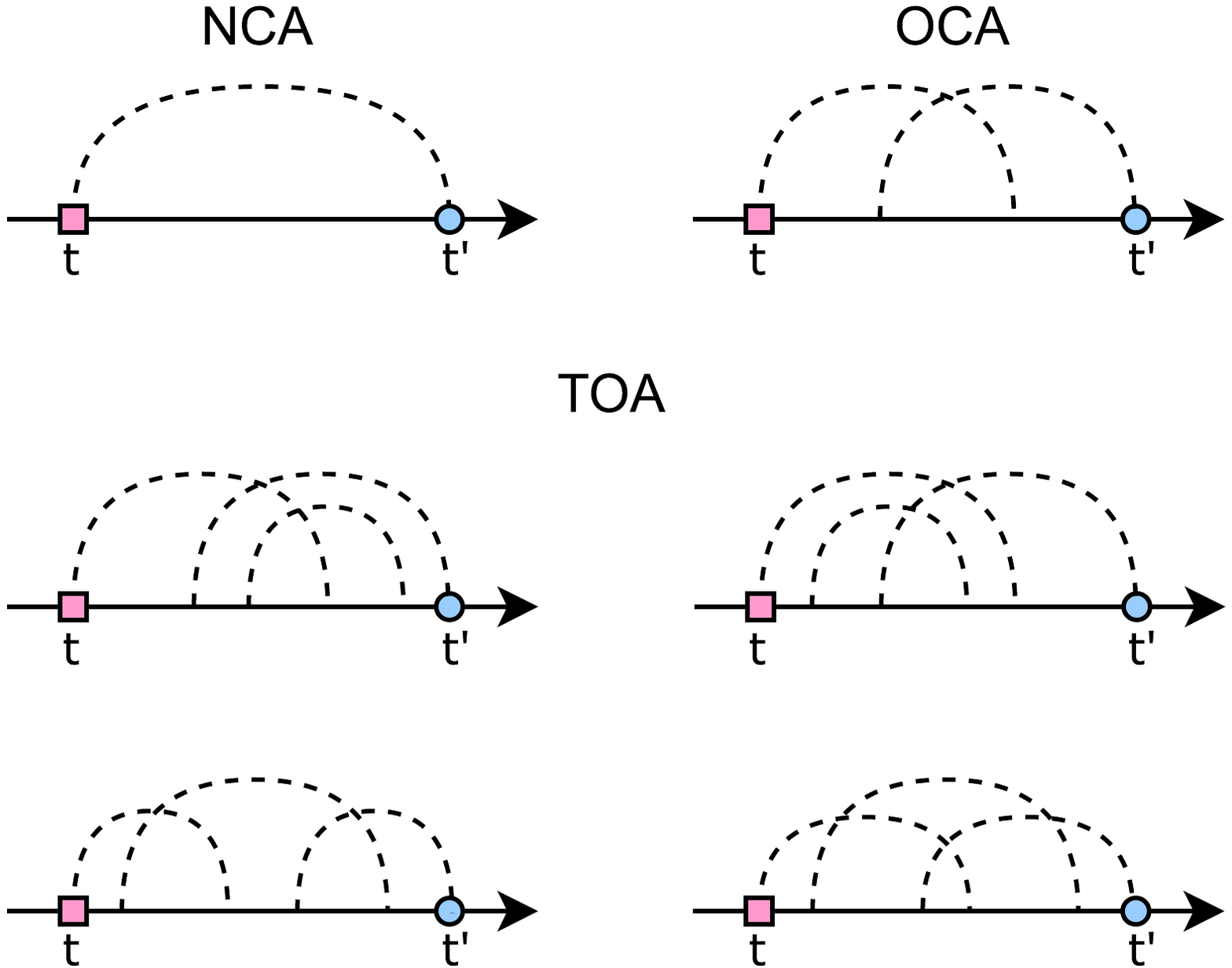}
  \caption{Topologies of the skeleton diagrams for the pseudo-particle self-energy $\Sigma_\text{pp}(t',t)$ at the NCA, OCA, and TOA level. Arrows indicate the direction along the Keldysh contour. Red squares represent the first operator at time $t$, while blue circles denote the last operator at time $t'$. Dashed lines represent hybridization functions $\Delta$, which can have two possible directions. 
	  The two ends of a dashed line correspond to a pair of impurity creation and annihilation operators. The solid lines connecting $t$ and $t'$ are projected pseudo-particle Green's functions $\mathcal{G}$, which always follow the direction of the Keldysh contour.
  }
  \label{fig2}
\end{figure}

\subsubsection{Strong-coupling impurity solver}

After introducing the pseudo-particle formalism, we now turn to the strong-coupling impurity solver itself. This solver is based on a systematic expansion of the partition function in powers of the hybridization function, so that only the hybridization term is treated perturbatively, while the local many-body interactions are incorporated non-perturbatively from the outset. This makes the method well-suited for strongly correlated systems, where hybridization effects are relatively weak.

The resulting expansion gives rise to a diagrammatic representation of the Luttinger-Ward functional $\Phi[\mathcal{G},\Delta]$, which consists of all closed skeleton diagrams built from interacting pseudo-particle Green's functions and hybridization lines. The first three orders of this expansion correspond to the NCA, OCA and TOA, respectively.

Once the functional $\Phi[\mathcal{G},\Delta]$ is specified, the pseudo-particle self-energy can be obtained by the functional derivative with respect to the pseudo-particle Green's function:
\begin{equation}
\Sigma_\text{pp}(t,t') = \frac{\delta \Phi[\mathcal{G},\Delta]}{\delta \mathcal{G}(t',t)}.
\label{eq:ppse_from_Z}
\end{equation}
Diagrammatically, this corresponds to removing a single pseudo-particle propagator line $\mathcal{G}$ from each skeleton diagram in $\Phi[\mathcal{G},\Delta]$, thereby generating all one-particle-irreducible diagrams contributing to the self-energy $\Sigma_\text{pp}$. The specific self-energy diagrams for NCA, OCA, and TOA are shown in Fig.~\ref{fig2}. It has been demonstrated that such a skeleton expansion yields conserving approximations of Baym and Kadanoff~\cite{baym1961conservation}. With the self-energy in hand, the pseudo-particle Green's function can then be obtained self-consistently via a pseudo-particle Dyson equation~\cite{aoki2014nonequilibrium}.

\begin{figure}[t]
  \centering
  \includegraphics[width=0.45\textwidth]{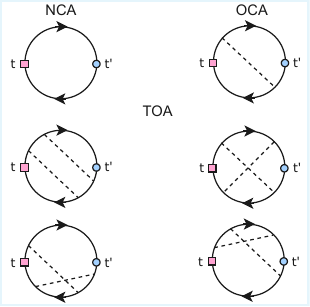}
  \caption{Topologies of the skeleton diagrams for the physical Green's function $G(t,t')$ within the NCA, OCA, and TOA approximations. Here, red squares correspond to annihilation operators, while blue circles correspond to creation operators.
  }
  \label{fig3}
\end{figure}

In addition to the diagrammatic representation of the pseudo-particle self-energy, the physical Green's function can also be expressed within the skeleton diagram formalism. According to the definition of the physical Green's function in Eq.~\eqref{eq3} and the pseudo-particle partition function in Eq.~\eqref{eq:pp_partition}, it can be written as a functional derivative with respect to the hybridization function:
\begin{equation}
G^\sigma(t,t') = -\frac{\delta \ln Z[\Delta]}{\delta \Delta_\sigma(t',t)}.
\label{eq:GF_physical}
\end{equation}
Here, $\ln Z[\Delta] = -\beta \Omega$ in the grand canonical ensemble, where $\Omega$ is the grand potential. The grand potential is related to the Luttinger-Ward functional $\Phi[\mathcal{G},\Delta]$ through
\begin{equation}
\Omega = \text{Tr}\left[\ln(-\mathcal{G})\right] + \text{Tr}(\Sigma_\text{pp} \mathcal{G}) + \Phi[\mathcal{G},\Delta].
\label{eq:LnZ_phi_relation}
\end{equation}
It is important to note that the term $\text{Tr}(\Sigma_\text{pp} \mathcal{G})$ shares the same topological skeleton structure as the Luttinger-Ward functional $\Phi[\mathcal{G},\Delta]$, up to a numerical factor. Consequently, the diagrammatic representation of the physical Green's function can be obtained by removing a single hybridization line from the skeleton diagrams of $\Phi[\mathcal{G},\Delta]$. The corresponding skeleton diagrams for the physical Green's function up to TOA are shown in Fig.~\ref{fig3}. The diagrammatic rules for evaluating the combinatorial prefactors of the self-energy and physical Green's function diagrams are discussed in Refs.~\cite{eckstein2010nonequilibrium,aoki2014nonequilibrium}.

All the diagrams discussed above should be defined on the Keldysh contour. For the numerical evaluation of the corresponding expressions within the NCA, OCA, and TOA methods, we adopt the ring-shaped real-time contour discussed in Ref.~\cite{kim2025strong}, where the starting point of the forward branch and the endpoint of the backward branch are connected to form a closed loop. Notably, this connection introduces a branch-cut point at the leftmost position of the ring, since the forward and backward branches are not truly connected at that point. However, when the diagrammatic contributions are located far from this branch cut, its influence becomes negligible. Therefore, we conventionally place the first operator of each diagram at the rightmost position on the ring. 

\begin{figure}[ht]
  \centering
  \includegraphics[width=0.49\textwidth]{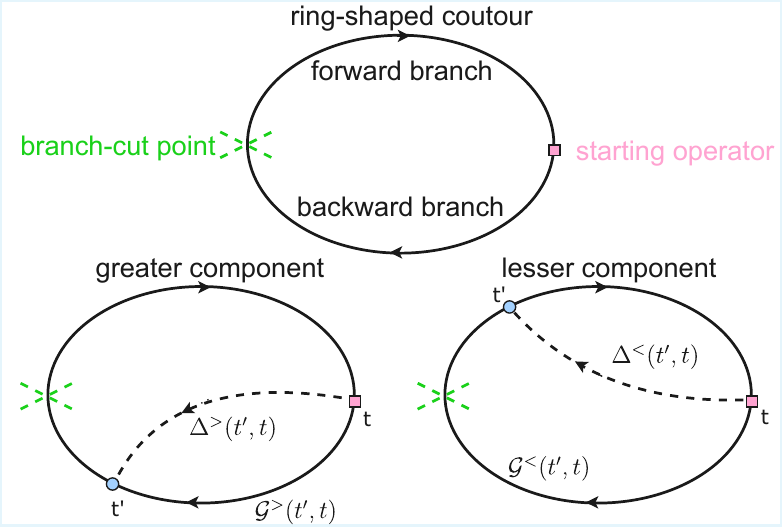}
  \caption{The ring-shaped Keldysh contour employed in this study. The branch-cut point connects the end of the backward branch to the beginning of the forward branch. If the time arguments of a Green's function, self-energy, or hybridization function cross this branch-cut point, the corresponding component is identified as the lesser component; otherwise, it is classified as the greater component.
  }
  \label{fig4}
\end{figure}

Fig.~\ref{fig4} shows an illustration of the ring-shaped contour. The lesser or greater component of a Green's function or hybridization function is determined by whether or not the final operator at $t'$ is placed on the backward or forward branch of the contour. Using this convention, the pseudo-particle self-energy diagrams in Fig.~\ref{fig2} can be interpreted as diagrams defined on this ring-shaped contour. Taking the NCA self-energy as an example, we can fix the starting operator at time $t$, as illustrated in Fig.~\ref{fig4}. The greater component then reads:
\begin{equation}
  \begin{split}
	  \left[\Sigma^>_{_\text{pp}}\right]_{mm}\hspace{-0.3em}(t',t)=\mathrm{i}\sum_{n,\sigma}&\left[\Delta_{\sigma}^>(t',t)\bar{F}^{\sigma}_{mn}\mathcal{G}^>_{nn}(t',t)F^{\sigma}_{nm}\right.\\
&\left.-\Delta_{\sigma}^<(t,t')F^{\sigma}_{mn}\mathcal{G}^>_{nn}(t',t)\bar{F}^{\sigma}_{nm}\right],\\
  \end{split}
  \label{eq:nca_se_greater}
\end{equation}
which consists of two terms corresponding to opposite hybridization line directions. The first term in Eq.~\eqref{eq:nca_se_greater} matches the diagram shown in the bottom left panel of Fig.~\ref{fig4}, while the second term represents the diagram with the opposite hybridization direction. Similarly, the lesser component of the NCA self-energy is given by
\begin{equation}
  \begin{split}
	  \left[\Sigma^<_{_\text{pp}}\right]_{mm}\hspace{-0.3em}(t',t)=-\mathrm{i}\sum_{n,\sigma}&\left[\Delta_{\sigma}^<(t',t)\bar{F}^{\sigma}_{mn}\mathcal{G}^<_{nn}(t',t)F^{\sigma}_{nm}\right.\\
&\left.-\Delta_{\sigma}^>(t,t')F^{\sigma}_{mn}\mathcal{G}^<_{nn}(t',t)\bar{F}^{\sigma}_{nm}\right],
  \end{split}
  \label{eq:nca_se_lesser}
\end{equation}
which can likewise be identified within the diagram shown in the bottom right panel of Fig.~\ref{fig4}.

In contrast to the self-energy, diagrams for the physical Green's function involve the full ring, as well as a trace over all four pseudo-particle sectors $|m\rangle$. Taking the NCA physical Green's function shown in Fig.~\ref{fig3} as an example, we fix the creation operator at time $t$, as in Fig.~\ref{fig4}. The greater component can then be expressed as
\begin{equation}
	G^>_{\sigma}(t',t)=-\mathrm{i}\sum_{m,n}\mathcal{G}^<_{mm}(t,t')F^{\sigma}_{mn}\mathcal{G}^>_{nn}(t',t)\bar{F}^{\sigma}_{nm}.
  \label{eq:nca_pg_greater}
\end{equation}
Similarly, the lesser component becomes
\begin{equation}
G^<_{\sigma}(t',t)=\mathrm{i}\sum_{m,n}\mathcal{G}^>_{mm}(t,t')F^{\sigma}_{mn}\mathcal{G}^<_{nn}(t',t)\bar{F}^{\sigma}_{nm}.
  \label{eq:nca_pg_lesser}
\end{equation}

Due to the fixed position of the starting operator, we can only directly compute $\Sigma_\text{pp}(t',t)$ and $G(t',t)$ for the case $t' < t$. The remaining half of the time domain can be reconstructed using the symmetry relations
\begin{equation}
  \begin{split}
G^{</>}(t,t') = -\left[G^{</>}(t',t)\right]^{\dagger}, \hspace{1mm} \Sigma_\text{pp}^{</>}(t,t') = -\left[\Sigma_\text{pp}^{</>}(t',t)\right]^{\dagger}.
  \end{split}
\label{eq:symmetry}
\end{equation}

\subsubsection{DMFT loop}

Employing the formalism introduced above, the DMFT self-consistency loop can now be reformulated as follows:
\begin{enumerate}
	\item Starting from an initial guess for the pseudo-particle Green's functions $\mathcal{G}(t,t')$ and the hybridization function $\Delta(t, t')$, the pseudo-particle self-energy $\Sigma_{\text{pp}}(t, t')$ is computed using the selected diagrammatic approximation shown in Fig.~\ref{fig2}.
  
  \item The Dyson equations for the pseudo-particles are solved in frequency space to get the pseudo-particle Green's function from the pseudo-particle self-energy~\cite{kim2025strong,li2021nonequilibrium}:
  \begin{align}
  \mathcal{G}^R(\omega) &= \left[\omega -H_{\text{loc}}-\Sigma_\text{pp}^R(\omega)\right]^{-1}, \label{eq:dyson_a} \\
  \mathcal{G}^<(\omega) &= \mathcal{G}^R(\omega) \Sigma_\text{pp}^<(\omega) \mathcal{G}^A(\omega), \label{eq:dyson_b} \\
  \mathcal{G}^>(\omega) &= \mathcal{G}^R(\omega) \left[\Sigma_\text{pp}^R(\omega) - \Sigma_\text{pp}^A(\omega)\right] \mathcal{G}^A(\omega), \label{eq:dyson_c}
  \end{align}
  where $H_{\text{loc}}$ is the local Hamiltonian, which expressed in the basis of impurity eigenstates reads $\text{diag}\{E_{0}, E_{\uparrow}, E_{\downarrow}, E_{\uparrow\downarrow}\}$.
  
  \item The physical Green's function for each spin $G_\sigma(t, t')$ is reconstructed via the updated pseudo-particle Green's function and the hybridization function, according to the diagrammatic rules shown in Fig.~\ref{fig3}. 
  
  \item The DMFT self-consistency condition is imposed. Since the Bethe lattice is used, the hybridization function is updated directly using the physical Green's function (Eq.~\eqref{eq7}).

  \item Steps (2)-(4) are iterated until the self-consistency is reached.
\end{enumerate}

It is worth noting that a numerical stabilization technique may need to be employed in the self-consistent DMFT loop, by adding a small frequency-dependent imaginary component to the pseudo-particle self-energy. This correction is optional for metallic systems, where the pseudo-particles naturally acquire sufficiently short lifetimes through the hybridization. However, it is crucial for insulating systems, where sharp spectral features can otherwise lead to numerical instabilities.

Following Ref.~\cite{li2021nonequilibrium}, we introduce additional self-energy contributions for the retarded and lesser components as
\begin{align}
\Delta\Sigma_\text{pp}^R(\omega) &= -\mathrm{i}\frac{\eta}{2} \left[\tanh\left(\frac{\beta(\omega - \mu_{pp})}{2}\right) + 1\right], \label{eq:delta_se_a} \\
\Delta\Sigma_\text{pp}^<(\omega) &= \mathrm{i} \eta \left[1 - \tanh\left(\frac{\beta(\omega - \mu_{pp})}{2}\right)\right], \label{eq:delta_se_b}
\end{align}
where $\eta$ is a small positive constant that controls the overall broadening, and $\mu_{pp}$ is the chemical potential of the pseudo-particles. 
This artificial broadening helps regulate the Dyson equation and ensures convergence of the self-consistency loop.

\subsection{QTCI-Based Numerical Implementation}\label{sec2c}

\subsubsection{QTT representation}

In the previous section, we reviewed the theoretical foundation of the strong-coupling expansion method and its implementation on the Keldysh contour for computing NESSs. As discussed, the central numerical challenge is the evaluation of the skeleton diagrams. For an expansion up to order $N$, the corresponding Feynman diagrams involve integrations over $2N-1$ independent time variables on the contour. As the diagram order increases, the evaluation of these high-dimensional integrals becomes the computational bottleneck of the solver, and it quickly becomes prohibitive.

In this section, we focus on the numerical strategies employed to accelerate the evaluation of such integrals using the QTCI method \cite{ritter2024quantics}. The core idea stems from the TCI framework \cite{nunez2022learning}, where a high-dimensional function is approximated as a low-rank tensor to reduce both the storage requirements and computational cost of the summation. Unlike traditional singular value decomposition (SVD) based tensor decomposition methods, TCI does not attempt to find the most compact basis representation. Instead, it identifies and stores a selected set of representative function values (referred to as pivots) and reconstructs the full function via interpolation. This strategy significantly enhances the efficiency at the cost of introducing a controllable approximation error~\cite{nunez2022learning,fernandez2024learning}.

Once an efficient low-rank approximation algorithm is selected, choosing an appropriate coordinate encoding and tensor representation of the data becomes critical for the good performance.
The recently developed QTCI method combines TCI with the quantics tensor train (QTT) representation~\cite{ritter2024quantics}. Specifically, one approximates a high-dimensional tensor as a low-rank tensor train, where each input variable $v_i$ is decomposed into $R$ binary digits as
\begin{equation}
v_i = \sum_{j=1}^{R}v_{ij}\frac{v_{i \max}}{2^j},
\label{eq:binary_bit}
\end{equation}
where $v_{ij}$ only takes the value of 0 or 1, and $v_{i \max}$ is the maximum value of the variable. In the standard method, the same bits from different variables are grouped and placed adjacently in the tensor train. This reordering greatly reduces the required bond dimensions and leads to more compact tensor representations when the structures for different variables are correlated at the same scales. If the same bits for different variables are merged into higher-order tensor blocks, we refer to the method as the bit-fused version of QTCI. In both approaches, the error of the tensor train representation is controlled by the maximum bond dimension between the different blocks.

In addition to the above QTCI scheme, as implemented in the xfac library~\cite{fernandez2024learning}, alternative encoding strategies can also be considered in principle. One such variant is to first decompose each variable into its binary digits, then arrange the bits of each individual variable consecutively (i.e., bit-by-bit within each variable), and finally concatenate the variables. 
This variant will be called the variable-separated QTCI scheme. The variable-separated approach may be advantageous in scenarios where the function exhibits strong intra-variable correlations rather than inter-variable bit correlations. 

\begin{figure}[t]
  \centering
  \includegraphics[width=0.49\textwidth]{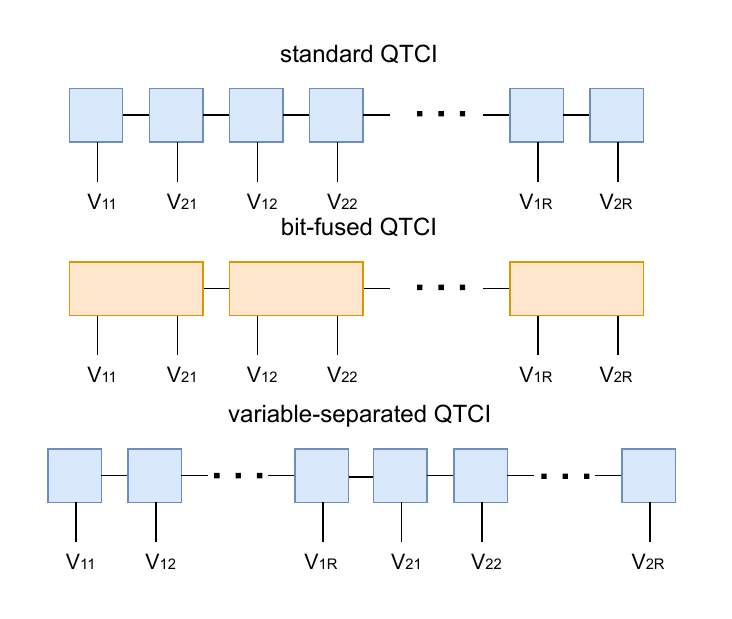}
  \caption{Illustration of three different QTCI-based encoding schemes for a two-variable function, where each variable $v_i$ ($i=1,2$) is represented by its binary digits $v_{ij}$.}
  \label{fig5}
\end{figure}

A schematic comparison of the bit arrangements for the different methods -- standard QTCI, bit-fused QTCI, and the proposed variable-separated QTCI -- is illustrated in Fig.~\ref{fig5}. To evaluate the efficiency of the different encoding schemes, we benchmarked their performance in a realistic context, under the same maximum bond dimension constraint. Among the three, the standard QTCI method enabled the fastest computations, while the bit-fused QTCI yielded the highest accuracy in reconstructing the target function. In contrast, the variable-separated QTCI, at least when benchmarked for the OCA solver, failed to consistently converge to the correct result. Although it could reproduce the overall structure of the target function, it struggled to accurately fit points that are far from the initial pivot set. Based on these observations, we adopt the bit-fused QTCI method in the rest of this work.

\subsubsection{Parametrization of the integrand}

Although we have determined how to decompose the variables into binary digits and how to arrange them in the tensor train, the choice of variables to parameterize the Feynman diagrams still needs to be discussed. While the diagrammatic structure determines the number of independent integration variables, these variables can be transformed via linear combinations into new sets of coordinates. The specific choice of coordinates can significantly affect the structure of the function to be tensorized, and thereby determine whether it can be efficiently represented by a low-rank tensor train.

We considered three types of parameterizations. In all three cases, we fix the starting operator at the rightmost point of the backward branch on the ring-shaped Keldysh contour. Below, we present detailed descriptions of these three schemes:

\begin{enumerate}
  \item \textbf{Scheme I:} We regard the entire ring-shaped contour as a single time line, beginning at the starting operator. Each movable diagrammatic vertex is then assigned a variable $v_i$, defined as the time interval between that vertex and its predecessor along the time line. Each $v_i \in [0, 2t_{\max}]$. This scheme corresponds to the parameterization used in Ref.~\cite{kim2025strong}.

  \item \textbf{Scheme II:} We first classify diagrams based on the number of operators on the forward branch, resulting in $2N - 1$ categories for an $N $-th order diagram. Within each category, each movable vertex is assigned a variable $v_i$, defined as its temporal distance to the the right-adjacent vertex or endpoint on the same branch. Each $v_i \in [0, t_{\text{max}}]$. This scheme is called cyclic parametrization in Ref.~\cite{eckstein2024solving}.

  \item \textbf{Scheme III:} All operators are first sorted by physical time, which is independent of the branch, and then categorized according to whether they reside on the forward or backward branch, resulting in $2^{N-1}$ categories. Within each class, the variables $v_i$ are defined as distances to the next operator in physical time. Again, each \( v_i \in [0, t_{\text{max}}] \). This scheme is called Keldysh parametrization in Ref.~\cite{eckstein2024solving}.
\end{enumerate}

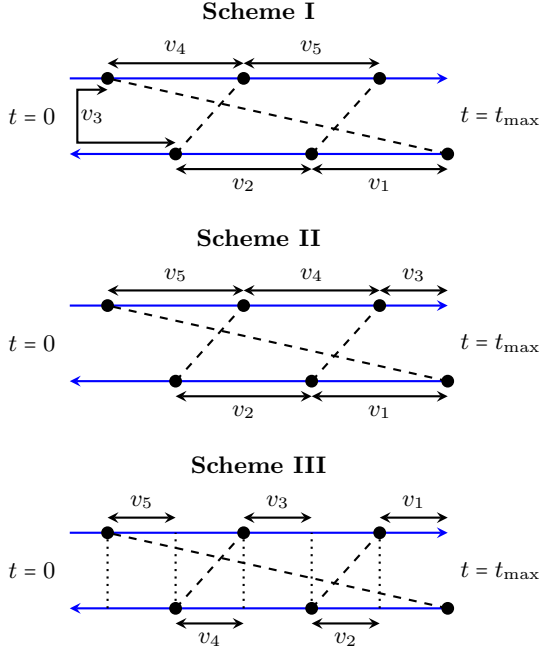
\begin{figure}[t]
  \centering
    \begin{tikzpicture}[thick, >=stealth]
    \node at (2.5,1.4) {\textbf{Scheme I}};
    \draw[->,blue] (0,0.5) -- (5,0.5);
    \draw[->,blue] (5,-0.5) -- (0,-0.5);

    \draw[dashed] (5,-0.5) -- (0.5,0.5);
    \draw[dashed] (3.2,-0.5) -- (4.1,0.5);
    \draw[dashed] (1.4,-0.5) -- (2.3,0.5);
    \node at (5.7,0) {$t=t_{\max}$};
    \node at (-0.5,0) {$t=0$};
    \filldraw (5,-0.5) circle (2pt);
    \filldraw (3.2,-0.5) circle (2pt);
    \filldraw (1.4,-0.5) circle (2pt);
    \filldraw (0.5,0.5) circle (2pt);
    \filldraw (2.3,0.5) circle (2pt);
    \filldraw (4.1,0.5) circle (2pt);
    \draw[<->, thick] (3.2,-0.7) -- node[below] {\(v_1\)} (5.0,-0.7);
    \draw[<->, thick] (1.4,-0.7) -- node[below] {\(v_2\)} (3.2,-0.7);
    \draw[<->, thick] (0.5, 0.35) -- (0.1, 0.35) -- (0.1, -0.35) -- (1.4, -0.35);
    \node at (0.3,0) {\(v_3\)};
    \draw[<->, thick] (0.5,0.7) -- node[above] {\(v_4\)} (2.3,0.7);
    \draw[<->, thick] (2.3,0.7) -- node[above] {\(v_5\)} (4.1,0.7);
    \end{tikzpicture}

    \vspace{0.25cm}

    \begin{tikzpicture}[thick, >=stealth]
    \node at (2.5,1.4) {\textbf{Scheme II}};
    \draw[->,blue] (0,0.5) -- (5,0.5);
    \draw[->,blue] (5,-0.5) -- (0,-0.5);

    \draw[dashed] (5,-0.5) -- (0.5,0.5);
    \draw[dashed] (3.2,-0.5) -- (4.1,0.5);
    \draw[dashed] (1.4,-0.5) -- (2.3,0.5);
    \node at (5.7,0) {$t=t_{\max}$};
    \node at (-0.5,0) {$t=0$};
    \filldraw (5,-0.5) circle (2pt);
    \filldraw (3.2,-0.5) circle (2pt);
    \filldraw (1.4,-0.5) circle (2pt);
    \filldraw (0.5,0.5) circle (2pt);
    \filldraw (2.3,0.5) circle (2pt);
    \filldraw (4.1,0.5) circle (2pt);
    \draw[<->, thick] (3.2,-0.7) -- node[below] {\(v_1\)} (5.0,-0.7);
    \draw[<->, thick] (1.4,-0.7) -- node[below] {\(v_2\)} (3.2,-0.7);
    \draw[<->, thick] (4.1,0.7) -- node[above] {\(v_3\)} (5,0.7);
    \draw[<->, thick] (0.5,0.7) -- node[above] {\(v_5\)} (2.3,0.7);
    \draw[<->, thick] (2.3,0.7) -- node[above] {\(v_4\)} (4.1,0.7);
    \end{tikzpicture}

    \vspace{0.25cm}

    \begin{tikzpicture}[thick, >=stealth]
    \node at (2.5,1.4) {\textbf{Scheme III}};
    \draw[->,blue] (0,0.5) -- (5,0.5);
    \draw[->,blue] (5,-0.5) -- (0,-0.5);

    \draw[dashed] (5,-0.5) -- (0.5,0.5);
    \draw[dashed] (3.2,-0.5) -- (4.1,0.5);
    \draw[dashed] (1.4,-0.5) -- (2.3,0.5);
    \draw[dotted] (4.1,-0.5) -- (4.1,0.5);
    \draw[dotted] (3.2,-0.5) -- (3.2,0.5);
    \draw[dotted] (2.3,-0.5) -- (2.3,0.5);
    \draw[dotted] (1.4,-0.5) -- (1.4,0.5);
    \draw[dotted] (0.5,-0.5) -- (0.5,0.5);
    \node at (5.7,0) {$t=t_{\max}$};
    \node at (-0.5,0) {$t=0$};
    \filldraw (5,-0.5) circle (2pt);
    \filldraw (3.2,-0.5) circle (2pt);
    \filldraw (1.4,-0.5) circle (2pt);
    \filldraw (0.5,0.5) circle (2pt);
    \filldraw (2.3,0.5) circle (2pt);
    \filldraw (4.1,0.5) circle (2pt);
    \draw[<->, thick] (4.1,0.7) -- node[above] {\(v_1\)} (5.0,0.7);
    \draw[<->, thick] (3.2,-0.7) -- node[below] {\(v_2\)} (4.1,-0.7);
    \draw[<->, thick] (2.3,0.7) -- node[above] {\(v_3\)} (3.2,0.7);
    \draw[<->, thick] (1.4,-0.7) -- node[below] {\(v_4\)} (2.3,-0.7);
    \draw[<->, thick] (0.5,0.7) -- node[above] {\(v_5\)} (1.4,0.7);
    \end{tikzpicture}

    \caption{Sketch of the three parameterization schemes, taking one of the TOA self-energy diagrams in Fig.~\ref{fig2} as an example. Since for each scheme the target function is defined as the sum over different hybridization directions, we do not indicate explicit directions for the hybridization lines.
    }
  \label{fig6}
\end{figure}

As a concrete example, we illustrate in Fig.~\ref{fig6} all three parameterization schemes using one configuration of a TOA self-energy diagram. 

In the discussion above, we have mainly focused on the operator ordering and time variable definitions in the three parameterization schemes. However, in actual diagrammatic evaluations, hybridization lines must also be included. For each operator configuration class defined in the three schemes, there are four distinct ways to connect the hybridization functions at the TOA level, corresponding to the four skeleton diagrams illustrated in Figs.~\ref{fig2} and~\ref{fig3}. Each hybridization line carries a specific direction, but in our implementation, the target function is defined as the sum over all possible hybridization directions, thereby simplifying the parameterization. 

In practice, we fit and evaluate the integrals for different topologies individually, and then add their contributions in frequency space. This procedure reduces the complexity of each fit, keeps the bond dimension low, and correctly accounts for the distinct weight functions associated with different diagrams. Taking Scheme III for model \eqref{eq2} as an example, we have four pseudo-particle flavors. At the TOA level, for each of the $2^5 = 32$ categories of operator configurations, there exist four distinct diagram topologies, leading to a total of $4 \times 4 \times 32 = 512$ target functions to process. For the TOA physical Green's function, we have two spin components, resulting in 256 target functions in total. Note that we do not manually distinguish between the lesser and greater components; instead, both are handled within a unified framework, where the specific component is determined by the location of the end operator on the cyclic contour, i.~e., whether it lies on the forward or backward branch. In Schemes II and III, it is straightforward to observe that only one class of configurations contributes to the greater component, while all others contribute exclusively to the lesser component. 

All of the target functions mentioned above are independently fitted using QTCI and then used to obtain the corresponding frequency-domain results through a weighted summation. The explicit weighting and summation formulas used in this process will be introduced in the next subsection. Once the frequency-domain results are obtained, we first sum over the contributions from the four topological classes of diagrams. On top of this, the $2^{N-1}$ lesser components corresponding to different operator configurations are added up to obtain the total lesser component.

It is important to note that the numerically defined domain of the target functions forms a simplex which satisfies the condition $\sum^{}_{i}v_i<t_{\text{max}}$, whereas the QTCI variable domain must be a hypercube with $0<v_i<t_{\text{max}}$~. 
Consequently, part of the hypercube's parameter space extend beyond the simplex. In such cases, we assign a value of zero to the corresponding function entries.

In practice, the integrand functions are composed of hybridization lines, pseudo-particle Green's functions and the matrix elements $F$. Since the hybridization functions typically decay much faster than the pseudo-particle propagators, they dominate the decay structure of the full integrand. 
As a result, the function only has non-zero values within a reasonably confined area. That means if $t_{\max}$ is large enough, the target function can  be safely assumed to be zero beyond our finite contour's domain of definition. In the scheme I, it has non-negligible values only near the two ends of the simplex, when the operators connected to the starting operator via hybridization are close to it on both branches. In the scheme II and III, since each operator is assigned a specific branch, the support of the function is mostly concentrated in the ``corner region'' where all variables are small. To reduce the computational cost and memory consumption, and take advantage of the distinct decay scales of the pseudo-particle and physical Green's functions in the time domain, we employ different maximum time domains for the two types of Green's functions. (For the pseudo-particle self-energy functions, since they involve the hybridization functions, we align their time grids with those of the physical Green's functions.) 
In our calculations, the $t_\text{max}$ for the physical Green's functions is chosen to be $1/32$ of that used for the pseudo-particle Green's functions. 
Within this shorter time window, the physical Green's functions can be safely assumed to decay to zero. Since this choice leads to different frequency resolutions for the two types of Green's functions in the frequency domain, we employ cubic spline interpolation to resample the frequency-domain physical Green's functions onto the same frequency grid as the pseudo-particle Green's functions, facilitating the relevant calculations.

In the TCI algorithm, initial pivots are required to start the fitting process. Due to the multi-centered structure of the non-zero regions in Scheme I, it is necessary in practice to manually place multiple initial pivots at different locations in order to ensure that the TCI algorithm can accurately capture the function throughout the domain~\cite{kim2025strong}. In contrast, in Schemes II and III, where the function is largely localized near the origin, placing the first pivot at the origin is typically sufficient for an accurate TCI representation.

\subsubsection{Practical implementation}

In a NESS, the self-energy or physical Green's function of interest is a function of a single-time variable. Therefore, after constructing the multi-variable tensor-train representation in the QTCI framework, we must integrate out the remaining variables to obtain the desired quantity. Taking the TOA self-energy diagram of Scheme III in Fig.~\ref{fig6} as an example, let us denote the tensor train under this parameterization by \(\sigma_1(v_1, v_2, v_3, v_4, v_5)\). (Note that this configuration represents only a subset of the lesser component contributions by this diagram topology.) Its corresponding time-domain self-energy can then be expressed as
\begin{align}
	  \Sigma_{1,\text{pp}}^<(t) =& \int_0^{t_{\max}}\hspace{-1.0em}dv_1 \int_0^{t_{\max}}\hspace{-1.0em}dv_2 \int_0^{t_{\max}}\hspace{-1.0em} dv_3 \int_0^{t_{\max}}\hspace{-1.0em} dv_4\int_0^{t_{\max}}\hspace{-1.0em} dv_5\nonumber\\
&\times \delta(t+v_1)~\sigma_1(v_1, v_2, v_3, v_4, v_5).
\label{eq:QTCI_int}
\end{align}

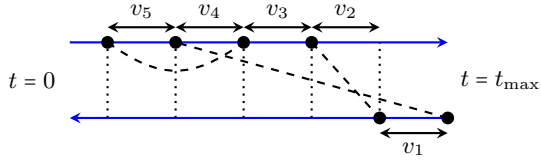
\begin{figure}[t]
  \centering
    \begin{tikzpicture}[thick, >=stealth]
    \draw[->,blue] (0,0.5) -- (5,0.5);
    \draw[->,blue] (5,-0.5) -- (0,-0.5);

    \draw[dashed] (5,-0.5) -- (1.4,0.5);
    \draw[dashed] (3.2,0.5) -- (4.1,-0.5);
    \draw[dashed] (0.5,0.5) .. controls (1.2,0) and (1.6,0) .. (2.3,0.5);
    \draw[dotted] (4.1,-0.5) -- (4.1,0.5);
    \draw[dotted] (3.2,-0.5) -- (3.2,0.5);
    \draw[dotted] (2.3,-0.5) -- (2.3,0.5);
    \draw[dotted] (1.4,-0.5) -- (1.4,0.5);
    \draw[dotted] (0.5,-0.5) -- (0.5,0.5);
    \node at (5.7,0) {$t=t_{\max}$};
    \node at (-0.5,0) {$t=0$};
    \filldraw (5,-0.5) circle (2pt);
    \filldraw (3.2,0.5) circle (2pt);
    \filldraw (1.4,0.5) circle (2pt);
    \filldraw (0.5,0.5) circle (2pt);
    \filldraw (2.3,0.5) circle (2pt);
    \filldraw (4.1,-0.5) circle (2pt);
    \draw[<->, thick] (4.1,-0.7) -- node[below] {\(v_1\)} (5.0,-0.7);
    \draw[<->, thick] (3.2,0.7) -- node[above] {\(v_2\)} (4.1,0.7);
    \draw[<->, thick] (2.3,0.7) -- node[above] {\(v_3\)} (3.2,0.7);
    \draw[<->, thick] (1.4,0.7) -- node[above] {\(v_4\)} (2.3,0.7);
    \draw[<->, thick] (0.5,0.7) -- node[above] {\(v_5\)} (1.4,0.7);
    \end{tikzpicture}
    \caption{An additional contribution to the lesser self-energy within Scheme III, illustrating an alternative operator configuration which is topologically equivalent to the Feynman diagram shown in Fig.~\ref{fig6}.
    }
  \label{fig7}
\end{figure}

In total, there are 31 contributions to the lesser component of the self-energy under Scheme III. For clarity, we show in Fig.~\ref{fig7} an alternative operator configuration that is topologically equivalent to the diagram in Fig.~\ref{fig6}. We denote this tensor train by \(\sigma_2(v_1, v_2, v_3, v_4, v_5)\), and the corresponding time-domain lesser self-energy is given by
\begin{align}
\Sigma_{2,\text{pp}}^<(t) =& \int_0^{t_{\max}}\hspace{-1.0em}dv_1 \int_0^{t_{\max}}\hspace{-1.0em}dv_2 \int_0^{t_{\max}}\hspace{-1.0em} dv_3 \int_0^{t_{\max}}\hspace{-1.0em} dv_4\int_0^{t_{\max}}\hspace{-1.0em} dv_5\nonumber\\
&\times \delta(t+v_1+v_2)~\sigma_2(v_1, v_2, v_3, v_4, v_5).
\label{eq:QTCI_int1}
\end{align}
All the contributions from these different configurations or diagrams are summed up in the end.

In the practical DMFT self-consistency loop, we are only interested in the frequency-domain self-energy $\Sigma(\omega)$, see Eqs.~\eqref{eq:dyson_a}-\eqref{eq:dyson_c}. For this reason, it is useful to incorporate the Fourier kernel into the integrand~\cite{kim2025strong}. Considering the symmetry \eqref{eq:symmetry}, this yields a five-dimensional integration,
\begin{align}
	\Sigma_\text{pp}^<(\omega) =& 2\mathrm{i}\text{Im}\bigg[\int_0^{t_{\max}}\hspace{-1.0em}dv_1 \int_0^{t_{\max}}\hspace{-1.0em} dv_2 \int_0^{t_{\max}}\hspace{-1.0em} dv_3 \int_0^{t_{\max}}\hspace{-1.0em} dv_4\int_0^{t_{\max}}\hspace{-1.0em}dv_5\nonumber\\
	&\times ~\Big(\prod_{ij} W^\omega_{ij}\Big)~\sigma(v_1, v_2, v_3, v_4, v_5)\bigg]~,
\label{eq:QTCI_int2}
\end{align}
which can be viewed as a summation over the tensor train with the weight function $W^\omega_{ij}$. The weights can be distributed across the bits of each variable according to Eq.~\eqref{eq:binary_bit}, allowing an efficient evaluation of the Fourier transformation within the QTCI framework. For the case of Eq.~\eqref{eq:QTCI_int}, the weight can be written as
\begin{equation}
	W^\omega_{ij} = \left\{
		\begin{array}{lr}
			\exp({-\mathrm{i}\omega t_{\text{max}}\frac{v_{ij}}{2^j}}) & (i=1)\\
			1 & (i \neq 1)
		\end{array}
	\right.~,
	\label{eqn:Wwij}
\end{equation}
while in the case of Eq.~\eqref{eq:QTCI_int1}, it becomes 

\begin{equation}
	W^\omega_{ij} = \left\{
		\begin{array}{lr}
			\exp({-\mathrm{i}\omega t_{\text{max}}\frac{v_{ij}}{2^j}}) & (i=1~\text{or}~i=2)\\
			1 & (i \neq 1~\text{and}~i\neq 2)
		\end{array}
	\right.~.
	\label{eqn:Wwij2}
\end{equation}

We have implemented all three parameterization schemes and systematically compared their performance. At the OCA level, all three schemes converge to the correct solution, but the required bond dimension $\chi$ differs significantly among them. Taking the paramagnetic DMFT calculation with $U=2$ and $\beta=10$ as an example, Scheme~I requires approximately $\chi \approx 150$ to achieve full convergence, whereas Scheme~II converges with $\chi \approx 50$, and Scheme~III needs only $\chi \approx 30$. In more demanding TOA-level calculations, only Scheme~III allows to reach high-accuracy solutions, with a bond dimension around $\chi \approx 50$, while the other two schemes fail to converge reliably.

These differences reflect the fact that the three schemes decompose the complex multivariable function into different numbers of subdomains with a simpler structure. From Scheme I to Scheme III, the number of such subdomains increases, resulting in a simpler function to be fitted and, hence, a reduced bond dimension required in the tensor-train representation. However, this comes at the cost of an increased number of individual tensor trains that need to be constructed and integrated. The tensor trains we need to handle are mutually independent and can be processed in parallel. Since the TCI fitting procedure does not scale efficiently with the number of CPU cores per task, we assign a single core to each tensor train. 

For TOA-level computations, the total cost of Scheme~III remains manageable. For an individual diagram of order $N$, the computational cost of the integral after fitting can be estimated as $\mathcal{O}\bigl(N_{\omega}(2N-1)R\chi^{3}\bigr)$~\cite{fernandez2024learning}, where $N_{\omega}$ is the number of sampled frequency points, $(2N-1)$ is the number of internal integrations for an $N$th-order diagram, $R$ is the number of digits in the binary encoding, and $\chi$ is the maximal bond dimension of the tensor train. In comparison, the cost of the QTCI fitting with the full-pivot option scales as $\mathcal{O}\bigl((2N-1)^{2}R\chi^{3}\bigr)$~\cite{fernandez2024learning}. This estimate provides a simple guideline for the required computational resources. At the TOA level, the QTCI fitting typically dominates the total cost due to its large prefactor and the relatively small number of sampled frequencies $N_{\omega}$.

\subsubsection{Higher-order integration schemes}

So far, we have described the use of QTCI-based strong-coupling impurity solvers within a standard DMFT loop. For paramagnetic DMFT calculations, this approach yields accurate results at a moderate computational cost. However, when applying the same method to antiferromagnetic DMFT calculations, one may encounter significant convergence problems. After careful inspection, we identified that this issue arises from the integration scheme employed in the evaluation of Eq.~\eqref{eq:QTCI_int2}. Specifically, the implementation described above (Eqs.~\eqref{eqn:Wwij} and \eqref{eqn:Wwij2}) uses the simplest rectangular rule for the numerical integrations. As mentioned earlier, the 
values of the integrand
are typically peaked near the boundaries of the domain. The rectangular rule tends to introduce large integration errors in these boundary regions, even for the small time steps employed in QTCI, and this can strongly affect the final result. 
To systematically improve the integration accuracy, especially for long-time integrals, higher-order integration schemes should be implemented. 

In practice, a simple remedy is to switch to the trapezoidal rule, which can be viewed as correcting the rectangular rule by subtracting appropriately weighted contributions from the boundary values. These corrections involve partial integrations over $(2N-2)$-, $(2N-3)$-, ..., 0-dimensional boundary surfaces. A brute-force yet practical method to obtain these boundary corrections is to reuse the original tensor train and perform multiple weighted summations, each time assigning nonzero weights only to the subregions corresponding to the relevant boundary. By repeating this process for each boundary layer and subtracting the resulting corrections from the original estimate, we obtain significantly improved results. This simple modification allows us to accurately describe antiferromagnetic DMFT solutions within the QTCI framework.

\begin{figure}[]
	\centering
	\includegraphics[width=0.49\textwidth]{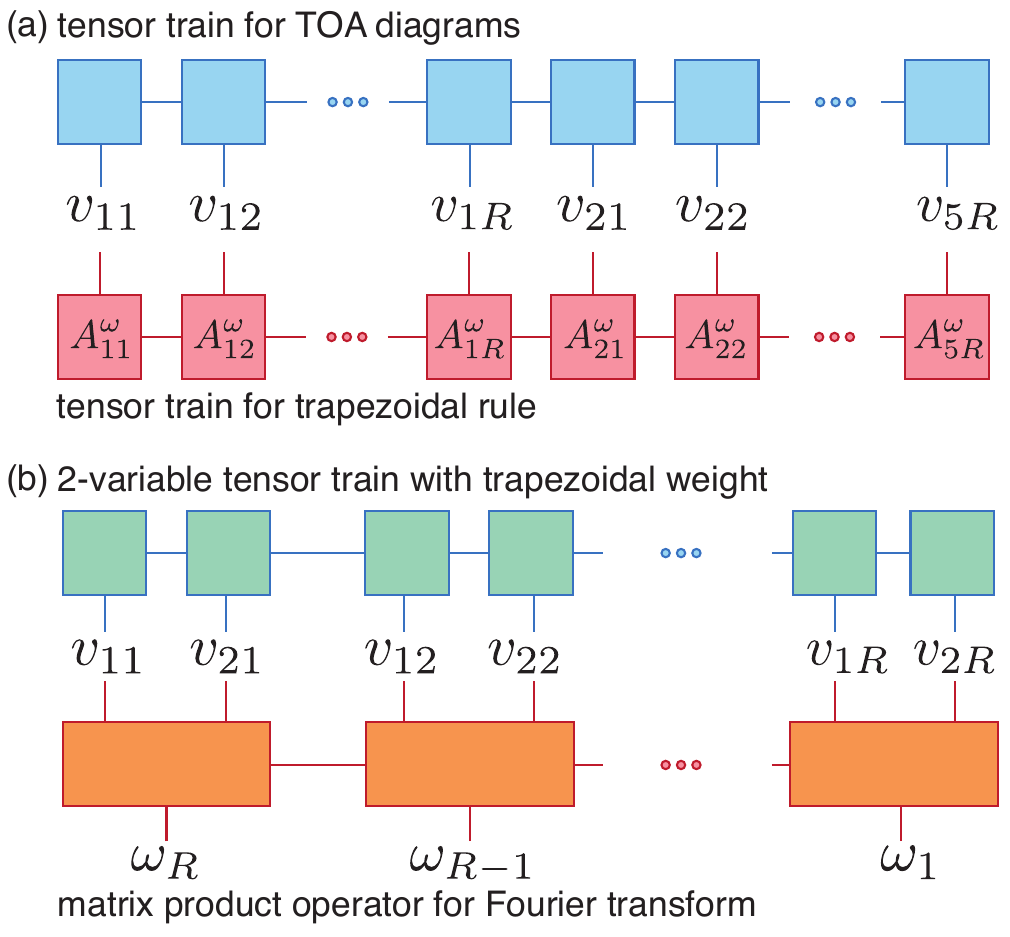}
	\caption{
		(a) Contraction between the $\sigma$ 
		TT (blue squares) and the Fourier TT with the trapezoidal rule (red squares), leading to the TOA contribution to the PP self-energy at a single fixed frequency $\omega$.    
		(b) Contraction between the $\tilde{\sigma}$ TT with trapezoidal rule (green squares) and the matrix product operator for the Fourier transform (orange squares). The resulting TT contains the full frequency dependence of the Fourier transformed function.
	}
	\label{fig:trapezoidal_mps}
\end{figure}

More elegantly, one can carry out the trapezoidal-rule integration using another QTT which implements its quadrature rule.
For example, the TOA QTT in the variable-separated representation can be summed up by contracting it with the red tensor in Fig.~\ref{fig:trapezoidal_mps}(a), whose explicit expression is 
\begin{equation}
	A^\omega_{ij} = W^\omega_{ij}\left\{
		\begin{array}{lr}
			\left( \begin{array}{ccc} 1 & -\delta_{v_{i1,0}}/2 & -\delta_{v_{i1,1}}/2 \end{array} \right) & (j=1)\\
			\left( \begin{array}{ccc} 1 & \delta_{v_{iR,0}} & \delta_{v_{iR,1}} \end{array} \right)^\intercal & (j=R)\\
			\text{diag}\left( \begin{array}{ccc} 1 & \delta_{v_{i1,0}} & \delta_{v_{i1,1}} \end{array} \right) & (\text{otherwise})
		\end{array}
	\right.~.
	\label{eqn:Awij}
\end{equation}
Since the trapezoidal rule is independently applied to each variable, the corresponding QTT can be completely factorized for the five different variables as in Eq.~(\ref{eqn:Awij}).
Within a variable, the trapezoidal rule for the quantics bits treats all-off bits and all-on bits in a special way, giving them half of the integral weight, compared to the others. The rank-3 matrices for a given variable in Eq.~(\ref{eqn:Awij}) have two additional \textit{channels} besides the simple sum that can subtract half of the integral weight for all off bits and all on bits.
So, the maximum bond dimension of 3 for the trapezoidal quadrature does not depend on the number of variables and quantics bits.
As one goes from the simple trapezoidal rule to higher-order Gregory quadrature, which involves a larger number of boundary corrections, the maximum bond dimension of the TT increases linearly with the Gregory order.

\begin{figure}[b]
	\centering
	\includegraphics[width=0.5\textwidth]{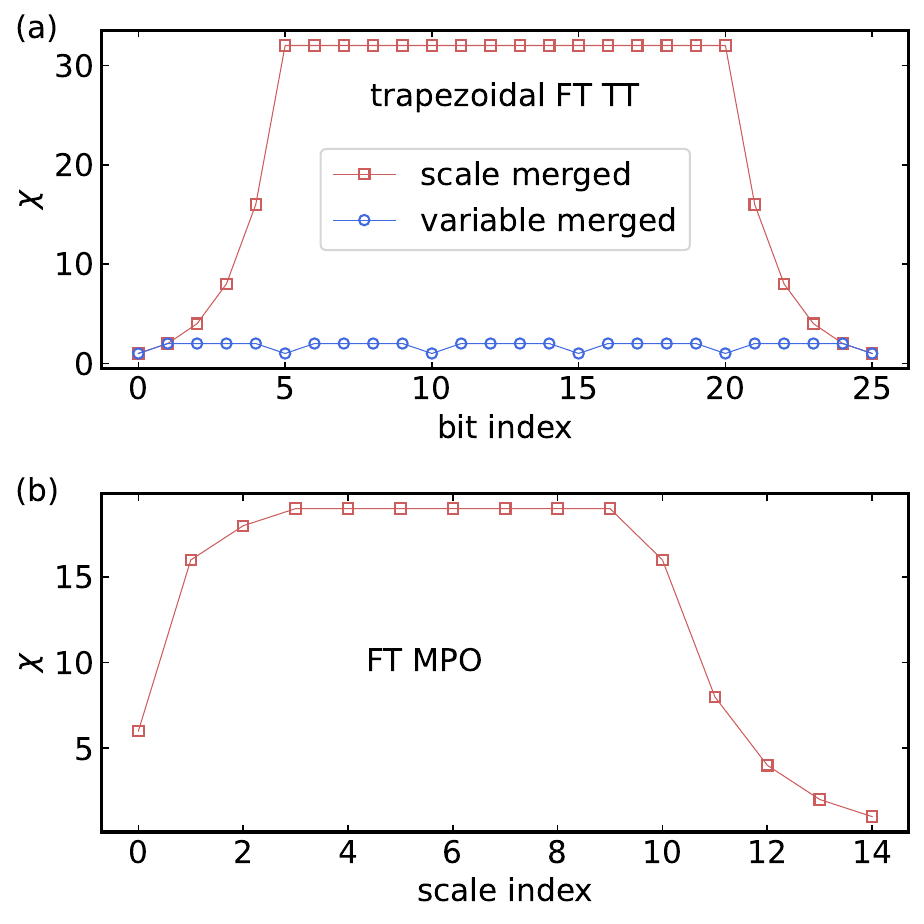}
	\caption{
		(a) Bond dimension along the 5-bit, 5-variable (TOA) Fourier TT for the trapezoidal rule. Only half of the trapezoidal rule (for all-off bits) is considered.
		(b) Bond dimension along the 15-bit, 5-variable (TOA) Fourier MPO without trapezoidal rule [orange tensor in Fig.~\ref{fig:trapezoidal_mps}(b)]. 
	}
	\label{fig:FTweightBondDim}
\end{figure}

However, in the scale-separated representation, although the Fourier TT has a saturated bond dimension, the maximum value increases exponentially as the number of variables increases. In constrast to the variable-separated scheme, where two channels (all-off or not) per variable are sequentially connected along the tensor trains, $2^{2N-1}$ simultaneous channels are possible for a given scale due to the different combinations of channels from all $2N-1$ variables.
Figure~\ref{fig:FTweightBondDim}(a) presents the bond dimension of the Fourier TT for both the variable- and scale-separated representations.
Here, we only consider the trapezoidal rule weights for the all-off bits, since the target function is assumed to be negligibly small for all-on bits.
The variable-separated case shows a maximum bond dimension of 2 within each variable and 1 between different variables, indicating complete factorization,
while the scale merged case has an exponentially increasing (decreasing) bond dimension for the first (last) 5 carriages (due to the number of variables) of the tensor train, resulting in a maximum bond dimension of $2^{2\times 3-1}=32$.
Although in the TOA the computational cost for the trapezoidal rule might still be manageable, considering the Kronecker product with the original TT, the generalization of this scheme to higher-order approximations will be numerically challenging, due to the exponentially increasing bond dimensions.

We found that this exponentially increasing bond dimension of the Fourier TT can be mitigated by transferring the trapezoidal rule weights to the $\sigma$ function 
and introducing the matrix product operator (MPO) for the Fourier transform, as shown in Fig.~\ref{fig:trapezoidal_mps}(b).
In the scale-separated representation, the Fourier kernel (without trapezoidal rule) from multiple time variables $\{v_i\}$ to a single frequency $\omega$ can be efficiently decomposed into an MPO with low bond dimensions. 
Although the Fourier kernel has a different form depending on the vertex configuration in the parametrization scheme III [Fig.~\ref{fig6}], the maximum bond dimension remains below $\sim 19$, independent of the number of train carriages [Fig.~\ref{fig:FTweightBondDim}(b)].
The trick is to pair the variable of the larger frequency scale with the one of the smaller time scale~\cite{fernandez2024learning}.

Now the trapezoidal weight should be assigned to the function $\sigma$ leading to a new target function to be QTCI decomposed:
$\tilde{\sigma}(v_1,v_2,v_3,v_4,v_5) = \left\{\prod_{i}(1-[\prod_j\delta_{v_{ij},0}]/2)\right\}\sigma(v_1,v_2,v_3,v_4,v_5)$.
Due to the abrupt changes at the boundary of the $\tilde{\sigma}$ function, it is natural that the QTCI requires larger bond dimensions for the same accuracy.
But it turns out that the increase in the bond dimensions is modest, compared to the exponential increase in the Fourier TT.
Furthermore, by introducing an MPO, the Fourier transformation itself can be performed in an efficient way.

\section{Results}\label{sec3}
\subsection{Single-Impurity Anderson Model}\label{sec3a}

\begin{figure}[htbp]
  \centering
  \includegraphics[width=0.49\textwidth]{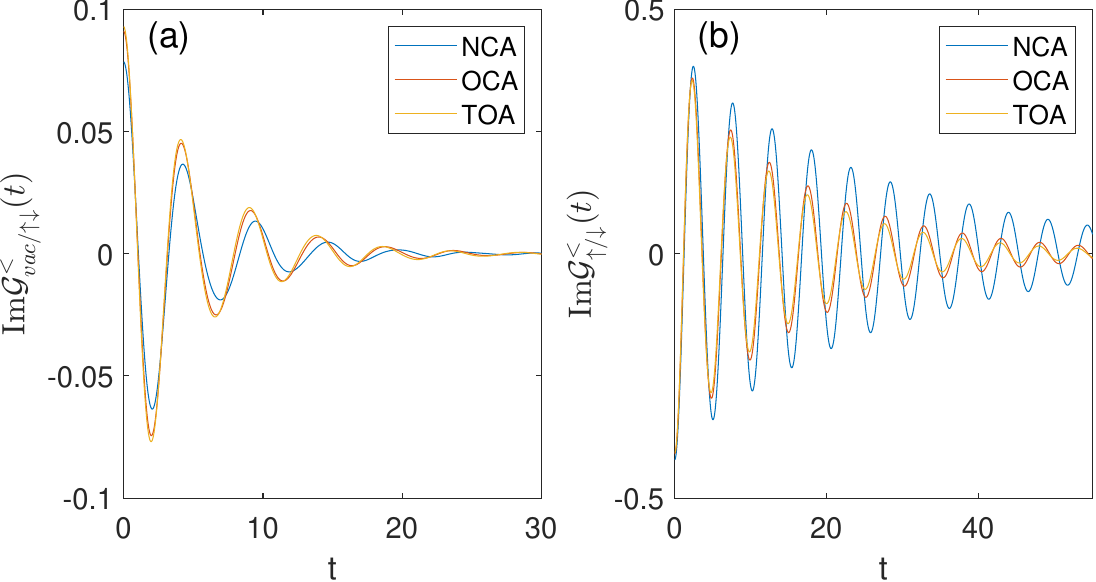}
  \caption{(a) Time evolution of the imaginary part of the lesser pseudo-particle Green's function $\mathcal{G}^<(t)$ for the doubly occupied and empty channels. (b) Time evolution of the imaginary part of the lesser pseudo-particle Green's function $\mathcal{G}^<(t)$ for the spin-up and spin-down channels. In both panels, results are shown for NCA, OCA and TOA.}
  \label{fig:AIM_ppg}
\end{figure}

To study the performance of the strong-coupling impurity solvers introduced in the previous section, we first consider a simple test case: a single impurity coupled to a semicircular noninteracting bath. Compared to the full DMFT framework, this setup is significantly simpler, as it does not require a self-consistent evaluation of the physical Green's function. As such, it provides a convenient way for assessing both the impact of higher-order strong-coupling corrections and different bond dimensions in the QTCI representation.

We adopt the same model parameters as in Ref.~\cite{kim2025strong}, which reported numerical results for the NCA and OCA. Specifically, we use the interaction strength $U = 2$, temperature $T = 0.1$, and chemical potential $\mu = 1$. The bath has a semicircular density of states with half-bandwidth 1, and the impurity-bath coupling is set to $g = 0.5$, leading to a hybridization function $\Delta^R(\omega) = g^2 \int d\epsilon\, \frac{\rho_{\text{bath}}(\epsilon)}{\omega - \epsilon+\mathrm{i}0_+}$. 
The numerical calculations are performed with a time step of $\Delta t = 0.01227$ and $2^{21}$ time grid points ($t_\text{max}=25732$ for the pseudo-particle Green's function), which ensures that both the physical and pseudo-particle Green's functions decay sufficiently to zero at long times. The bond dimension used in the QTT compression is set to $\chi = 30$, which provides a good balance between accuracy and computational efficiency. In this work, we compute results using the TOA and compare them to the previously reported NCA and OCA data.

Fig.~\ref{fig:AIM_ppg} presents the time evolution of the imaginary parts of the lesser pseudo-particle Green's functions for each channel, as obtained from NCA, OCA, and TOA calculations. Due to spin symmetry, the spin-up and spin-down Green's functions are identical, as are those for the empty and doubly occupied pseudo-particles. In the chosen parameter regime, higher-order self-energy corrections suppress the amplitude of the spinful pseudo-particle Green's functions while enhancing the amplitude of the empty and doubly occupied ones, indicating a redistribution of spectral weight among the pseudo-particle channels. It is also evident that the TOA correction is significantly smaller than the OCA correction, highlighting the rapid convergence of the strong-coupling approach with higher-order terms in this regime.

\begin{figure}[t]
  \centering
  \includegraphics[width=0.49\textwidth]{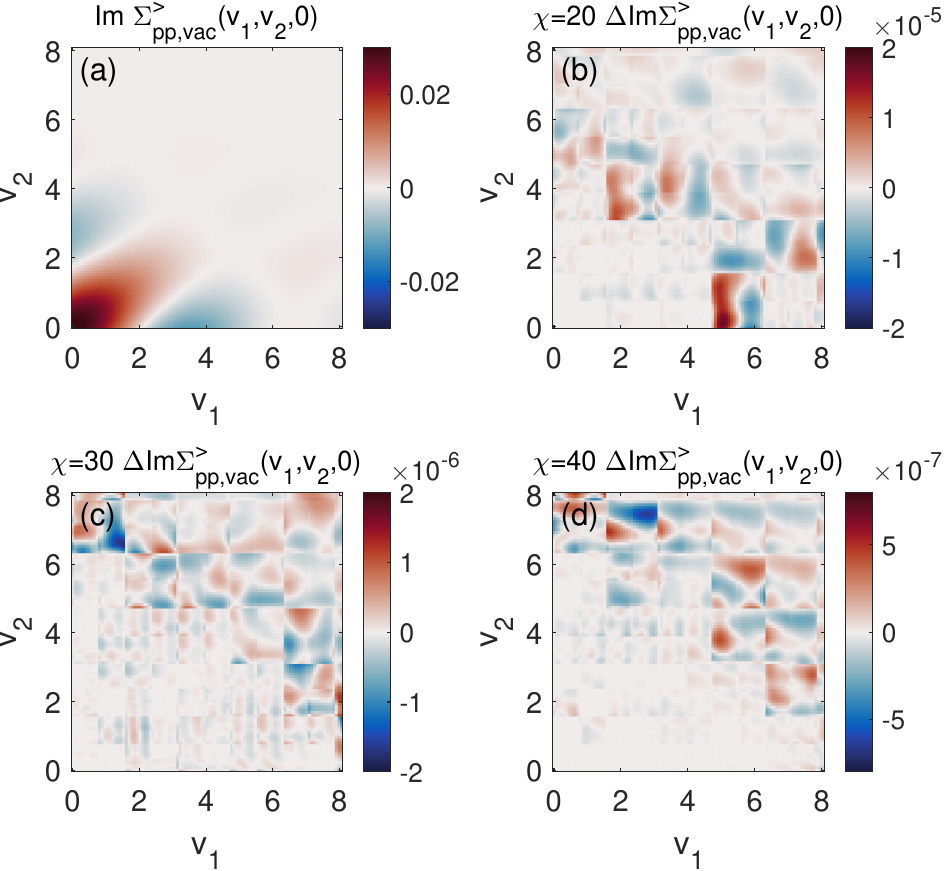}
  \caption{Test of the fitting accuracy for a OCA pseudo-particle self-energy slice. 
	  (a) Target function $\text{Im} \Sigma_{\text{pp},\mathrm{vac}}^>(v_1,v_2,0)$ in the space of $v_1$ and $v_2$. (b)--(d) Fitting error $\Delta \text{Im} \Sigma_{\text{pp},\mathrm{vac}}^>(v_1,v_2,0)=\text{Im} \Sigma_{\text{pp},\mathrm{vac}}^{>,\text{QTCI}}(v_1,v_2,0) - \text{Im} \Sigma_{\text{pp},\mathrm{vac}}^>(v_1,v_2,0)$ between the original function and its QTT fitting results with bond dimensions $\chi = 20$, 30, and 40, respectively.
    }
  \label{fig:AIM_SE_dif_OCA}
\end{figure}

\begin{figure}[t]
  \centering
  \includegraphics[width=0.49\textwidth]{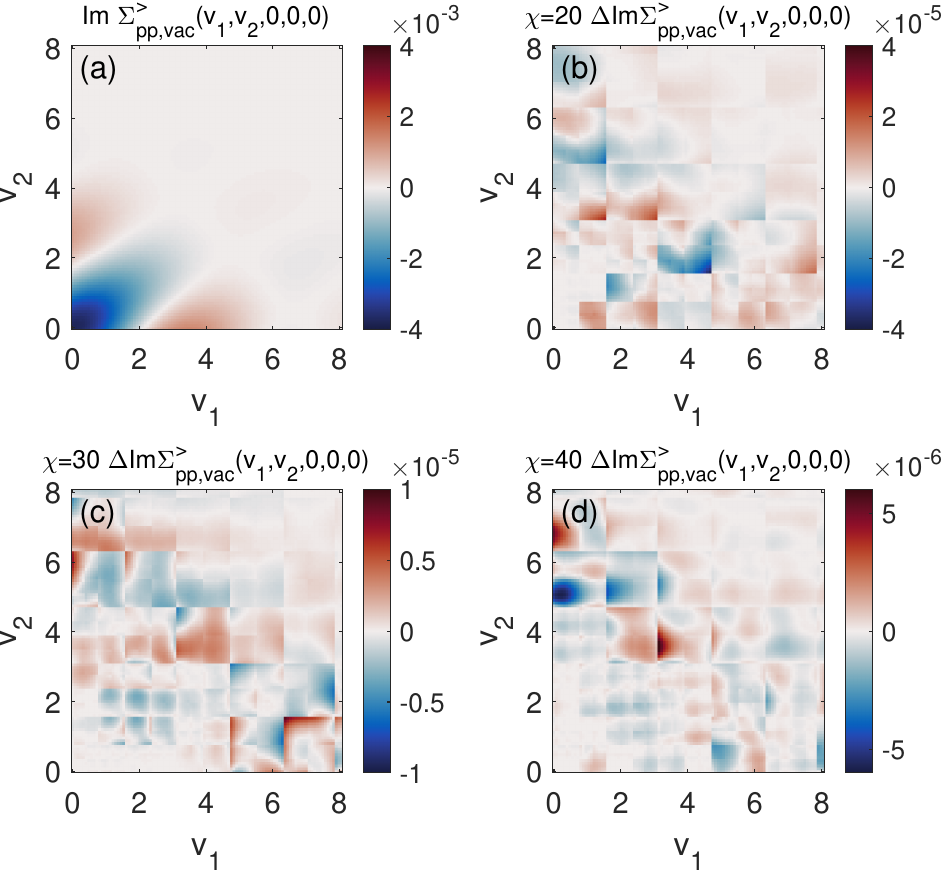}
  \caption{Test of the fitting accuracy for a TOA pseudo-particle self-energy slice (contribution from one diagram only). 
    (a) Target function $\text{Im} \Sigma_{\text{pp},\mathrm{vac}}^>(v_1,v_2,0,0,0)$ in the space of $v_1$ and $v_2$. (b)--(d) Fitting error $\Delta \text{Im} \Sigma_{\text{pp},\mathrm{vac}}^>(v_1,v_2,0,0,0)$ between the original function and its QTT fitting results with bond dimensions $\chi = 20$, 30, and 40, respectively.
    }
  \label{fig:AIM_SE_dif_TOA}
\end{figure}

To assess the effect of the QTT bond dimension on the accuracy of the pseudo-particle self-energy representation, we perform one-shot QTT reconstructions of the full self-energy for both the OCA and TOA. 
For the OCA fitting process, we first obtain a fully converged NCA solution, and then compute the OCA greater pseudo-particle self-energy, using the NCA pseudo-particle Green's functions and hybridization function as inputs. The exact result for a slice $\text{Im} \Sigma_{pp,\text{vac}}^>(v_1,v_2,0)$, along with the absolute difference between the exact and QTT-reconstructed functions for various bond dimensions, is shown in Fig.~\ref{fig:AIM_SE_dif_OCA}. A similar procedure is followed for the TOA fitting process, where we start from a converged OCA solution and consider only one of the four TOA diagrams (the bottom-left diagram of Fig.~\ref{fig2}). The results for the TOA greater pseudo-particle self-energy slice $\text{Im} \Sigma_{\text{pp},\text{vac}}^>(v_1,v_2,0,0,0)$ are shown in Fig.~\ref{fig:AIM_SE_dif_TOA}. Throughout this analysis, scheme~III is employed for the QTT fitting, ensuring that each block of the self-energy can be represented with high accuracy, even at moderate bond dimensions.

Let us first analyze the similarities between Fig.~\ref{fig:AIM_SE_dif_OCA} and Fig.~\ref{fig:AIM_SE_dif_TOA}. In both cases, the fitting error decreases as the bond dimension $\chi$ increases. Furthermore, the errors tend to be smaller in regions where the target self-energy functions have larger amplitudes,  
likely because the fitting algorithm preferentially places the pivots there.
Overall, the fitting accuracy reaches a satisfactory level even with moderate bond dimensions, demonstrating the robustness and efficiency of the QTCI-based implementation in capturing complex, high-dimensional time-dependent structures.

In addition to these similarities, notable differences also emerge. Specifically, for a fixed $\chi$, the relative error in the TOA is larger than for the OCA. And the rate at which the fitting error decreases with increasing $\chi$ is slower for the TOA than for the OCA. This behavior reflects the increased complexity of the TOA self-energy integrands, which poses greater challenges for low-rank tensor compression.

\subsection{DMFT: Paramagnetic Phase}\label{sec3b}

In contrast to the previous subsection, where we focused on an impurity model with a fixed bath, we now turn to the lattice problem described by the Hubbard model. Throughout this section, we consider paramagnetic equilibrium states, since this provides the most common and conceptually simple setting for DMFT. The Hubbard model is mapped onto a self-consistent impurity problem via DMFT as described in Sec.~\ref{sec2a}, and we solve this impurity problem using strong-coupling impurity solvers of various expansion orders. We set the rescaled hopping amplitude to $v = 0.5$, corresponding to a semicircular density of states with a half-bandwidth of unity in the non-interacting case.

\begin{figure}[t]
  \centering
  \includegraphics[width=0.49\textwidth]{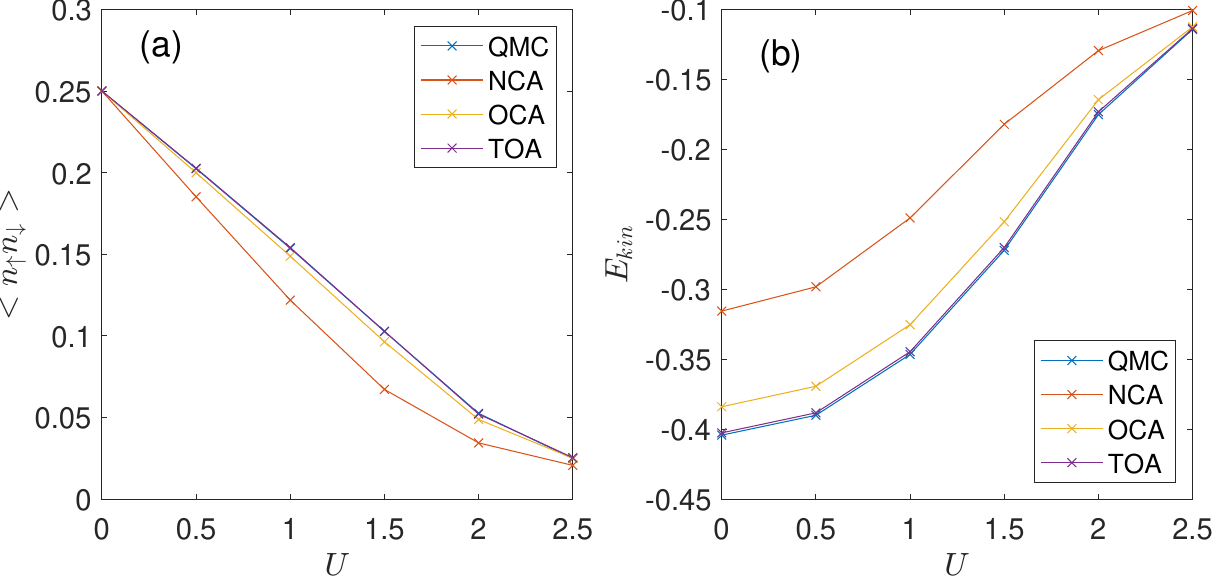}
  \caption{Comparison of the double occupancy and kinetic energy obtained using the strong-coupling impurity solvers and QMC simulations at inverse temperature $\beta = 10$. Panel (a) shows the double occupancy as a function of interaction strength $U$, while panel (b) presents the kinetic energy. }
  \label{fig:energy_convergence}
\end{figure}

To check the validity of our strong-coupling solvers, it is instructive to consider physical observables that are directly accessible using QMC on the imaginary axis, which hence can provide reliable benchmarks without the need for analytic continuation. In particular, we compare the energy results to assess the quantitative accuracy of the strong-coupling solvers. For the Hubbard model, the total energy can be decomposed into potential ($E_{\text{pot}}$) and kinetic ($E_{\text{kin}}$) contributions. The potential energy is directly related to the double occupancy, and can be expressed as $E_{\text{pot}} = U \langle n_{\uparrow} n_{\downarrow} \rangle$, where the double occupancy $\langle n_{\uparrow} n_{\downarrow} \rangle$ can be extracted from the corresponding pseudo-particle Green's function in our framework. The kinetic energy $E_{\text{kin}}$, on the other hand, is determined by the physical Green's function. It can be evaluated via a reformulation of Eq.~(112) in Ref.~\cite{aoki2014nonequilibrium},
\begin{equation}
E_{\text{kin}} = \int_{-\infty}^{\infty} d\omega\, \text{Re}\Delta^R(\omega) \,f(\omega) A(\omega),
\label{eq:kinetic_energy}
\end{equation}
where $A(\omega)$ is the spectral function, $f(\omega)$ is the Fermi-Dirac distribution and the hybridization function $\Delta(\omega)$ can be obtained from the physical Green's function via Eq.~\eqref{eq7}.

In Fig.~\ref{fig:energy_convergence}(a) and (b), we compare the double occupancy and kinetic energy, obtained using NCA, OCA, and TOA, with the QMC results, for the range of interaction strengths $0\le U \le 2.5$ at fixed inverse temperature $\beta = 10$. For the double occupancy, we observe that the deviation between NCA/OCA and QMC is not monotonic: as $U$ decreases from $U=2.5$, the discrepancy initially grows and then decreases. This behavior arises because in the non-interacting limit, all methods yield the same value $ \langle n_{\uparrow} n_{\downarrow} \rangle= 0.25$, since all four pseudo-particles have the same occupation. Notably, TOA shows excellent agreement with QMC across the entire interaction range, with deviations too small to be discerned visually in the figure. 

\begin{figure}[]
  \centering
  \includegraphics[width=0.49\textwidth]{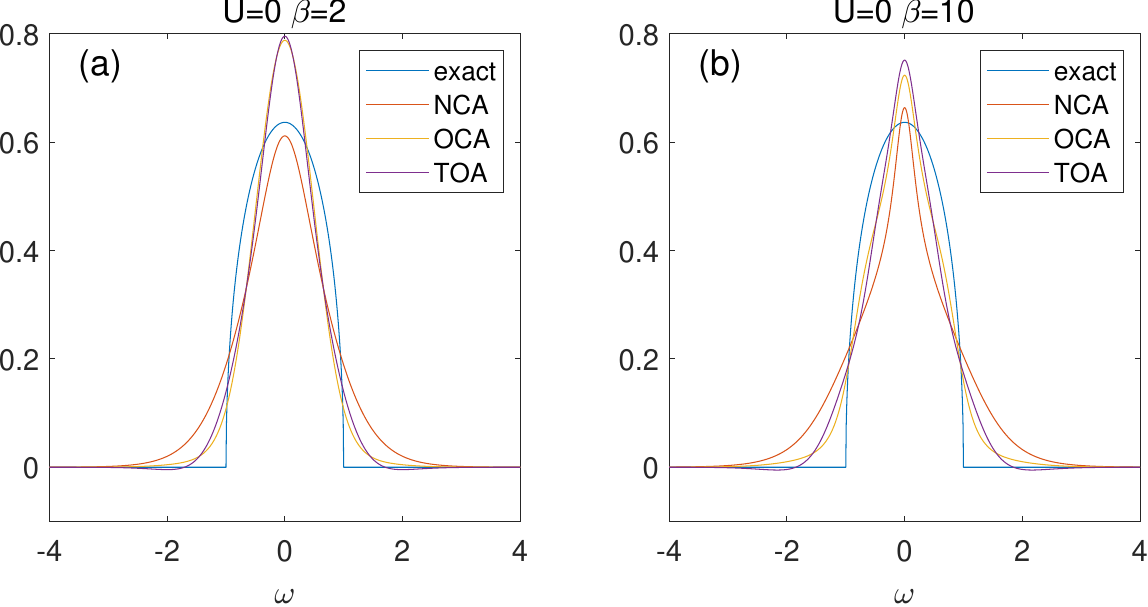}
  \caption{Comparison of the spectral functions from the strong-coupling solvers to the exact semi-elliptical one at inverse temperature $\beta = 2$ (panel (a)) and $\beta = 10$ (panel (b)). The interaction strength is fixed as $U=0$.}
  \label{fig:U0_spectrum}
\end{figure}

For the kinetic energy shown in Fig.~\ref{fig:energy_convergence}(b), the deviation between NCA/OCA and QMC increases monotonically as $U$ decreases, although the rate of increase remains relatively modest. Both OCA and TOA significantly improve upon NCA, and TOA remains in close agreement with the QMC results even in the weak-coupling limit. 

It is important to note, however, that a good agreement for the energy does not necessarily imply an accurate reproduction of the spectral function. As shown in Fig.~\ref{fig:U0_spectrum}, the TOA spectral function at $U = 0$ still deviates significantly from the exact non-interacting spectral function $A(\omega)=1/(2\pi v^2)\sqrt{4v^2-\omega^2}$ for both low and high temperatures, and that it even exhibits unphysical negative values in the spectral tails. It is well known that the NCA can yield unphysical results which violate causality in the weak-correlation regime~\cite{eckstein2010nonequilibrium,Pruschke1993}. Apparently, similar problems also plague the higher-order implementations of the self-consistent strong-coupling expansion. 

The observation of consistent energies and non-consistent spectral functions underscores an important point: integrated quantities can appear deceptively well converged even when the underlying spectral features are not faithfully captured. This highlights the importance of not relying solely on integrated quantities when benchmarking impurity solvers; instead, one should also examine frequency-dependent physical quantities such as the spectral function.

As shown in Fig.~\ref{fig:U0_spectrum}, our spectral functions exhibit substantial deviations from the exact semi-elliptical result in the weak-interaction limit, including regions of unphysical negative spectral weight. Such behavior can be interpreted as a manifestation of the breakdown of Fermi-liquid (FL) behavior in self-consistent strong-coupling expansions such as NCA and OCA. 

It is well established that for the particle-hole symmetric Anderson impurity model, the NCA solution is qualitatively incorrect at any temperature and effectively describes an overscreened Kondo effect, leading to a residual impurity entropy~\cite{sposetti2016qualitative}. While OCA partially remedies this pathology and restores the correct ground-state entropy, it still exhibits deviations from FL predictions at low temperatures and in the weak-coupling regime. As emphasized in Refs.~\cite{kuramoto1984self,muller1984self}, the charge-fluctuation regime is particularly problematic: deviations from FL theory become severe, and vertex corrections—neglected in NCA and OCA—develop singular contributions that spoil the $1/N$ expansion. 

Our findings are consistent with these earlier observations. A possible route to address these deficiencies is the inclusion of infinite classes of vertex corrections, as achieved in the conserving T-matrix approximation (CTMA), which has been shown to recover FL properties in the Anderson impurity model~\cite{Kroha1997,kroha2004fermi,kirchner2002self}. Whether CTMA or similar resummation techniques can serve as reliable impurity solvers within DMFT for correlated lattice problems remains an interesting question for future investigations.

\begin{figure*}[t]
  \centering
  \includegraphics[width=0.8\textwidth]{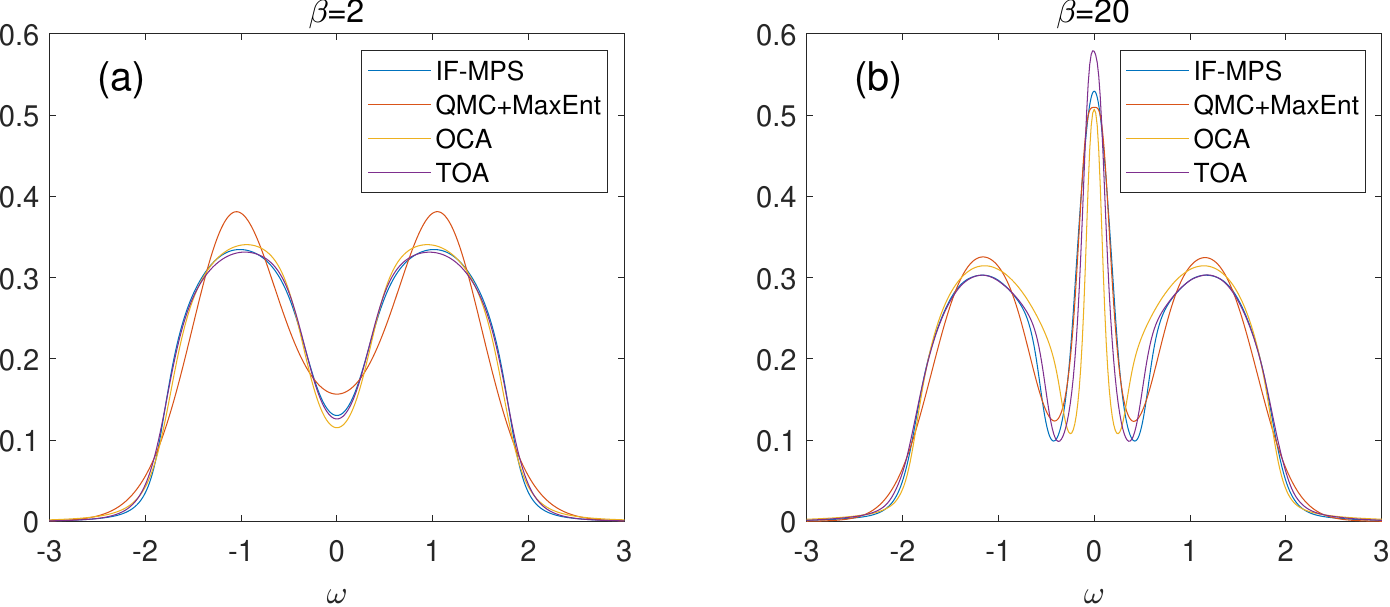}
  \caption{Comparison of the paramagnetic spectral functions obtained from various methods: QMC+MaxEnt, IF-MPS and our steady-state strong-coupling solvers (OCA and TOA). The interaction strength is fixed at $U = 2$, while panels (a) and (b) show results at inverse temperatures $\beta = 2$ and 20, respectively.
  }
  \label{fig:method_comparison}
\end{figure*}

To examine the validity of our higher-order strong-coupling solvers for the spectral function of a correlated metallic system, we show in Fig.~\ref{fig:method_comparison} the OCA and TOA spectral functions at fixed $U=2$ and inverse temperatures $\beta = 2$ and $20$. These results are benchmarked against data obtained using the influence functional matrix product state (IF-MPS) method in Ref.~\cite{nayak2025steady}, as well as analytically continued QMC results obtained via the maximum entropy (MaxEnt) \cite{Jarrell1996} method.

For both temperatures, the spectral functions derived from analytic continuation appear smooth but lack the detailed structure of the Hubbard bands, deviating significantly at high energies from the real-frequency results produced by the other methods. At high temperature ($\beta = 2$), all real-frequency methods yield broadly consistent spectra, with the TOA showing an almost quantitative agreement with the IF-MPS result, which can be regarded as the numerically exact reference in this regime. Also at lower temperatures, the TOA result more closely matches the IF-MPS spectrum compared to OCA. This agreement provides supporting evidence for the reliability of both the TOA solver and the IF-MPS method in the intermediate coupling regime. 

A closer inspection reveals that the high-energy features in the TOA spectrum align well with those in the IF-MPS result, while notable differences remain in the low-energy region. The overall weight of the quasiparticle peak appears to be rather consistent, but the height of the quasiparticle peak ($\omega = 0$ value) differs. A previous study~\cite{nayak2025steady} showed that the quasiparticle peak obtained from IF-MPS agrees well with results from NRG, time-evolving block decimation (TEBD), and MaxEnt analytically continued QMC data, thereby validating the accuracy of the IF-MPS method in reproducing the quasiparticle peak structure in the low-frequency regime. Hence, the discrepancy in peak height suggests that the strong-coupling solvers do not accurately capture the quasi-particle features, even at the TOA level. Interestingly, the height of the quasiparticle peak in the OCA spectrum happens to be closer to the MaxEnt and IF-MPS results than the TOA prediction. This shows that, at least for certain observables like the peak height, the strong-coupling expansion does not converge monotonically with increasing order.

\begin{figure*}[t]
  \centering
  \includegraphics[width=\textwidth]{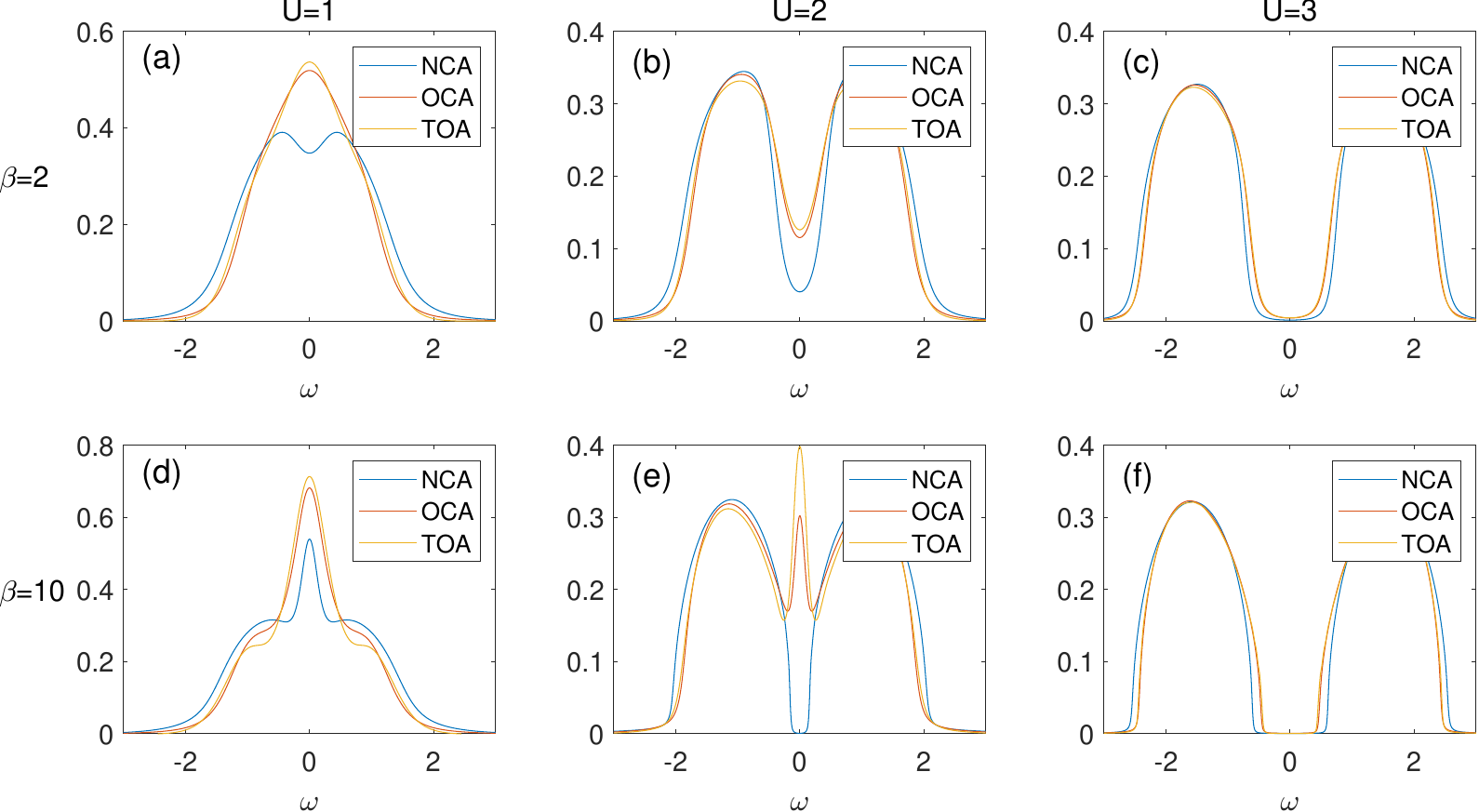}
  \caption{Spectral functions of the Hubbard model in the paramagnetic equilibrium state, computed using the NCA, OCA, and TOA.  Panels (a)--(c) show results at inverse temperature $\beta = 2$, while panels (d)--(f) show results at $\beta = 10$. Each column corresponds to a different interaction strength: $U = 1$, 2, and 3 from left to right. }
  \label{fig:NCAOCATOA_comparison}
\end{figure*}

To systematically investigate the convergence behavior of the strong-coupling expansion and delineate the applicability of different orders across a broad parameter space, we directly compute and compare the real-frequency spectral functions obtained using the NCA, OCA, and TOA methods. Fig.~\ref{fig:NCAOCATOA_comparison} presents the spectral functions for three interaction strengths: $U = 1$, $2$, and $3$, which correspond to the metallic regime, the crossover region, and the Mott insulating regime, respectively. For each value of $U$, we consider both a high-temperature case ($\beta = 2$) and a low-temperature case ($\beta = 10$). In the strongly correlated regime ($U=3$), all three methods yield similar results, with TOA providing only small corrections compared to OCA. This is consistent with the physical expectation that the strong-coupling expansion converges rapidly when $U$ is large, and in particular in the Mott regime, whose spectral function is qualitatively similar to that of the atomic limit. In contrast, at weak coupling ($U=1$), the discrepancies among the three methods become substantial, and the convergence behavior is more difficult to assess, reflecting the limitations of the expansion in this regime. At intermediate coupling, the DMFT solution of the Hubbard model undergoes an insulator-metal transition as temperature is lowered, and the phase transition (or crossover) line shifts substantially between different approximations \cite{eckstein2010nonequilibrium}. Hence, for $U=2$ and low temperature, we find significant differences between NCA and the higher-order methods: the NCA predicts a Mott insulator, while OCA and TOA yield a metallic solution with a low-frequency quasiparticle peak. At high temperature, in the metal-insulator crossover region, where thermal fluctuations suppress coherent quasiparticle features, the spectral functions from all three methods become qualitatively similar.

\begin{figure}[t]
  \centering
  \includegraphics[width=0.47\textwidth]{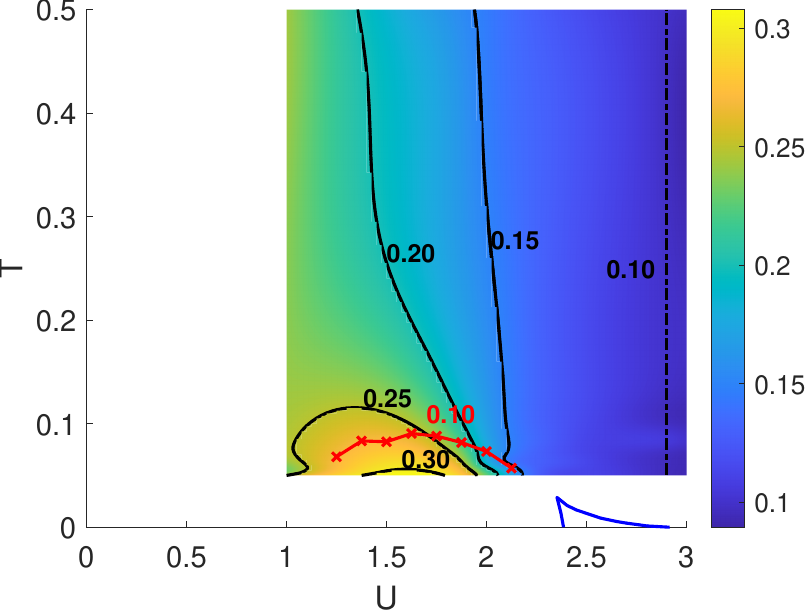}
  \caption{Color map of the spectral function distance defined in Eq.~\eqref{eq:SF_distance} between the NCA and OCA results, shown as a function of temperature $T$ and interaction strength $U$. Black contour lines indicate distance values of 0.10, 0.15, 0.20, 0.25, and 0.30. The red line marks the contour corresponding to a spectral distance of 0.10 between the OCA and TOA results. The blue lines represents the metal-insulator coexistence region obtained by QMC calculations in Ref.~\cite{blumer2003mott}.
  }
  \label{fig:U_T_diagram}
\end{figure}

After showing representative spectral functions computed using the different strong-coupling impurity solvers and a comparison against results from other established methods, we now turn to a quantitative assessment of the differences between two spectral functions obtained from different strong-coupling methods. To this end, we define the ``distance'' $D$ between two spectral functions $A_1$ and $A_2$ as the integral of the absolute difference between them:
\begin{equation}
D[A_1, A_2] = \int_{-\infty}^{\infty} d\omega\, |A_1(\omega) - A_2(\omega)|.
  \label{eq:SF_distance}
\end{equation}
Since each spectral function is normalized to unity, the maximum possible distance is bounded by $D \leq 2$. To investigate how the spectral functions change with the strong-coupling expansion order across different regions of parameter space, we compute the distance between the NCA and OCA spectra over a broad range of interaction strengths and temperatures ($U = 1 \sim 3$, $T = 0.05 \sim 0.5$). For weaker interaction strengths, as previously discussed, although the higher-order corrections appear small, the spectra do not converge toward the correct result. This casts doubt on the reliability of the strong-coupling methods in this regime, and therefore, we do not consider even weaker interactions in our analysis. The resulting distance map is shown in Fig.~\ref{fig:U_T_diagram}, where black contour lines indicate constant values of the spectral distance. 

It is important to note that in the Mott insulating regime, where the low-frequency spectral weight vanishes, the Fermi-Dirac distribution becomes less relevant for the electronic distribution, and the spectral function is relatively insensitive to temperature. In this regime, however, a small broadening parameter $\eta$ must be introduced to stabilize the numerical calculation, which can slightly affect the computed distance. As a result, some artificial fluctuations appear in the lower-right region of Fig.~\ref{fig:U_T_diagram}. To mitigate this, we replace the noisy 0.1 contour line, which fluctuates around $U=2.9$, with a dashed line at $U=2.9$.

From this analysis, we observe that the distance between the NCA and OCA spectral functions generally decreases with increasing temperature and increasing interaction strength, consistent with the intuition that strong-coupling methods converge more readily in the high-$U$, high-$T$ regime. However, the dependence is not strictly monotonic. In particular, we observe a pronounced peak in the distance around $U = 1.5 \sim 2.0$ at low temperatures (below $T=0.15$). As mentioned before, the exact DMFT solution for the paramagnetic state exhibits a metal insulator phase transition at low temperatures. The high-temperature end point of this line is near $U = 2.33$, $T = 0.029$ and the zero-temperature end point near $U = 2.9$~\cite{blumer2003mott}. In Fig.~\ref{fig:U_T_diagram}, the blue line indicates the coexistence region (from Ref.~\cite{blumer2003mott}) between the metallic and insulating solutions. In the NCA solution, the transition line becomes a crossover and is shifted to substantially lower $U$. The largest deviations between the NCA and OCA spectra originate from this mismatch in the prediction of the Mott transition line or crossover. 

To relate the examples shown in Fig.~\ref{fig:NCAOCATOA_comparison} to Fig.~\ref{fig:U_T_diagram}, we list the NCA-OCA spectral distances for the six panels of Fig.~\ref{fig:NCAOCATOA_comparison}: (a)~0.2428, (b)~0.1455, (c)~0.0966, (d)~0.2459, (e)~0.1685, and (f)~0.0948. These examples help to build an intuitive understanding of what different spectral distance magnitudes correspond to in terms of actual spectral differences. In particular, we see that distances $D\gtrsim 0.15$ can mean qualitatively different solutions (e. g. one metallic and the other insulating). Distances $D\lesssim 0.1$ are needed in order to claim that the NCA solution is a good approximation to the OCA result. The dashed line in Fig.~\ref{fig:U_T_diagram} thus shows the expected result: NCA is accurate only in the Mott insulating regime. 

In addition, we also computed the spectral distance between the TOA and OCA results for selected parameters. Using interpolation, we extracted the contour corresponding to a spectral distance of 0.1, which is plotted with a red line in Fig.~\ref{fig:U_T_diagram}. Since a distance of 0.1 indicates a fairly close agreement between two spectra, the red contour can be viewed as an approximate boundary for the validity of the OCA solver. In particular, the validity of OCA extends throughout the metal-insulator crossover regime at $T>0.1$, and throughout the finite-temperature Mott regime. We do not show results for
$U \leq 1.25$, since the spectral difference loses its practical usefulness for the characterization of the convergence behavior. 

In the strongly correlated low-temperature metal regime, OCA is not quantitatively accurate, while TOA should provide a significant improvement. To support this point and to gain a quantitative sense of the accuracy of the TOA, we computed the spectral function distances between the results obtained from OCA/TOA and the benchmark IF-MPS data in the two representative cases shown in Fig.~\ref{fig:method_comparison}. For the case in Fig.~\ref{fig:method_comparison}(a), the spectral function distance between OCA and IF-MPS is 0.0424, while the distance between TOA and IF-MPS is 0.0204. Similarly, for the case in Fig.~\ref{fig:method_comparison}(b), the spectral function distances for OCA and TOA with IF-MPS are 0.1607 and 0.0633, respectively. Hence, at $U=2$, TOA yields accurate spectra at least down to $T=1/20$.

\subsection{DMFT: Antiferromagnetic Phase}\label{sec3c}

In the previous section, we focused on equilibrium DMFT calculations for the paramagnetic phase. However, it is well known that at low temperatures, and for $0<U<\infty$, the ground state of the half-filled Hubbard model within DMFT is antiferromagnetically ordered~\cite{georges1996dynamical}. It is thus relevant to examine how strong-coupling impurity solvers with different orders perform in the antiferromagnetic (AFM) regime.

The self-consistency condition~\eqref{eq7} used in the paramagnetic calculations assumes uniformity across all lattice sites. In the half-filled AFM state on a bipartite lattice, however, the spins exhibit a staggered pattern across the lattice, necessitating the introduction of sublattices $A$ and $B$. The local Green's function on one sublattice serves as the bath for the other, leading to the self-consistency condition $\Delta_{A,\sigma} = v^2 G_{B,\sigma}$. Due to the condition of zero total magnetization in the AFM state, one has $G_{A,\sigma} = G_{B,\bar{\sigma}}$, allowing us to simplify the DMFT self-consistency condition to \cite{georges1996dynamical}
\begin{equation}
  \Delta_{\sigma} = v^2 G_{\bar{\sigma}},
\end{equation}
which avoids the explicit treatment of $A$ and $B$ sublattices in the numerical implementation.

\begin{figure}[t]
  \centering
  \includegraphics[width=0.47\textwidth]{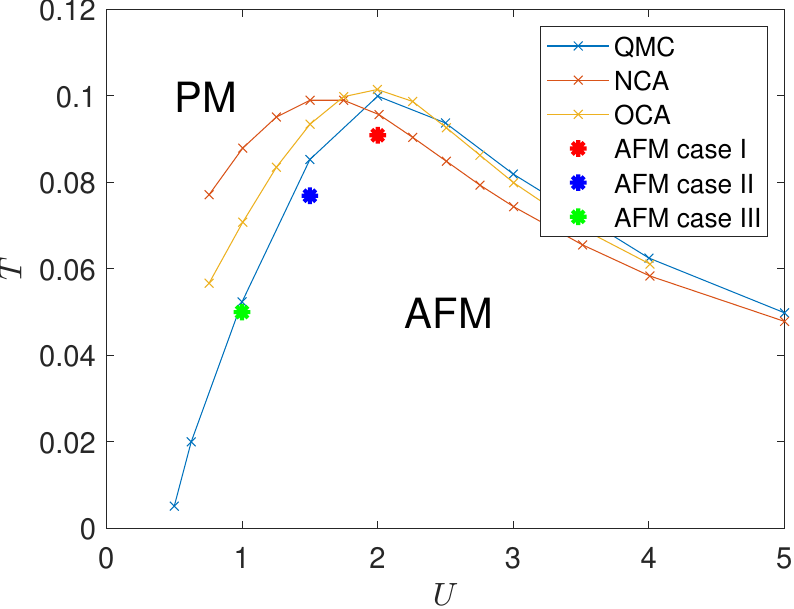}
  \caption{The $U$-$T$ phase diagram showing the paramagnetic (PM) to antiferromagnetic (AFM) transition boundary obtained using QMC (blue line), NCA (red line) and OCA (yellow line) impurity solvers, with data taken from Ref.~\cite{werner2012nonthermal}. The red, blue and green stars indicate the locations of case I, case II and case III, respectively, which are analyzed in detail in this work.
  }
  \label{fig:AFM_phase_diagram}
\end{figure}

\begin{figure*}[t]
  \centering
  \includegraphics[width=\textwidth]{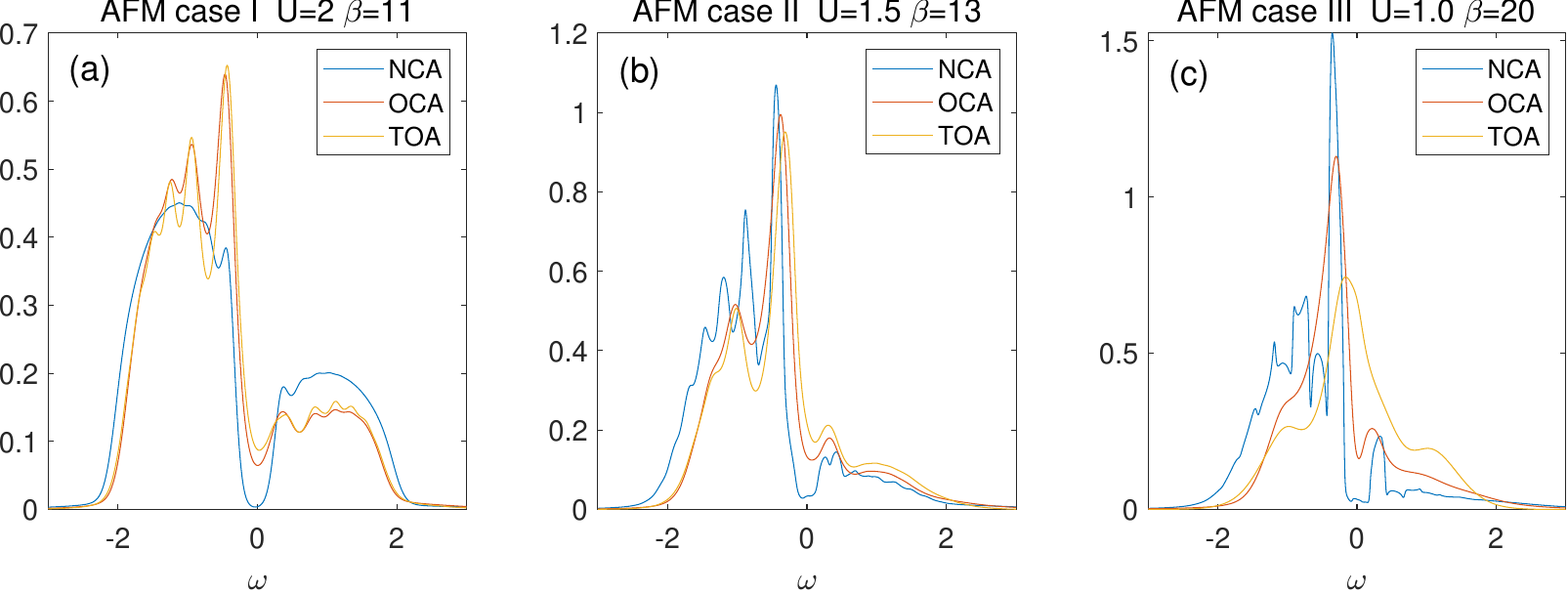}
  \caption{Antiferromagnetic spectral functions computed using different strong-coupling impurity solvers: NCA, OCA, and TOA. Panel (a) shows the results for case I ($U = 2$, $\beta = 11$), panel (b) shows the results for case II ($U = 1.5$, $\beta = 13$), and panel (c) shows the results for case III ($U = 1.0$, $\beta = 20$).
  }
  \label{fig:AFM_SF}
\end{figure*}

To locate our calculated examples within the phase diagram, we display in Fig.~\ref{fig:AFM_phase_diagram} the $U$-$T$ phase diagram obtained with QMC, NCA and OCA impurity solvers from Ref.~\cite{werner2012nonthermal}, and indicate with stars the three representative cases which we will analyze in more detail. For these three cases, we compute the real-frequency spectral functions using NCA, OCA, and TOA impurity solvers, with the results shown in Fig.~\ref{fig:AFM_SF}. Due to the spin symmetry breaking in the AFM phase, there exist two spectral functions for the two opposite spins, which at half-filling are symmetric about $\omega=0$. We only show the majority-spin component (with the higher occupation) for clarity.

Several notable features distinguish the AFM spectra from the paramagnetic counterparts. Apart from the asymmetry with respect to $\omega = 0$, which reflects the presence of a local magnetization, in the strongly correlated regime, a series of sharp quasiparticle peaks appears in the Hubbard band region. These peaks correspond to quasi-bound states arising from the restricted motion of the electrons in the AFM background, reflecting discrete excitation structures induced by the coupling between the electron motion and spin fluctuations \cite{martinez1991,strack1992}. For large $U$, the peak spacing scales approximately with the exchange energy $J_\text{ex} \propto t^2/U$, and serves as a characteristic fingerprint of the Heisenberg-AFM regime. In contrast, in the weakly correlated (Slater-AFM) regime, we find only one prominent peak and one sideband in the spectral function, in agreement with calculations based on weak-coupling diagrammatic expansions \cite{tsuji2013nonequilibrium}.

Comparing the three cases in Fig.~\ref{fig:AFM_SF}, we find contrasting behaviors in the influence of higher-order corrections. In case I, increasing the solver order leads to sharper quasiparticle peaks and an enhanced local magnetization, while in case II and III, the higher-order corrections reduce the peak sharpness and suppress the magnetization. This trend is consistent with the comparison between the NCA, OCA and QMC phase boundaries shown in Fig.~\ref{fig:AFM_phase_diagram} and illustrates that the impact of higher-order strong-coupling corrections depends on the parameter regime. Furthermore, for case I and II, the corrections from TOA relative to OCA are significantly smaller than those from OCA relative to NCA, indicating good convergence properties of the strong-coupling expansion in these regimes. Similar to the paramagnetic results, we observe that OCA and TOA tend to favor a metallic state in the crossover region, while NCA predicts an insulating solution.

It is worth noting that the three cases considered here are close to the AFM phase boundary. Although, in general, lowering the temperature leads to more pronounced oscillatory structures in the time-domain self-energy and Green's functions, which increases the fitting complexity, in the paramagnetic case, especially in the insulating regime, the spectral function does not change dramatically with temperature. As a result, the required bond dimension grows only very slowly as $\beta$ increases. In contrast, in the present AFM cases, the spectral function becomes significantly sharper upon lowering the temperature, corresponding to more weakly damped hybridization functions in the time domain and increasingly complex high-dimensional target functions for the TT representation. Consequently, the required bond dimension in the QTCI fitting process grows rapidly and can quickly exceed practical computational limits. All the paramagnetic calculations shown in this paper converge with a bond dimension around $40$, while the AFM calculations require a bond dimension of approximately $80$ to achieve similar convergence.

\begin{table*}[t]
\centering
\caption{
Comparison of the local magnetization ($m$) and double occupancy ($d$) for case I, case II and case III computed using QMC, NCA, OCA, and TOA impurity solvers. 
}
\begin{tabular}{lcccccc}
\hline\hline
Method & $m$ (case I) & $d$ (case I) & $m$ (case II) & $d$ (case II) & $m$ (case III) & $d$ (case III)\\
\hline
NCA  & 0.3681 & 0.0354 & 0.6904 & 0.0654 & 0.7256 & 0.1045\\
OCA  & 0.5137 & 0.0501 & 0.5923 & 0.0912 & 0.5026 & 0.1416\\
TOA  & 0.4947 & 0.0532 & 0.4890 & 0.1003 & 0.1419 & 0.1626\\
QMC  & 0.4827 & 0.0528 & 0.4776 & 0.1004 & 0.2155 & 0.1595\\
\hline\hline
\end{tabular}
\label{tab:AFM_comparison}
\end{table*}

After analyzing the antiferromagnetic spectral functions obtained from different orders of strong-coupling impurity solvers, we proceed to benchmark these methods against QMC by comparing two physical quantities that are directly accessible on the imaginary-time axis. Specifically, for the three cases discussed above, we compare the double occupancy and local magnetization, both of which can be extracted from the equal-time values of the lesser pseudo-particle Green's functions in our real-time formalism. The results are summarized in Table~\ref{tab:AFM_comparison}.

Let us first consider the double occupancy. Similar to the paramagnetic case, the results from the strong-coupling solvers exhibit a systematic improvement with increasing expansion order. In all three cases, the TOA results closely match the QMC ones, demonstrating excellent accuracy for the double occupation in the considered parameter regime. 
The behavior of the local magnetization is more complex. For case II, the magnetization shows a monotonic convergence toward the QMC result as the expansion order increases. However, the TOA result still exhibits a noticeable deviation, indicating that the magnetization is a more sensitive and challenging quantity to converge compared to the double occupancy, especially close to the N\'eel temperature. This is also evident from the large shifts in the phase boundaries between NCA, OCA and QMC in Fig.~\ref{fig:AFM_phase_diagram}. In case I, the convergence is no longer monotonic: the NCA underestimates the magnetization, while both OCA and TOA slightly overestimate it relative to QMC. The opposite behavior is found in case III. This observation can be understood by looking at the trends in the AFM phase boundaries obtained from the different solvers (see Fig.~\ref{fig:AFM_phase_diagram}), where NCA overestimates the Néel temperature in the weak-coupling regime and underestimate it in the strong-coupling regime. As a result, in the crossover region, one may encounter situations where the NCA or OCA yields magnetization values closer to the exact result than higher-order methods, explaining the non-monotonic convergence.

\subsection{DMFT: Photo-Doped States}\label{sec3d}

\begin{figure*}[t]
  \centering
  \includegraphics[width=\textwidth]{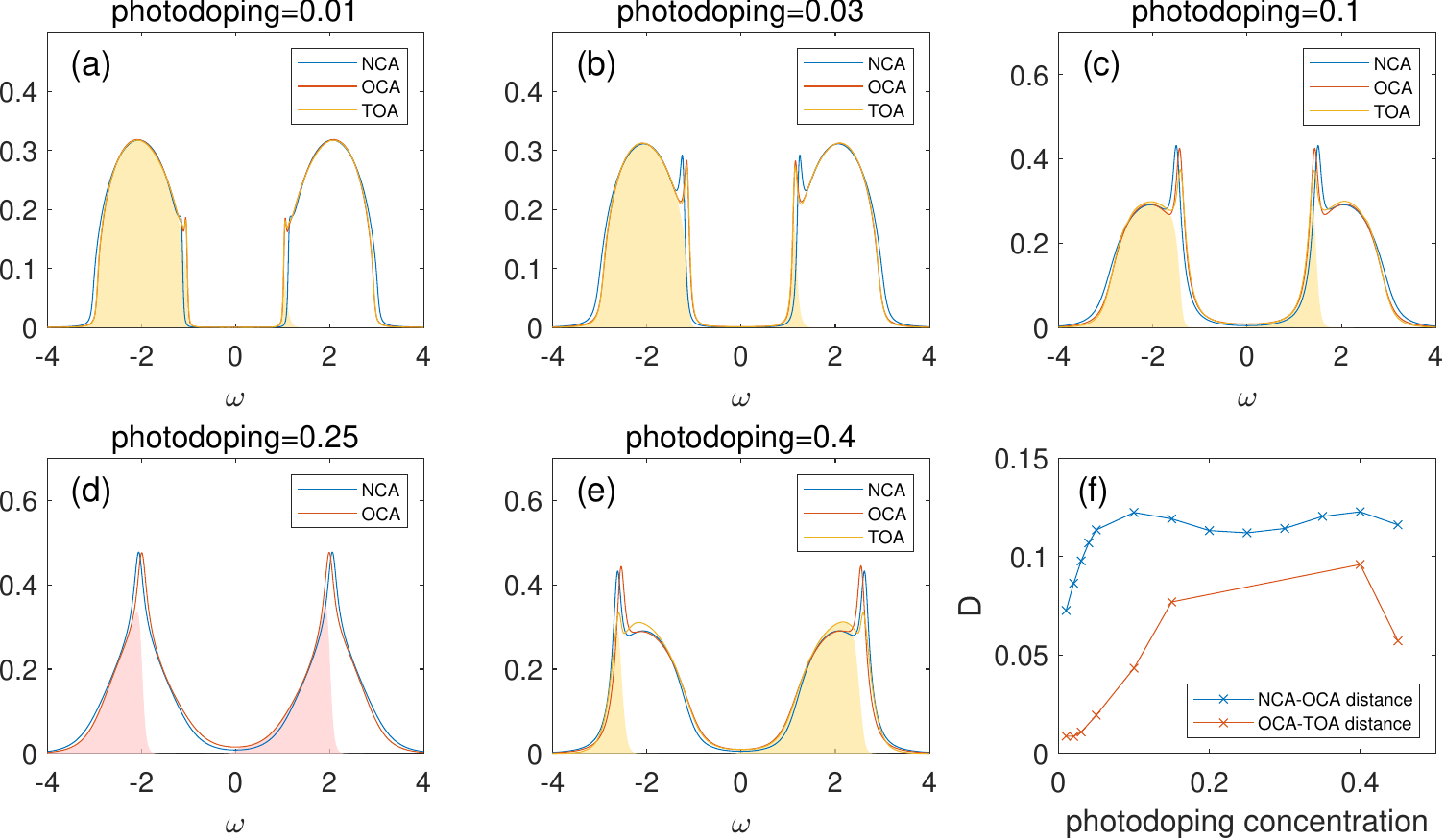}
  \caption{(a)-(e) Spectral functions computed using NCA, OCA, and TOA at photodoping concentrations of 0.01, 0.03, 0.1, 0.25 and 0.4, respectively. The occupied parts of the spectra obtained from the TOA calculations are indicated by the yellow shading. The TOA data are missing in panel (d) because the computational cost required for fitting exceeded our capabilities; accordingly, the occupied part shown is taken from the OCA calculation and indicated by the red shading. (f) The spectral function distances between NCA and OCA (blue) and between OCA and TOA (red) as a function of the photodoping concentration. 
  }
  \label{fig:photodope_beta20}
\end{figure*}

As noted earlier, a key advantage of performing real-frequency DMFT calculations is the ability to access steady states beyond thermal equilibrium, where the distribution function differs from the Fermi-Dirac distribution but remains stable over long timescales. A paradigmatic example in strongly correlated systems is the photodoped Mott insulator: under laser irradiation, electrons initially residing in the lower Hubbard band are excited into the upper Hubbard band, creating electron- and hole-like charge carriers (doublons and holons). In case of a large Mott gap, these photodoped carriers cannot thermalize rapidly and instead undergo prethermalization within the Hubbard bands, forming a distribution characterized by an effective temperature near the band edges \cite{murakami2023photo}. This behavior has been observed in time-dependent DMFT calculations using three-branch nonequilibrium Green's functions~\cite{aoki2014nonequilibrium,eckstein2013}, but the heavy computational and memory costs associated with three-branch calculations makes it difficult to study long timescales. The two-branch steady-state framework leverages time-translation invariance to significantly reduce the numerical cost, making it well-suited for exploring such metastable photodoped states \cite{li2021nonequilibrium}.

Previous studies have demonstrated that photodoped Mott insulators can exhibit a rich landscape of phases with order parameters, including spin and orbital orders, $\eta$-pairing states, and excitonic orders~\cite{ueda2025exotic,murakami2023spin,werner2018enhanced,bittner2020photoenhanced,ray2024photoinduced,ray2023nonthermal}, many of which have not yet been thoroughly investigated. Most of the existing studies employed NCA impurity solvers to describe these photodoped states, which typically are metallic. This raises concerns regarding the reliability of the NCA description of these nonthermal states and phases. It is therefore crucial to assess the convergence and accuracy of the NCA results by employing higher-order strong-coupling solvers.

The computational procedure for simulating photodoped Mott insulators is essentially the same as that used in the equilibrium calculations discussed previously, but with one key modification: in constructing the lesser Green's function from the retarded Green's function, we do not use a Fermi-Dirac distribution but instead assume that the distribution function within each Hubbard band adopts a given effective temperature distribution corresponding to $T_\text{eff}=1/\beta_\text{eff}$. To connect the two distributions in the Mott gap region, we use a smooth function, following the procedure described in Ref.~\cite{kunzel2024numerically}. To enforce calculations at a fixed total photodoping density, we iteratively adjust the chemical potentials for the doublons and holons during the DMFT self-consistency loop.

We focus on the case of $U=4$ and $\beta_\text{eff}=20$ at half-filling in this study, computing the spectral functions for various photodoping concentrations using NCA, OCA, and TOA impurity solvers. The results are displayed in Fig.~\ref{fig:photodope_beta20}. Panels (a)-(e) show representative spectra at different photodoping concentrations (photodoping=$x$ means a fraction of $x$ photodoped doublons and $x$ photodoped holons). As the photodoping concentration increases, the quasiparticle peaks at the upper edge of the lower Hubbard band and the lower edge of the upper Hubbard band initially become more pronounced and shift towards the band center, consistent with the trends observed in Ref.~\cite{kunzel2024numerically}. However, once the photodoping exceeds 0.25, these peaks gradually shift to the upper/lower edge of the upper/lower Hubbard band. Physically, this means that the states associated with the quasi-particle peak switch from doublons/holons to singlons.  

Examining the results across different impurity solvers, we find that the corrections introduced by the higher-order methods initially grow with increasing photodoping. For low-level photodoping, these corrections primarily shift the quasiparticle peak toward the Fermi level, in line with our earlier conclusion that higher-order methods mitigate the overestimation of the effective interaction strength $U$. To quantitatively measure the high-order corrections for the spectral functions, we compute the spectral function distance between the NCA and OCA results, and between the OCA and TOA results, using the previous definition in Eq.~\eqref{eq:SF_distance}. These distances are plotted as a function of photodoping concentration in Fig.~\ref{fig:photodope_beta20}(f). 

For the NCA-OCA distance, we observe a rapid increase at low doping levels, saturating around 0.05, followed by a shallow minimum around 0.25 and a second maximum around 0.4. We speculate that this nonmonotonic behavior is related to the shifting position of the quasiparticle peak with doping. In the large-$U$ limit, the quasiparticle peak position should be symmetric around 0.25 doping, and the observed symmetry in the spectral function distance reflects this property. For the OCA-TOA distance, we observe a significantly different trend. For small photodoping concentrations (less than 0.03), the distance remains nearly constant and the small value indicates a rapid convergence toward the exact spectral function. However, as the doping concentration increases further, the OCA-TOA distance reaches values comparable to the NCA-OCA distance, indicating that the convergence of the strong-coupling expansion becomes poor for photodoping levels near 0.25. For very large photodoping levels, the OCA-TOA distance shows some decline; however, it remains significantly larger than that of the mirrored points about photodoping concentration 0.25. This is because $U=4$ is far from the large-$U$ limit and the TOA result is less correlated than the OCA and NCA spectra. We have also examined this asymmetry at larger $U$, and indeed find that the spectral curves become more symmetric as $U$ increases. 

Additionally, at intermediate doping levels, when the system becomes highly metallic, the calculation requires a large bond dimension (beyond 100) in the QTCI fitting of the TOA target functions to achieve convergence. As a result of this, we were unable to obtain reliable TOA spectra in this regime, explaining the missing data in Figs.~\ref{fig:photodope_beta20}(d) and (f). 

\begin{figure}[t]
  \centering
  \includegraphics[width=0.45\textwidth]{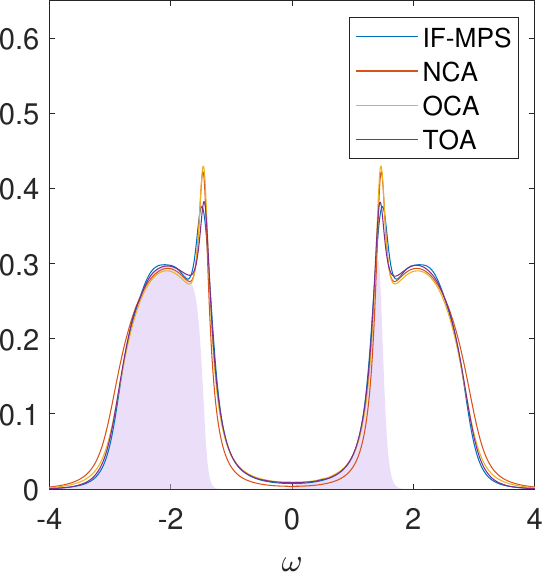}
  \caption{Comparison of the spectral functions obtained using three strong-coupling expansion methods (NCA, OCA, TOA) with chemical potentials $\mu_\pm=\pm1.5$ and inverse temperature $\beta_\text{eff}=20$ for the holons/doublons in the lower/upper Hubbard bands, respectively, with the IF-MPS result from Ref.~\cite{nayak2025steady}. The purple shading indicates the occupied part of the spectrum computed using the TOA.
  }
  \label{fig:photodope_compari}
\end{figure}

In the photodoped case, we cannot obtain any reference data from imaginary-axis QMC simulations. To cross-check the validity of our methods under photodoping, we compare our spectra with those obtained using the IF-MPS approach in Ref.~\cite{nayak2025steady}. In this comparison, we fix the effective chemical potentials in the lower and upper Hubbard bands at $\mu=\pm1.5$ rather than fixing the photodoping concentration as in Fig.~\ref{fig:photodope_beta20}, while keeping the inverse effective temperature at $\beta_\text{eff}=20$. This setup corresponds to Fig.~10(f) in Ref.~\cite{nayak2025steady}, and the comparison is presented in Fig.~\ref{fig:photodope_compari}.

We observe that with increasing expansion order, our results systematically approach the IF-MPS benchmark, demonstrating clear convergence behavior. However, this convergence does not occur uniformly across the whole frequency range. Taking the IF-MPS results as a reference, we find that OCA significantly improves the accuracy of the spectral function near the Hubbard band edges compared to NCA, while producing limited improvements (and occasionally larger deviations) near the quasiparticle peak and the Hubbard band maxima. The TOA, in contrast, not only preserves the improved accuracy at the band edges but also substantially enhances the precision around the peak regions, notably reducing the height of the quasiparticle peak. As a result, the TOA and IF-MPS spectra show a near-perfect agreement. 

These results highlight the necessity of using higher-order strong-coupling expansion methods to accurately capture the spectral properties in photodoped systems, demonstrating that within this parameter regime, the TOA already provides quantitatively reliable results suitable for detailed analyses. 

\section{Conclusions}\label{sec4}

In this work, we extended the real-frequency strong-coupling impurity solver up to the third order using the QTCI technique, and systematically examined its convergence properties for the Anderson impurity model and as a solver within DMFT. We explored equilibrium paramagnetic and antiferromagnetic states of the Hubbard model, as well as photodoped steady states, and demonstrated the practical performance of the high-order strong-coupling methods.

Our results show that, with proper parameterization schemes, QTCI can efficiently fit the target functions of the third-order diagrams, significantly accelerating calculations while maintaining high accuracy. For the typical cases studied here, where the TCI pivot error satisfies $\epsilon < 10^{-5}$ (ensuring that the resulting spectral functions are visually indistinguishable from converged results), a full DMFT calculation with 30 iterations on a 128-core node completes within minutes for NCA, within about half an hour for OCA, and within roughly two days for TOA. Benchmarks against other methods revealed that the higher-order corrections indeed improve the accuracy of the results in most regimes. NCA is reliable only in the strong-coupling limit where the interaction strength significantly exceeds the bandwidth, while OCA alleviates the overestimation of $U$ present in NCA, providing qualitatively correct results even at intermediate interactions down to temperatures of about 0.1 times the bandwidth. 

For the paramagnetic case, we compared the kinetic and potential energies obtained from different solvers with QMC results. While TOA yields quantitatively accurate energies even in the weak-coupling limit, this agreement does not imply that the spectral functions converge correctly. In fact, while the higher-order methods still exhibit convergence behavior in the weak-coupling limit, they do not converge toward the correct result, but deviate significantly from the exact solution and even violate causality. It appears that for interactions significantly smaller than the bandwidth, the self-consistent strong-coupling expansion encounters a problem of misleading convergence, similar to the one which has been extensively discussed for the self-consistent weak-coupling expansion at large $U$. Even higher order results are however needed to fully clarify this issue. 

To quantitatively assess the importance of higher-order corrections across different parameter regimes, we introduced the spectral function distance as a diagnostic and evaluated both the NCA-OCA and OCA-TOA distances. The spectral function distance for adjacent-order methods provides a practical reference for the applicability of different solvers, particularly in the intermediate to strong-coupling regimes, where the strong-coupling method systematically converges, and the distance effectively reflects the convergence quality of each method. At lower $U$, due to the aforementioned problem of misleading convergence, this distance analysis becomes ambiguous. The precise range of applicability and convergence of the strong-coupling expansion methods thus remains an open question for future investigations.

In the antiferromagnetic case, we found that boundary corrections in QTCI, corresponding to higher-order integration schemes, are essential to obtain physically correct results. We have introduced efficient techniques to compute the OCA and TOA diagrams in the AFM state. However, since the higher-order diagrams have opposite effects in the weak-coupling and strong-coupling regimes, one encounters a slow convergence with expansion order close to the AFM phase boundary in the intermadiate-coupling regime.  

Finally, we investigated photodoped Mott insulators as a typical example of strongly correlated nonequilibrium steady states. As the photodoping concentration increases, the metallicity of the system is enhanced, leading to larger higher-order corrections. In the case of a moderate Mott gap (comparable to the bandwidth), the OCA corrections quickly saturate at a large value, while the TOA corrections are initially small but continue to increase with further doping. The comparison with an IF-MPS spectrum showed that higher-order corrections are important in photo-doped systems, even at relatively large $U$. The TOA was however found to accurately capture the spectral features. It should thus be a very powerful method for the study of nonequilibrium phases in photo-doped Mott insulators, including $\eta$-pairing states. 

In this work, we focused on the single-band Hubbard model with and without spontaneous symmetry breaking, providing benchmarks for local observables and spectral functions across different impurity solvers and parameter regimes. The systematic comparison and quantitative analysis presented here allow to assess the reliability and limitations of the self-consistently resummed hybridization expansion in real-frequency DMFT calculations. 
In cases where rapid convergence is observed, the data presented here can also serve as reliable benchmarks for other real-frequency solvers. 

As a next step, we plan to apply higher-order strong-coupling impurity solvers to multiband Hubbard models and systems with electron-boson interactions. While the extension to multi-orbital cases is in principle straightforward, the computational cost grows rapidly with orbital number. Our tests indicate that OCA calculations for two-orbital problems are feasible without symmetry-based reduction in the number of diagrams, whereas calculations for $n>2$ orbitals will likely require additional symmetry considerations and more efficient parallelization strategies to remain computationally tractable. 
Still, the combination of strong-coupling expansions with QTCI provides a promising strategy for obtaining quantitatively accurate results for these complex systems at a manageable computational cost.

\acknowledgements

We thank M. Nayak for providing IF-MPS results shown in this work, and M. \'{S}roda, Y. N. Fern\'{a}ndez and Y. Murakami for helpful discussions. This work was supported by SNSF Grant No. 200021-196966 and 
the DGIST Start-up Fund Program of the Ministry of Science and ICT (2025010006) and Basic Science Research Program through the National Research Foundation of Korea (NRF) funded by the Ministry of Education (2025090055).
The calculations were run on the Beo06 cluster at the University of Fribourg.

\appendix
\section{Effect of Time-Grid Resolution and Integration Schemes}
\label{app_A}
\begin{figure*}[htb]
  \centering
  \includegraphics[width=\textwidth]{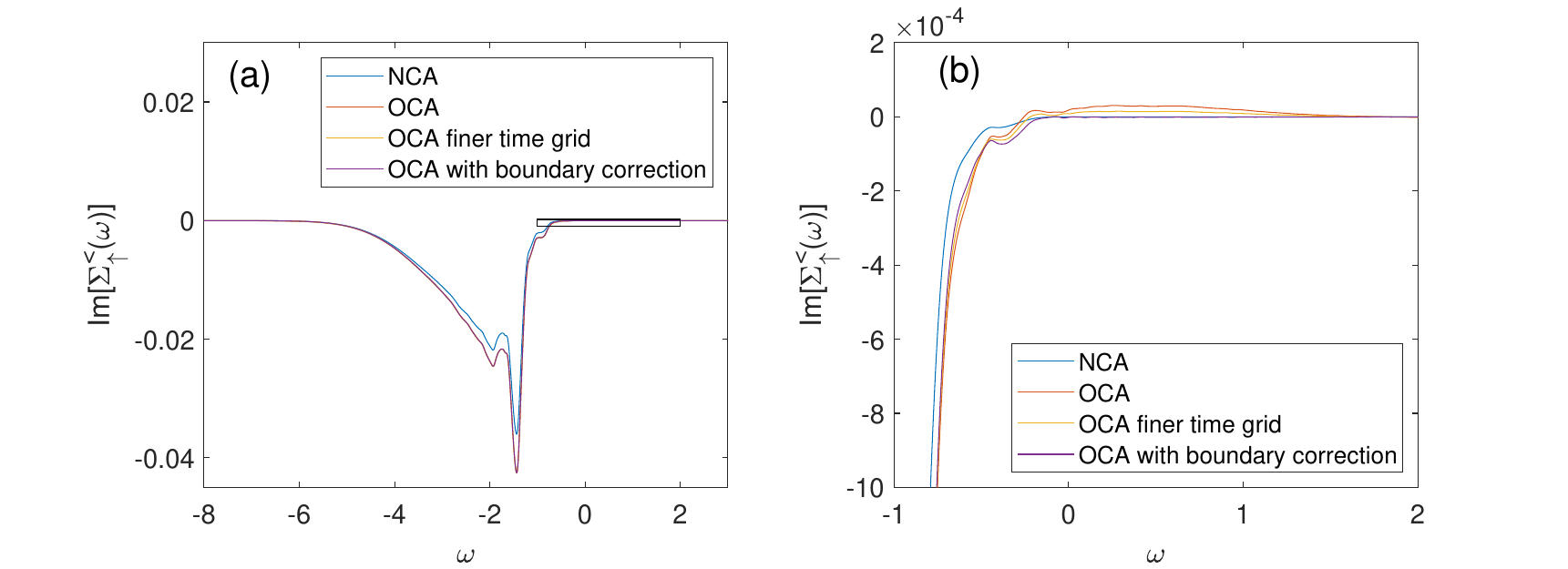}
  \caption{Effect of the time-grid resolution and boundary corrections
  on the OCA self-energy in case~I.
  (a) Comparison between NCA, OCA without corrections, OCA with doubled grid resolution,
  and OCA with boundary corrections.
  (b) Zoom-in of the boxed region in (a), where an unphysical positive bump
  in the uncorrected OCA result can be observed.}
  \label{fig.appendixA}
\end{figure*}

In this appendix, we explicitly illustrate how the choice of time-grid resolution
and integration scheme affects the results of the high-dimensional integrals
involved in the strong-coupling expansion.
As a representative example, we consider case~III in Sec.~\ref{sec3c},
and compare one-shot OCA calculations employing different schemes.

Figure~\ref{fig.appendixA} summarizes our results.
Panel~(a) shows the overall comparison between NCA, OCA without boundary corrections,
OCA with doubled time-grid resolution but without boundary corrections, and OCA with boundary corrections.
Panel~(b) presents a magnified view of the region highlighted by the box in panel~(a). 
Here, the difference between the methods becomes more apparent.
While the NCA curve does not exhibit any positive values,
the OCA result without boundary corrections shows a small unphysical positive bump.
This artifact becomes weaker as the time grid becomes finer, 
and is effectively removed by including the boundary corrections.

Although this artifact is very small in the one-shot calculation shown here,
it can have a significant impact in a fully self-consistent calculation.
Even a tiny violation of causality can eventually drive the solution
from an antiferromagnetic to a paramagnetic state.
Therefore, using improved integration schemes, such as
the Gregory boundary corrections adopted here, is crucial for obtaining
physically meaningful results in antiferromagnetic systems.

\section{Examples of TOA diagrammatic expressions}

\begin{figure}[H]
  \centering
    \begin{tikzpicture}[thick, >=stealth]
    \draw[->,blue] (0,0.5) -- (5,0.5);
    \draw[->,blue] (5,-0.5) -- (0,-0.5);

    \draw[dashed,arrowmid] (5,-0.5) -- (1.4,0.5);
    \draw[dashed,arrowmid] (3.2,0.5) -- (4.1,-0.5);
    \draw[dashed,arrowmid] (0.5,0.5) .. controls (1.2,0) and (1.6,0) .. (2.3,0.5);
    \draw[dotted] (4.1,-0.5) -- (4.1,0.5);
    \draw[dotted] (3.2,-0.5) -- (3.2,0.5);
    \draw[dotted] (2.3,-0.5) -- (2.3,0.5);
    \draw[dotted] (1.4,-0.5) -- (1.4,0.5);
    \draw[dotted] (0.5,-0.5) -- (0.5,0.5);
    \node at (5.7,0) {$t=t_{\max}$};
    \node at (-0.5,0) {$t=0$};
    \filldraw (5,-0.5) circle (2pt);
    \filldraw (3.2,0.5) circle (2pt);
    \filldraw (1.4,0.5) circle (2pt);
    \filldraw (0.5,0.5) circle (2pt);
    \filldraw (2.3,0.5) circle (2pt);
    \filldraw (4.1,-0.5) circle (2pt);
    \draw[<->, thick] (4.1,-0.7) -- node[below] {\(v_1\)} (5.0,-0.7);
    \draw[<->, thick] (3.2,0.7) -- node[above] {\(v_2\)} (4.1,0.7);
    \draw[<->, thick] (2.3,0.7) -- node[above] {\(v_3\)} (3.2,0.7);
    \draw[<->, thick] (1.4,0.7) -- node[above] {\(v_4\)} (2.3,0.7);
    \draw[<->, thick] (0.5,0.7) -- node[above] {\(v_5\)} (1.4,0.7);
    \end{tikzpicture}
    \caption{A representative diagram for the lesser self-energy in Scheme III.
    }
  \label{appeendixBfig1}
\end{figure}
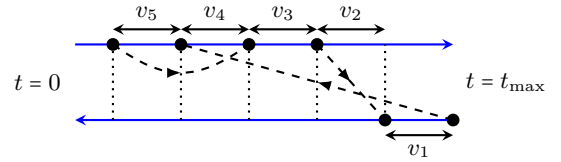

In this section, we provide explicit expressions for several TOA diagrams to illustrate the integration rules and sign conventions used in this work. 
As an example of the lesser pseudo-particle self-energy, we consider one specific configuration of hybridization-line directions out of Fig.~\ref{fig7}, as shown in Fig.~\ref{appeendixBfig1}.

The corresponding contribution to the lesser self-energy diagram can be written as  

\begin{align}
	\Sigma_\text{pp}^<(\omega) =& \, 2\mathrm{i}~\text{Im}\biggl\{\int_0^{t_{\mathrm{max}}}\hspace{-1.5em}dv_1 \int_0^{t_{\mathrm{max}}}\hspace{-1.5em} dv_2 \int_0^{t_{\max}}\hspace{-1.5em} dv_3\int_0^{t_{\max}}\hspace{-1.5em} dv_4\int_0^{t_{\max}}\hspace{-1.5em} dv_5\nonumber\\
		\times&\exp[-\mathrm{i}\omega(v_1+v_2)]\nonumber\\
	\times&\kappa\hspace{0.3em}\Delta^<(-v_1-v_2-v_3-v_4)\Delta^>(v_4+v_5)\Delta^>(v_2)\nonumber\\
	\times&F\mathcal{G}^>(v_3)\bar{F}\mathcal{G}^>(v_4)\bar{F}\mathcal{G}^>(v_5)F\nonumber\\
	\times&\mathcal{G}^<(-v_2-v_3-v_4-v_5)\bar{F}\mathcal{G}^>(-v_1)F \biggl\}, \label{appeendixBeq1}
\end{align}

which includes the the Fourier weight $W^\omega$ defined in Eq.~\eqref{eq:QTCI_int2} and the multi-variable target function of QTCI, $\sigma(v_1,\cdots,v_5)$:

\begin{align}
	\prod_{ij}W^\omega_{ij} =& e^{-i\omega(v_1+v_2)}~,\\
	\sigma(v_1,\cdots,v_5) =& \kappa\hspace{0.3em}\Delta^<(-v_1-v_2-v_3-v_4)\Delta^>(v_4+v_5)\Delta^>(v_2)\nonumber\\
	\times& F\mathcal{G}^>(v_3)\bar{F}\mathcal{G}^>(v_4)\bar{F}\mathcal{G}^>(v_5)F\nonumber\\
	\times& \mathcal{G}^<(-v_2-v_3-v_4-v_5)\bar{F}\mathcal{G}^>(-v_1)F.
	\label{eqn:W_sigma}
\end{align}

Equation~\eqref{eqn:W_sigma} contains a global factor $\kappa$ as well as a product of hybridization functions $\Delta$ and the matrix products of $F$ ($\bar{F}$) and $\mathcal{G}$, which are represented by dashed lines, dots and solid lines in Fig.~\ref{appeendixBfig1}, respectively. 


The prefactor $\kappa$ is determined by the following diagrammatic rules, consistent with Ref.~\cite{eckstein2010nonequilibrium}. 
(i) $\kappa$ contains a factor $\mathrm{i}^{N}$, where $N$ is the diagram order. 
This imaginary factor arises from the interaction-picture expansion of Eq.~(\ref{eq:action}).
For the third-order diagram shown in Fig.~\ref{appeendixBfig1}, this gives $\mathrm{i}^{3}$. 
(ii) Each crossing between hybridization lines contributes a minus sign, giving an overall $(-1)^s$ factor, where $s$ is the total number of crossings; for Fig.~\ref{appeendixBfig1}, $s=2$. 
(iii) The hybridization lines oriented opposite to the pseudo-particle backbone contribute $(-1)^{f}$, where $f$ is the number of such lines. 
In Fig.~\ref{appeendixBfig1}, two lines are parallel to the pseudo-particle propagators, while $\Delta^{<}(-v_{1}-v_{2}-v_{3}-v_{4})$ runs in the opposite direction, yielding $(-1)^{1}$. 
Rules (ii) and (iii) both originate from the contour ordering of the fermionic operators.
(iv) For each internal vertex on the lower branch, assign an additional minus sign.
In Fig.~\ref{appeendixBfig1}, out of four internal vertices, only the one at $t=t_{\mathrm{max}}-v_1$ satisfies this condition, giving another $(-1)^1$ factor. 
This factor 
comes from the difference in integral directions between the Keldysh-contour integral and the integrals in Eq.~(\ref{appeendixBeq1}).
Note that the Fourier transform for the external variable is performed along the positive direction, not along the contour direction.
Combining all these factors, we get $\kappa=-\mathrm{i}$ for Fig.~\ref{appeendixBfig1}.

To be compatible with some previous literature \cite{aoki2014nonequilibrium}, one could introduce a pseudo-particle dependent sign factor [$+$ ($-$) for the singly occupied (doubly occupied and empty) state] for the lesser pseudo-particle Green's function, which leads to a modified rule (iv). With this modification, the sign factors of the two formulations are the same.

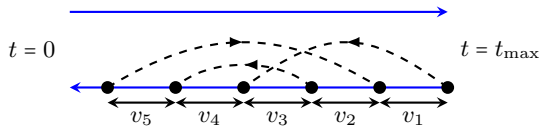
\begin{figure}[H]
  \centering
    \begin{tikzpicture}[thick, >=stealth]
    \draw[->,blue] (0,0.5) -- (5,0.5);
    \draw[->,blue] (5,-0.5) -- (0,-0.5);

    \draw[dashed,arrowmid] (5,-0.5) .. controls (3.9,0.3) and (3.4,0.3) .. (2.3,-0.5);
    \draw[dashed,arrowmid] (0.5,-0.5) .. controls (1.8,0.3) and (2.8,0.3) .. (4.1,-0.5);
    \draw[dashed,arrowmid] (3.2,-0.5) .. controls (2.8,-0.1) and (1.8,-0.1) .. (1.4,-0.5);
    \draw[dotted] (4.1,-0.5) -- (4.1,-0.5);
    \draw[dotted] (3.2,-0.5) -- (3.2,-0.5);
    \draw[dotted] (2.3,-0.5) -- (2.3,-0.5);
    \draw[dotted] (1.4,-0.5) -- (1.4,-0.5);
    \draw[dotted] (0.5,-0.5) -- (0.5,-0.5);
    \node at (5.7,0) {$t=t_{\max}$};
    \node at (-0.5,0) {$t=0$};
    \filldraw (5,-0.5) circle (2pt);
    \filldraw (3.2,-0.5) circle (2pt);
    \filldraw (1.4,-0.5) circle (2pt);
    \filldraw (0.5,-0.5) circle (2pt);
    \filldraw (2.3,-0.5) circle (2pt);
    \filldraw (4.1,-0.5) circle (2pt);
    \draw[<->, thick] (4.1,-0.7) -- node[below] {\(v_1\)} (5.0,-0.7);
    \draw[<->, thick] (3.2,-0.7) -- node[below] {\(v_2\)} (4.1,-0.7);
    \draw[<->, thick] (2.3,-0.7) -- node[below] {\(v_3\)} (3.2,-0.7);
    \draw[<->, thick] (1.4,-0.7) -- node[below] {\(v_4\)} (2.3,-0.7);
    \draw[<->, thick] (0.5,-0.7) -- node[below] {\(v_5\)} (1.4,-0.7);
    \end{tikzpicture}
    \caption{A representative diagram for the retarded self-energy in Scheme III.
    }
  \label{appeendixBfig1ret}
\end{figure}

In addition to the lesser pseudo-particle self-energy, we also need to evaluate the retarded component for the pseudo-particle Green's function via Eqs.~(\ref{eq:dyson_a}-\ref{eq:dyson_c}). 
The retarded component can be obtained from the greater self-energy diagrams, where all operators can be located on the lower branch of the contour as shown in Fig.~\ref{appeendixBfig1ret}: this example is chosen to have the same diagram topology and direction of hybridization lines as Fig.~\ref{appeendixBfig1}.
To maintain consistency with the lesser self-energy diagrams, the starting operator is still placed at the lower-right corner.
We first compute $\Sigma_\text{pp}^{>}\left(-t\right)$, and then use Eq.~\eqref{eq:symmetry} to obtain $\Sigma_\text{pp}^{>}\left(t\right)$. 
Finally, the retarded self-energy is given by $\Sigma_\text{pp}^{\text{R}}(t) = \theta(t)\,\Sigma_\text{pp}^{>}(t)$, which is subsequently Fourier transformed. 
It should be emphasized that, due to the presence of the step function $\theta(t)$, the Fourier transform only involves an integration over the positive time axis, which leads to a finite real part in the resulting self-energy $\Sigma_\text{pp}^{\text{R}}(\omega)$.

The explicit expression for the retarded pseudo-particle self-energy shown in Fig.~\ref{appeendixBfig1ret} is
\begin{align}
	\Sigma_\text{pp}^{\text{R}}(\omega) &= \,\int_0^{t_{\max}}\hspace{-1.5em}dv_1 \int_0^{t_{\max}}\hspace{-1.5em}dv_2 \int_0^{t_{\max}}\hspace{-1.5em}dv_3 \int_0^{t_{\max}}\hspace{-1.5em}dv_4 \int_0^{t_{\max}}\hspace{-1.5em}dv_5\nonumber\\
&\times~\exp[\mathrm{i}\omega(v_1+v_2+v_3+v_4+v_5)]\nonumber\\
&\times~\kappa\Bigl\{-\Delta^>(-v_1-v_2-v_3)\Delta^<(v_2+v_3+v_4+v_5)\nonumber\\
&\times~\Delta^>(-v_3-v_4)F\mathcal{G}^>(-v_5)\bar{F}\mathcal{G}^>(-v_4)\bar{F}\nonumber\\
&\times~\mathcal{G}^>(-v_3)F\mathcal{G}^<(-v_2)\bar{F}\mathcal{G}^>(-v_1)F \Bigr\}^{\dagger},
\label{appeendixBeq1ret}
\end{align}

where the global factor $\kappa$ is determined as follows. Since this diagram is topologically equivalent to Fig.~\ref{appeendixBfig1}, it is still a third-order diagram with two crossings, contributing factors of $\mathrm{i}^{3}$ and $(-1)^{2}$, respectively. 
In addition, the hybridization line $\Delta^{<}(v_{2}+v_{3}+v_{4}+v_{5})$ is oriented opposite to the contour direction, resulting in an additional factor of $(-1)$. 
Finally, the four internal vertices on the lower branch contribute a $(-1)^4$ factor, resulting in $\kappa=\mathrm{i}$~.

\begin{figure}[H]
  \centering
    \begin{tikzpicture}[thick, >=stealth]
    \draw[->,blue] (0,0.5) -- (5,0.5);
    \draw[->,blue] (5,-0.5) -- (0,-0.5);

    \draw[dashed,arrowmid] (3.2,0.5) -- (4.1,-0.5);
    \draw[dashed,arrowmid] (0.5,0.5) .. controls (1.2,0) and (1.6,0) .. (2.3,0.5);
    \draw[dotted] (4.1,-0.5) -- (4.1,0.5);
    \draw[dotted] (3.2,-0.5) -- (3.2,0.5);
    \draw[dotted] (2.3,-0.5) -- (2.3,0.5);
    \draw[dotted] (1.4,-0.5) -- (1.4,0.5);
    \draw[dotted] (0.5,-0.5) -- (0.5,0.5);
    \node at (5.7,0) {$t=t_{\max}$};
    \node at (-0.5,0) {$t=0$};
    \filldraw (5,-0.5) circle (2pt);
    \filldraw (3.2,0.5) circle (2pt);
    \filldraw (1.4,0.5) circle (2pt);
    \filldraw (0.5,0.5) circle (2pt);
    \filldraw (2.3,0.5) circle (2pt);
    \filldraw (4.1,-0.5) circle (2pt);
    \draw[<->, thick] (4.1,-0.7) -- node[below] {\(v_1\)} (5.0,-0.7);
    \draw[<->, thick] (3.2,0.7) -- node[above] {\(v_2\)} (4.1,0.7);
    \draw[<->, thick] (2.3,0.7) -- node[above] {\(v_3\)} (3.2,0.7);
    \draw[<->, thick] (1.4,0.7) -- node[above] {\(v_4\)} (2.3,0.7);
    \draw[<->, thick] (0.5,0.7) -- node[above] {\(v_5\)} (1.4,0.7);
    \end{tikzpicture}
    \caption{A representative diagram for the lesser physical Green's function in Scheme III.
    }
  \label{appeendixBfig2}
\end{figure}
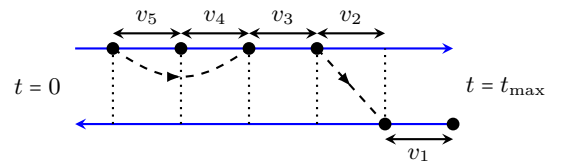

In the following, we present the expressions for the physical Green's functions. 
These can be obtained by removing a single hybridization line from the closed two-particle irreducible diagrams of the Luttinger-Ward functional. 
An example of such a diagram, with one hybridization line removed, is shown in Fig.~\ref{appeendixBfig2}. 
Here, we focus on evaluating the physical Green's function $G^{<}(-v_{1}-v_{2}-v_{3}-v_{4})$ in the time domain. 
According to Eq.~\eqref{eq:GF_physical}, the removed hybridization line must point from the operator at $t = t_{\max} -v_{1}-v_{2}-v_{3}-v_{4}$ on the upper branch to the starting operator at the right bottom corner. 
The resulting expression can be written as
\begin{align}
	G^<(\omega) &= \, 2\mathrm{i}~\text{Im}\biggl\{\int_0^{t_{\max}}\hspace{-1.5em}dv_1 \int_0^{t_{\max}}\hspace{-1.5em}dv_2 \int_0^{t_{\max}}\hspace{-1.5em}dv_3 \int_0^{t_{\max}}\hspace{-1.5em}dv_4\int_0^{t_{\max}}\hspace{-1.5em}dv_5~\nonumber\\
&\times \exp[-\mathrm{i}\omega(v_1+v_2+v_3+v_4)]\Delta^>(v_4+v_5)\Delta^>(v_2)\nonumber\\
&\times \kappa\hspace{0.3em}\text{Tr}\bigl[\mathcal{G}^>(v_1+v_2)F\mathcal{G}^>(v_3)\bar{F}\mathcal{G}^>(v_4)\bar{F}\nonumber\\
&\mathcal{G}^>(v_5)F\mathcal{G}^<(-v_2-v_3-v_4-v_5)\bar{F}\mathcal{G}^>(-v_1)F\bigr] \biggr\}.
\label{appeendixBeq2}
\end{align}
Here, the trace over the pseudo-particle space is evaluated in the presence of the pseudo-particle Green's function $\mathcal{G}^{>}(v_{1}+v_{2})$ covering the whole Keldysh contour. 
Regarding the prefactor $\kappa$, we must determine its value after reinserting the hybridization line. 
The diagram order still contributes a factor $\mathrm{i}^{3}$, and the two remaining crossings yield $(-1)^{2}$. 
All hybridization lines (including the reinserted one) in Fig.~\ref{appeendixBfig2} are oriented along the contour direction, so no additional sign factor arises from their orientation. 
Finally, the vertex on the lower branch at $t=t_{\mathrm{max}}-v_1$ gives an additional $(-1)$.
So, the overall factor is $\kappa=i$~.

\begin{figure}[H]
  \centering
    \begin{tikzpicture}[thick, >=stealth]
    \draw[->,blue] (0,0.5) -- (5,0.5);
    \draw[->,blue] (5,-0.5) -- (0,-0.5);
    \draw[dashed,arrowmid] (0.5,0.5) .. controls (2.4,0) .. (4.1,-0.5);
    \draw[dashed,arrowmid] (3.2,-0.5) .. controls (2.8,-0.1) and (1.8,-0.1) .. (1.4,-0.5);
    \draw[dotted] (4.1,-0.5) -- (4.1,-0.5);
    \draw[dotted] (3.2,-0.5) -- (3.2,-0.5);
    \draw[dotted] (2.3,-0.5) -- (2.3,-0.5);
    \draw[dotted] (1.4,-0.5) -- (1.4,0.5);
    \draw[dotted] (0.5,-0.5) -- (0.5,0.5);
    \node at (5.7,0) {$t=t_{\max}$};
    \node at (-0.5,0) {$t=0$};
    \filldraw (5,-0.5) circle (2pt);
    \filldraw (3.2,-0.5) circle (2pt);
    \filldraw (1.4,-0.5) circle (2pt);
    \filldraw (0.5,0.5) circle (2pt);
    \filldraw (2.3,-0.5) circle (2pt);
    \filldraw (4.1,-0.5) circle (2pt);
    \draw[<->, thick] (4.1,-0.7) -- node[below] {\(v_1\)} (5.0,-0.7);
    \draw[<->, thick] (3.2,-0.7) -- node[below] {\(v_2\)} (4.1,-0.7);
    \draw[<->, thick] (2.3,-0.7) -- node[below] {\(v_3\)} (3.2,-0.7);
    \draw[<->, thick] (1.4,-0.7) -- node[below] {\(v_4\)} (2.3,-0.7);
    \draw[<->, thick] (0.5,0.7) -- node[above] {\(v_5\)} (1.4,0.7);
    \end{tikzpicture}
    \caption{A representative diagram for the greater self-energy in Scheme III.
    }
  \label{appeendixBfig2gtr}
\end{figure}
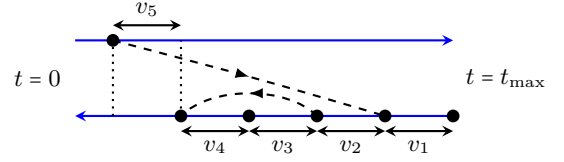

Our final example is a contribution to the greater physical Green's function. 
We consider a diagram that is topologically equivalent to Fig.~\ref{appeendixBfig2}, but with only one operator placed on the upper branch, see Fig.~\ref{appeendixBfig2gtr}. 
Its explicit expression is 

\begin{align}
G^>(\omega) =& 2\mathrm{i}~\text{Im}\biggl\{\int_0^{t_{\max}}\hspace{-1.5em}dv_1 \int_0^{t_{\max}}\hspace{-1.5em}dv_2 \int_0^{t_{\max}}\hspace{-1.5em}dv_3 \int_0^{t_{\max}}\hspace{-1.5em}dv_4\int_0^{t_{\max}}\hspace{-1.5em}dv_5\nonumber\\
\times& \exp[\mathrm{i}\omega(v_1+v_2+v_3)]\nonumber\\
\times& \kappa\hspace{0.5em}\Delta^>(-v_3-v_4)\Delta^>(v_2+v_3+v_4+v_5)\nonumber\\
\times& \text{Tr}\bigl[-\mathcal{G}^>(v_1+v_2+v_3+v_4+v_5)F\mathcal{G}^<(-v_5)\bar{F}\nonumber\\
&\mathcal{G}^>(-v_4)\bar{F}\mathcal{G}^>(-v_3)F\mathcal{G}^>(-v_2)\bar{F}\mathcal{G}^>(-v_1)F\bigr]^*\biggr\}.
\label{appeendixBeq2gtr}
\end{align}

After reinserting the hybridization line pointing from the operator at $t = t_{\max} - v_{1} - v_{2} - v_{3}$ to the one at $t = t_{\max}$, the resulting diagram remains a third-order diagram with two crossings of hybridization lines. 
The reinserted hybridization line is oriented opposite to the contour direction, contributing an additional factor $(-1)^{1}$. 
Furthermore, there are three internal variables on the lower branch.
The final global factor becomes $\kappa=-\mathrm{i}$~.

\bibliography{mybibtex}

@article{kuramoto1984self,
  title={Self-consistent perturbation theory for dynamics of valence fluctuations: III. Zero-temperature limit},
  author={Kuramoto, Y and Kojima, H},
  journal={Z. Phys. B Condens. Matter},
  volume={57},
  number={2},
  pages={95--105},
  year={1984},
  publisher={Springer}
}

@article{muller1984self,
  title={Self-consistent perturbation theory of the Anderson model: ground state properties},
  author={M{\"u}ller-Hartmann, E},
  journal={Z. Phys. B Condens. Matter},
  volume={57},
  number={4},
  pages={281--287},
  year={1984},
  publisher={Springer}
}

@article{Kroha1997,
  title = {Unified Description of Fermi and Non-Fermi Liquid Behavior in a Conserving Slave Boson Approximation for Strongly Correlated Impurity Models},
  author = {Kroha, J. and W\"olfle, P. and Costi, T. A.},
  journal = {Phys. Rev. Lett.},
  volume = {79},
  issue = {2},
  pages = {261--264},
  numpages = {0},
  year = {1997},
  month = {Jul},
  publisher = {American Physical Society},
  doi = {10.1103/PhysRevLett.79.261},
  url = {https://link.aps.org/doi/10.1103/PhysRevLett.79.261}
}

@incollection{kroha2004fermi,
  title={Fermi and Non-Fermi Liquid Behavior of Quantum Impurity Models: A Diagrammatic Pseudo-Particle Approach},
  author={Kroha, J and W{\"o}lfle, P},
  booktitle={Theoretical Methods for Strongly Correlated Electrons},
  pages={297--339},
  year={2004},
  publisher={Springer}
}

@article{kirchner2002self,
  title={Self-consistent conserving theory for quantum impurity systems: Renormalization group analysis},
  author={Kirchner, Stefan and Kroha, Johann},
  journal={J. Low Temp. Phys.},
  volume={126},
  number={3},
  pages={1233--1249},
  year={2002},
  publisher={Springer}
}

@article{profuno2015,
  title = {Quantum Monte Carlo for correlated out-of-equilibrium nanoelectronic devices},
  author = {Profumo, Rosario E. V. and Groth, Christoph and Messio, Laura and Parcollet, Olivier and Waintal, Xavier},
  journal = {Phys. Rev. B},
  volume = {91},
  issue = {24},
  pages = {245154},
  numpages = {18},
  year = {2015},
  month = {Jun},
  publisher = {American Physical Society},
  doi = {10.1103/PhysRevB.91.245154},
  url = {https://link.aps.org/doi/10.1103/PhysRevB.91.245154}
}

@Article{Chu2024,
	title={{Efficient construction of the Feynman-Vernon influence functional as matrix product states}},
	author={Chu Guo and Ruofan Chen},
	journal={SciPost Phys. Core},
	volume={7},
	pages={063},
	year={2024},
	publisher={SciPost},
	doi={10.21468/SciPostPhysCore.7.3.063},
	url={https://scipost.org/10.21468/SciPostPhysCore.7.3.063},
}

@article{Chen2024,
  title = {Grassmann time-evolving matrix product operators for quantum impurity models},
  author = {Chen, Ruofan and Xu, Xiansong and Guo, Chu},
  journal = {Phys. Rev. B},
  volume = {109},
  issue = {4},
  pages = {045140},
  numpages = {12},
  year = {2024},
  month = {Jan},
  publisher = {American Physical Society},
  doi = {10.1103/PhysRevB.109.045140},
  url = {https://link.aps.org/doi/10.1103/PhysRevB.109.045140}
}

@article{Karp2022,
  title = {Superconductivity and antiferromagnetism in ${\mathrm{NdNiO}}_{2}$ and ${\mathrm{CaCuO}}_{2}$: A cluster DMFT study},
  author = {Karp, Jonathan and Hampel, Alexander and Millis, Andrew J.},
  journal = {Phys. Rev. B},
  volume = {105},
  issue = {20},
  pages = {205131},
  numpages = {11},
  year = {2022},
  month = {May},
  publisher = {American Physical Society},
  doi = {10.1103/PhysRevB.105.205131},
  url = {https://link.aps.org/doi/10.1103/PhysRevB.105.205131}
}

@article{Nagai2020,
  title = {DMFT Reveals the Non-Hermitian Topology and Fermi Arcs in Heavy-Fermion Systems},
  author = {Nagai, Yuki and Qi, Yang and Isobe, Hiroki and Kozii, Vladyslav and Fu, Liang},
  journal = {Phys. Rev. Lett.},
  volume = {125},
  issue = {22},
  pages = {227204},
  numpages = {7},
  year = {2020},
  month = {Nov},
  publisher = {American Physical Society},
  doi = {10.1103/PhysRevLett.125.227204},
  url = {https://link.aps.org/doi/10.1103/PhysRevLett.125.227204}
}

@article{Brito2016,
  title = {Metal-Insulator Transition in ${\mathrm{VO}}_{2}$: A $\mathrm{DFT}+\mathrm{DMFT}$ Perspective},
  author = {Brito, W. H. and Aguiar, M. C. O. and Haule, K. and Kotliar, G.},
  journal = {Phys. Rev. Lett.},
  volume = {117},
  issue = {5},
  pages = {056402},
  numpages = {6},
  year = {2016},
  month = {Jul},
  publisher = {American Physical Society},
  doi = {10.1103/PhysRevLett.117.056402},
  url = {https://link.aps.org/doi/10.1103/PhysRevLett.117.056402}
}

@article{strack1992,
  title = {Dynamics of a hole in the t-J model with local disorder: Exact results for high dimensions},
  author = {Strack, Rainer and Vollhardt, Dieter},
  journal = {Phys. Rev. B},
  volume = {46},
  issue = {21},
  pages = {13852--13861},
  numpages = {0},
  year = {1992},
  month = {Dec},
  publisher = {American Physical Society},
  doi = {10.1103/PhysRevB.46.13852},
  url = {https://link.aps.org/doi/10.1103/PhysRevB.46.13852}
}

@article{martinez1991,
  title = {Spin polarons in the t-J model},
  author = {Martinez, Gerardo and Horsch, Peter},
  journal = {Phys. Rev. B},
  volume = {44},
  issue = {1},
  pages = {317--331},
  numpages = {0},
  year = {1991},
  month = {Jul},
  publisher = {American Physical Society},
  doi = {10.1103/PhysRevB.44.317},
  url = {https://link.aps.org/doi/10.1103/PhysRevB.44.317}
}

@article{Pruschke1993,
  title = {Hubbard model at infinite dimensions: Thermodynamic and transport properties},
  author = {Pruschke, Th. and Cox, D. L. and Jarrell, M.},
  journal = {Phys. Rev. B},
  volume = {47},
  issue = {7},
  pages = {3553--3565},
  numpages = {0},
  year = {1993},
  month = {Feb},
  publisher = {American Physical Society},
  doi = {10.1103/PhysRevB.47.3553},
  url = {https://link.aps.org/doi/10.1103/PhysRevB.47.3553}
}

@article{eckstein2013,
  title = {Photoinduced States in a Mott Insulator},
  author = {Eckstein, Martin and Werner, Philipp},
  journal = {Phys. Rev. Lett.},
  volume = {110},
  issue = {12},
  pages = {126401},
  numpages = {5},
  year = {2013},
  month = {Mar},
  publisher = {American Physical Society},
  doi = {10.1103/PhysRevLett.110.126401},
  url = {https://link.aps.org/doi/10.1103/PhysRevLett.110.126401}
}

@article{ray2023nonthermal,
  title = {Nonthermal superconductivity in photodoped multiorbital Hubbard systems},
  author = {Ray, Sujay and Murakami, Yuta and Werner, Philipp},
  journal = {Phys. Rev. B},
  volume = {108},
  issue = {17},
  pages = {174515},
  numpages = {11},
  year = {2023},
  month = {Nov},
  publisher = {American Physical Society},
  doi = {10.1103/PhysRevB.108.174515},
  url = {https://link.aps.org/doi/10.1103/PhysRevB.108.174515}
}

@article{ray2024photoinduced,
  title = {Photoinduced ferromagnetic and superconducting orders in multiorbital Hubbard models},
  author = {Ray, Sujay and Werner, Philipp},
  journal = {Phys. Rev. B},
  volume = {110},
  issue = {4},
  pages = {L041109},
  numpages = {6},
  year = {2024},
  month = {Jul},
  publisher = {American Physical Society},
  doi = {10.1103/PhysRevB.110.L041109},
  url = {https://link.aps.org/doi/10.1103/PhysRevB.110.L041109}
}

@article{murakami2023photo,
  title={Photo-induced nonequilibrium states in Mott insulators},
  author={Murakami, Yuta and Gole{\v{z}}, Denis and Eckstein, Martin and Werner, Philipp},
  journal={arXiv:2310.05201},
  url ={https://doi.org/10.48550/arXiv.2310.05201},
  year={2023}
}

@article{bittner2020photoenhanced,
  title = {Photoenhanced excitonic correlations in a Mott insulator with nonlocal interactions},
  author = {Bittner, Nikolaj and Gole\ifmmode \check{z}\else \v{z}\fi{}, Denis and Eckstein, Martin and Werner, Philipp},
  journal = {Phys. Rev. B},
  volume = {101},
  issue = {8},
  pages = {085127},
  numpages = {13},
  year = {2020},
  month = {Feb},
  publisher = {American Physical Society},
  doi = {10.1103/PhysRevB.101.085127},
  url = {https://link.aps.org/doi/10.1103/PhysRevB.101.085127}
}

@article{ueda2025exotic,
  title={Exotic Doublon-Holon Pairing State in Photodoped Mott Insulators},
  author={Ueda, Ryota and Sarkar, Madhumita and Lenar{\v{c}}i{\v{c}}, Zala and Gole{\v{z}}, Denis and Kuroki, Kazuhiko and Kaneko, Tatsuya},
  journal={arXiv:2504.03324},
  url={https://doi.org/10.48550/arXiv.2504.03324},
  year={2025}
}

@article{werner2018enhanced,
  title = {Enhanced pairing susceptibility in a photodoped two-orbital Hubbard model},
  author = {Werner, Philipp and Strand, Hugo U. R. and Hoshino, Shintaro and Murakami, Yuta and Eckstein, Martin},
  journal = {Phys. Rev. B},
  volume = {97},
  issue = {16},
  pages = {165119},
  numpages = {11},
  year = {2018},
  month = {Apr},
  publisher = {American Physical Society},
  doi = {10.1103/PhysRevB.97.165119},
  url = {https://link.aps.org/doi/10.1103/PhysRevB.97.165119}
}

@article{murakami2023spin,
  title = {Spin, Charge, and $\ensuremath{\eta}$-Spin Separation in One-Dimensional Photodoped Mott Insulators},
  author = {Murakami, Yuta and Takayoshi, Shintaro and Kaneko, Tatsuya and L\"auchli, Andreas M. and Werner, Philipp},
  journal = {Phys. Rev. Lett.},
  volume = {130},
  issue = {10},
  pages = {106501},
  numpages = {7},
  year = {2023},
  month = {Mar},
  publisher = {American Physical Society},
  doi = {10.1103/PhysRevLett.130.106501},
  url = {https://link.aps.org/doi/10.1103/PhysRevLett.130.106501}
}

@article{werner2012nonthermal,
  title = {Nonthermal symmetry-broken states in the strongly interacting Hubbard model},
  author = {Werner, Philipp and Tsuji, Naoto and Eckstein, Martin},
  journal = {Phys. Rev. B},
  volume = {86},
  issue = {20},
  pages = {205101},
  numpages = {7},
  year = {2012},
  month = {Nov},
  publisher = {American Physical Society},
  doi = {10.1103/PhysRevB.86.205101},
  url = {https://link.aps.org/doi/10.1103/PhysRevB.86.205101}
}

@article{Jarrell1996,
title = {Bayesian inference and the analytic continuation of imaginary-time quantum Monte Carlo data},
journal = {Phys. Rep.},
volume = {269},
number = {3},
pages = {133-195},
year = {1996},
issn = {0370-1573},
doi = {https://doi.org/10.1016/0370-1573(95)00074-7},
url = {https://www.sciencedirect.com/science/article/pii/0370157395000747},
author = {Mark Jarrell and J.E. Gubernatis}
}

@phdthesis{blumer2003mott,
  author      = {Nils Blümer},
  title       = {Mott-Hubbard Metal-Insulator Transition and Optical Conductivity in High Dimensions},
  school      = {University of Augsburg},
  year        = {2003},
  type        = {Ph.D. thesis},
  url         = {https://www.blogs.uni-mainz.de/fb08-komet337/files/2018/04/bluemer_color.pdf}
}

@article{georges1996dynamical,
  title = {Dynamical mean-field theory of strongly correlated fermion systems and the limit of infinite dimensions},
  author = {Georges, Antoine and Kotliar, Gabriel and Krauth, Werner and Rozenberg, Marcelo J.},
  journal = {Rev. Mod. Phys.},
  volume = {68},
  issue = {1},
  pages = {13--125},
  numpages = {0},
  year = {1996},
  month = {Jan},
  publisher = {American Physical Society},
  doi = {10.1103/RevModPhys.68.13},
  url = {https://link.aps.org/doi/10.1103/RevModPhys.68.13}
}

@article{metzner1989correlated,
  title = {Correlated Lattice Fermions in $d=\ensuremath{\infty}$ Dimensions},
  author = {Metzner, Walter and Vollhardt, Dieter},
  journal = {Phys. Rev. Lett.},
  volume = {62},
  issue = {3},
  pages = {324--327},
  numpages = {0},
  year = {1989},
  month = {Jan},
  publisher = {American Physical Society},
  doi = {10.1103/PhysRevLett.62.324},
  url = {https://link.aps.org/doi/10.1103/PhysRevLett.62.324}
}

@article{anderson1961localized,
  title = {Localized Magnetic States in Metals},
  author = {Anderson, P. W.},
  journal = {Phys. Rev.},
  volume = {124},
  issue = {1},
  pages = {41--53},
  numpages = {0},
  year = {1961},
  month = {Oct},
  publisher = {American Physical Society},
  doi = {10.1103/PhysRev.124.41},
  url = {https://link.aps.org/doi/10.1103/PhysRev.124.41}

}

@article{rozenberg1992mott,
  title = {Mott-Hubbard Transition in Infinite Dimensions},
  author = {Noack, R. M. and Gebhard, F.},
  journal = {Phys. Rev. Lett.},
  volume = {82},
  issue = {9},
  pages = {1915--1918},
  numpages = {0},
  year = {1999},
  month = {Mar},
  publisher = {American Physical Society},
  doi = {10.1103/PhysRevLett.82.1915},
  url = {https://link.aps.org/doi/10.1103/PhysRevLett.82.1915}
}

@article{shim2007modeling,
  author  = {J. H. Shim and K. Haule and G. Kotliar},
  title   = {Modeling the localized-to-itinerant electronic transition in the heavy fermion system CeIrIn$_5$},
  journal = {Science},
  volume  = {318},
  number  = {5856},
  pages   = {1615--1617},
  year    = {2007},
  doi     = {10.1126/science.1149064},
  url     = {https://www.science.org/doi/abs/10.1126/science.1149064}
}

@article{werner2020nickelate,
  title = {Nickelate superconductors: Multiorbital nature and spin freezing},
  author = {Werner, Philipp and Hoshino, Shintaro},
  journal = {Phys. Rev. B},
  volume = {101},
  issue = {4},
  pages = {041104},
  numpages = {5},
  year = {2020},
  month = {Jan},
  publisher = {American Physical Society},
  doi = {10.1103/PhysRevB.101.041104},
  url = {https://link.aps.org/doi/10.1103/PhysRevB.101.041104}
}

@article{rubtsov2005,
  title = {Continuous-time quantum Monte Carlo method for fermions},
  author = {Rubtsov, A. N. and Savkin, V. V. and Lichtenstein, A. I.},
  journal = {Phys. Rev. B},
  volume = {72},
  issue = {3},
  pages = {035122},
  numpages = {9},
  year = {2005},
  month = {Jul},
  publisher = {American Physical Society},
  doi = {10.1103/PhysRevB.72.035122},
  url = {https://link.aps.org/doi/10.1103/PhysRevB.72.035122}
}

@article{gull2011continuous,
  title = {Continuous-time Monte Carlo methods for quantum impurity models},
  author = {Gull, Emanuel and Millis, Andrew J. and Lichtenstein, Alexander I. and Rubtsov, Alexey N. and Troyer, Matthias and Werner, Philipp},
  journal = {Rev. Mod. Phys.},
  volume = {83},
  issue = {2},
  pages = {349--404},
  numpages = {0},
  year = {2011},
  month = {May},
  publisher = {American Physical Society},
  doi = {10.1103/RevModPhys.83.349},
  url = {https://link.aps.org/doi/10.1103/RevModPhys.83.349}
}

@article{werner2006continuous,
  title = {Continuous-Time Solver for Quantum Impurity Models},
  author = {Werner, Philipp and Comanac, Armin and de' Medici, Luca and Troyer, Matthias and Millis, Andrew J.},
  journal = {Phys. Rev. Lett.},
  volume = {97},
  issue = {7},
  pages = {076405},
  numpages = {4},
  year = {2006},
  month = {Aug},
  publisher = {American Physical Society},
  doi = {10.1103/PhysRevLett.97.076405},
  url = {https://link.aps.org/doi/10.1103/PhysRevLett.97.076405}

}

@article{amaricci2022edipack,
  title={EDIpack: A parallel exact diagonalization package for quantum impurity problems},
  author={Amaricci, Adriano and Crippa, Lorenzo and Scazzola, A and Petocchi, Francesco and Mazza, G and de Medici, Luca and Capone, M},
  journal={Comput. Phys. Commun.},
  volume={273},
  pages={108261},
  year={2022},
  publisher={Elsevier},
  doi={https://doi.org/10.1016/j.cpc.2021.108261}
}

@article{bulla2008numerical,
  title = {Numerical renormalization group method for quantum impurity systems},
  author = {Bulla, Ralf and Costi, Theo A. and Pruschke, Thomas},
  journal = {Rev. Mod. Phys.},
  volume = {80},
  issue = {2},
  pages = {395--450},
  numpages = {0},
  year = {2008},
  month = {Apr},
  publisher = {American Physical Society},
  doi = {10.1103/RevModPhys.80.395},
  url = {https://link.aps.org/doi/10.1103/RevModPhys.80.395}
}

@article{ganahl2015efficient,
  title = {Efficient DMFT impurity solver using real-time dynamics with matrix product states},
  author = {Ganahl, Martin and Aichhorn, Markus and Evertz, Hans Gerd and Thunstr\"om, Patrik and Held, Karsten and Verstraete, Frank},
  journal = {Phys. Rev. B},
  volume = {92},
  issue = {15},
  pages = {155132},
  numpages = {12},
  year = {2015},
  month = {Oct},
  publisher = {American Physical Society},
  doi = {10.1103/PhysRevB.92.155132},
  url = {https://link.aps.org/doi/10.1103/PhysRevB.92.155132}
}

@article{ng2023real,
  title = {Real-time evolution of Anderson impurity models via tensor network influence functionals},
  author = {Ng, Nathan and Park, Gunhee and Millis, Andrew J. and Chan, Garnet Kin-Lic and Reichman, David R.},
  journal = {Phys. Rev. B},
  volume = {107},
  issue = {12},
  pages = {125103},
  numpages = {8},
  year = {2023},
  month = {Mar},
  publisher = {American Physical Society},
  doi = {10.1103/PhysRevB.107.125103},
  url = {https://link.aps.org/doi/10.1103/PhysRevB.107.125103}
}

@article{eckstein2010nonequilibrium,
  title = {Nonequilibrium dynamical mean-field calculations based on the noncrossing approximation and its generalizations},
  author = {Eckstein, Martin and Werner, Philipp},
  journal = {Phys. Rev. B},
  volume = {82},
  issue = {11},
  pages = {115115},
  numpages = {13},
  year = {2010},
  month = {Sep},
  publisher = {American Physical Society},
  doi = {10.1103/PhysRevB.82.115115},
  url = {https://link.aps.org/doi/10.1103/PhysRevB.82.115115}

}

@article{kirchner2004dynamical,
  title = {Dynamical properties of the Anderson impurity model within a diagrammatic pseudoparticle approach},
  author = {Kirchner, S. and Kroha, J. and W\"olfle, P.},
  journal = {Phys. Rev. B},
  volume = {70},
  issue = {16},
  pages = {165102},
  numpages = {14},
  year = {2004},
  month = {Oct},
  publisher = {American Physical Society},
  doi = {10.1103/PhysRevB.70.165102},
  url = {https://link.aps.org/doi/10.1103/PhysRevB.70.165102}

}

@article{kroha2005conserving,
  title={Conserving diagrammatic approximations for quantum impurity models: NCA and CTMA},
  author={Kroha, Johann and W{\"o}lfle, Peter},
  journal={J. Phys. Soc. Jpn.},
  volume={74},
  number={1},
  pages={16--26},
  year={2005},
  doi={https://doi.org/10.1143/jpsj.74.16},
  publisher={The Physical Society of Japan}
}

@article{sposetti2016qualitative,
  title = {Qualitative breakdown of the noncrossing approximation for the symmetric one-channel Anderson impurity model at all temperatures},
  author = {Sposetti, C. N. and Manuel, L. O. and Roura-Bas, P.},
  journal = {Phys. Rev. B},
  volume = {94},
  issue = {8},
  pages = {085139},
  numpages = {8},
  year = {2016},
  month = {Aug},
  publisher = {American Physical Society},
  doi = {10.1103/PhysRevB.94.085139},
  url = {https://link.aps.org/doi/10.1103/PhysRevB.94.085139}

}

@article{costi1996spectral,
  title = {Spectral properties of the Anderson impurity model: Comparison of numerical-renormalization-group and noncrossing-approximation results},
  author = {Costi, T. A. and Kroha, J. and W\"olfle, P.},
  journal = {Phys. Rev. B},
  volume = {53},
  issue = {4},
  pages = {1850--1865},
  numpages = {0},
  year = {1996},
  month = {Jan},
  publisher = {American Physical Society},
  doi = {10.1103/PhysRevB.53.1850},
  url = {https://link.aps.org/doi/10.1103/PhysRevB.53.1850}

}

@article{oseledets2011tensor,
author = {Oseledets, I. V.},
title = {Tensor-Train Decomposition},
journal = {SIAM J. Sci. Comput.},
volume = {33},
number = {5},
pages = {2295-2317},
year = {2011},
doi = {10.1137/090752286},
url = {https://doi.org/10.1137/090752286}
}

@article{oseledets2010tt,
title = {TT-cross approximation for multidimensional arrays},
author={Oseledets, Ivan and Tyrtyshnikov, Eugene},
journal = {Linear Algebra Appl.},
volume = {432},
number = {1},
pages = {70-88},
year = {2010},
issn = {0024-3795},
doi = {https://doi.org/10.1016/j.laa.2009.07.024},
url = {https://www.sciencedirect.com/science/article/pii/S0024379509003747}
}

@article{nunez2022learning,
  title = {Learning Feynman Diagrams with Tensor Trains},
  author = {N\'u\~nez Fern\'andez, Yuriel and Jeannin, Matthieu and Dumitrescu, Philipp T. and Kloss, Thomas and Kaye, Jason and Parcollet, Olivier and Waintal, Xavier},
  journal = {Phys. Rev. X},
  volume = {12},
  issue = {4},
  pages = {041018},
  numpages = {30},
  year = {2022},
  month = {Nov},
  publisher = {American Physical Society},
  doi = {10.1103/PhysRevX.12.041018},
  url = {https://link.aps.org/doi/10.1103/PhysRevX.12.041018}

}

@article{ritter2024quantics,
  title = {Quantics Tensor Cross Interpolation for High-Resolution Parsimonious Representations of Multivariate Functions},
  author = {Ritter, Marc K. and N\'u\~nez Fern\'andez, Yuriel and Wallerberger, Markus and von Delft, Jan and Shinaoka, Hiroshi and Waintal, Xavier},
  journal = {Phys. Rev. Lett.},
  volume = {132},
  issue = {5},
  pages = {056501},
  numpages = {6},
  year = {2024},
  month = {Jan},
  publisher = {American Physical Society},
  doi = {10.1103/PhysRevLett.132.056501},
  url = {https://link.aps.org/doi/10.1103/PhysRevLett.132.056501}
}

@article{shinaoka2023multiscale,
  title = {Multiscale Space-Time Ansatz for Correlation Functions of Quantum Systems Based on Quantics Tensor Trains},
  author = {Shinaoka, Hiroshi and Wallerberger, Markus and Murakami, Yuta and Nogaki, Kosuke and Sakurai, Rihito and Werner, Philipp and Kauch, Anna},
  journal = {Phys. Rev. X},
  volume = {13},
  issue = {2},
  pages = {021015},
  numpages = {27},
  year = {2023},
  month = {Apr},
  publisher = {American Physical Society},
  doi = {10.1103/PhysRevX.13.021015},
  url = {https://link.aps.org/doi/10.1103/PhysRevX.13.021015}
}

@article{fernandez2024learning,
title={{Learning tensor networks with tensor cross interpolation: New algorithms and libraries}},
author={Yuriel N{\'u}{\~n}ez Fern{\'a}ndez and Marc K. Ritter and Matthieu Jeannin and Jheng-Wei Li and Thomas Kloss and Thibaud Louvet and Satoshi Terasaki and Olivier Parcollet and Jan von Delft and Hiroshi Shinaoka and Xavier Waintal},
journal={SciPost Phys.},
volume={18},
pages={104},
year={2025},
publisher={SciPost},
doi={10.21468/SciPostPhys.18.3.104},
url={https://scipost.org/10.21468/SciPostPhys.18.3.104},
}

@article{sroda2024high,
  title={High-resolution nonequilibrium $ GW $ calculations based on quantics tensor trains},
  author={{\'S}roda, Maksymilian and Inayoshi, Ken and Shinaoka, Hiroshi and Werner, Philipp},
  journal={arXiv:2412.14032},
  url={https://doi.org/10.48550/arXiv.2412.14032},
  year={2024}
}

@article{murray2024nonequilibrium,
  title = {Nonequilibrium diagrammatic many-body simulations with quantics tensor trains},
  author = {Murray, Matthias and Shinaoka, Hiroshi and Werner, Philipp},
  journal = {Phys. Rev. B},
  volume = {109},
  issue = {16},
  pages = {165135},
  numpages = {12},
  year = {2024},
  month = {Apr},
  publisher = {American Physical Society},
  doi = {10.1103/PhysRevB.109.165135},
  url = {https://link.aps.org/doi/10.1103/PhysRevB.109.165135}
}

@article{erpenbeck2023tensor,
  title = {Tensor train continuous time solver for quantum impurity models},
  author = {Erpenbeck, A. and Lin, W.-T. and Blommel, T. and Zhang, L. and Iskakov, S. and Bernheimer, L. and N\'u\~nez-Fern\'andez, Y. and Cohen, G. and Parcollet, O. and Waintal, X. and Gull, E.},
  journal = {Phys. Rev. B},
  volume = {107},
  issue = {24},
  pages = {245135},
  numpages = {11},
  year = {2023},
  month = {Jun},
  publisher = {American Physical Society},
  doi = {10.1103/PhysRevB.107.245135},
  url = {https://link.aps.org/doi/10.1103/PhysRevB.107.245135}
}

@article{kim2025strong,
  title = {Strong coupling impurity solver based on quantics tensor cross interpolation},
  author = {Kim, Aaram J. and Werner, Philipp},
  journal = {Phys. Rev. B},
  volume = {111},
  issue = {12},
  pages = {125120},
  numpages = {11},
  year = {2025},
  month = {Mar},
  publisher = {American Physical Society},
  doi = {10.1103/PhysRevB.111.125120},
  url = {https://link.aps.org/doi/10.1103/PhysRevB.111.125120}
}

@article{nayak2025steady,
  title = {Steady-state dynamical mean field theory based on influence functional matrix product states},
  author = {Nayak, Mithilesh and Thoenniss, Julian and Sonner, Michael and Abanin, Dmitry A. and Werner, Philipp},
  journal = {Phys. Rev. B},
  volume = {112},
  issue = {3},
  pages = {035103},
  numpages = {21},
  year = {2025},
  month = {Jul},
  publisher = {American Physical Society},
  doi = {10.1103/xsbn-jk16},
  url = {https://link.aps.org/doi/10.1103/xsbn-jk16}
}

@article{cao2021tree,
  title = {Tree tensor-network real-time multiorbital impurity solver: Spin-orbit coupling and correlation functions in ${\mathrm{Sr}}_{2}{\mathrm{RuO}}_{4}$},
  author = {Cao, X. and Lu, Y. and Hansmann, P. and Haverkort, M. W.},
  journal = {Phys. Rev. B},
  volume = {104},
  issue = {11},
  pages = {115119},
  numpages = {16},
  year = {2021},
  month = {Sep},
  publisher = {American Physical Society},
  doi = {10.1103/PhysRevB.104.115119},
  url = {https://link.aps.org/doi/10.1103/PhysRevB.104.115119}
}

@phdthesis{grundner2025tensor,
  author = {Martin Grundner},
  title = {Tensor Network Impurity Solvers},
  school = {Ludwig Maximilian University of Munich},
  year = {2025},
  type = {Ph.D. thesis},
  url = {https://edoc.ub.uni-muenchen.de/35102/1/Grundner_Martin.pdf}
}

@article{stadler2016interleaved,
  title = {Interleaved numerical renormalization group as an efficient multiband impurity solver},
  author = {Stadler, K. M. and Mitchell, A. K. and von Delft, J. and Weichselbaum, A.},
  journal = {Phys. Rev. B},
  volume = {93},
  issue = {23},
  pages = {235101},
  numpages = {16},
  year = {2016},
  month = {Jun},
  publisher = {American Physical Society},
  doi = {10.1103/PhysRevB.93.235101},
  url = {https://link.aps.org/doi/10.1103/PhysRevB.93.235101}

}

@article{yu2025inchworm,
  title={Inchworm Tensor Train Hybridization Expansion Quantum Impurity Solver},
  author={Yu, Yang and Erpenbeck, Andr{\'e} and Zgid, Dominika and Cohen, Guy and Parcollet, Olivier and Gull, Emanuel},
  journal={arXiv:2505.16117},
  url={https://doi.org/10.48550/arXiv.2505.16117},
  year={2025}
}

@article{matsuura2025tensor,
  title = {Tensor cross interpolation approach for quantum impurity problems based on the weak-coupling expansion},
  author = {Matsuura, Shuta and Shinaoka, Hiroshi and Werner, Philipp and Tsuji, Naoto},
  journal = {Phys. Rev. B},
  volume = {111},
  issue = {15},
  pages = {155150},
  numpages = {15},
  year = {2025},
  month = {Apr},
  publisher = {American Physical Society},
  doi = {10.1103/PhysRevB.111.155150},
  url = {https://link.aps.org/doi/10.1103/PhysRevB.111.155150}
}

@article{eidelstein2020multiorbital,
  title = {Multiorbital Quantum Impurity Solver for General Interactions and Hybridizations},
  author = {Eidelstein, Eitan and Gull, Emanuel and Cohen, Guy},
  journal = {Phys. Rev. Lett.},
  volume = {124},
  issue = {20},
  pages = {206405},
  numpages = {5},
  year = {2020},
  month = {May},
  publisher = {American Physical Society},
  doi = {10.1103/PhysRevLett.124.206405},
  url = {https://link.aps.org/doi/10.1103/PhysRevLett.124.206405}
}

@article{antipov2017currents,
  title = {Currents and Green's functions of impurities out of equilibrium: Results from inchworm quantum Monte Carlo},
  author = {Antipov, Andrey E. and Dong, Qiaoyuan and Kleinhenz, Joseph and Cohen, Guy and Gull, Emanuel},
  journal = {Phys. Rev. B},
  volume = {95},
  issue = {8},
  pages = {085144},
  numpages = {10},
  year = {2017},
  month = {Feb},
  publisher = {American Physical Society},
  doi = {10.1103/PhysRevB.95.085144},
  url = {https://link.aps.org/doi/10.1103/PhysRevB.95.085144}
}

@article{eckstein2024solving,
  title={Solving quantum impurity models in the non-equilibrium steady state with tensor trains},
  author={Eckstein, Martin},
  journal={arXiv:2410.19707},
  url={https://doi.org/10.48550/arXiv.2410.19707},
  year={2024}
}

@incollection{keldysh2024diagram,
  author = {L. V. Keldysh},
  title = {Diagram technique for nonequilibrium processes},
  journal = {Sov. Phys. JETP},
  volume = {20},
  pages = {1018--1026},
  year = {1965},
  note = {Zh. Eksp. Teor. Fiz. 47, 1515--1527 (1964)}
}

@article{kubo1957statistical,
  title={Statistical-mechanical theory of irreversible processes. I. General theory and simple applications to magnetic and conduction problems},
  author={Kubo, Ryogo},
  journal={J. Phys. Soc. Jpn.},
  volume={12},
  number={6},
  pages={570--586},
  year={1957},
  url={https://doi.org/10.1143/JPSJ.12.570},
  publisher={The Physical Society of Japan}
}

@article{martin1959theory,
  title = {Theory of Many-Particle Systems. I},
  author = {Martin, Paul C. and Schwinger, Julian},
  journal = {Phys. Rev.},
  volume = {115},
  issue = {6},
  pages = {1342--1373},
  numpages = {0},
  year = {1959},
  month = {Sep},
  publisher = {American Physical Society},
  doi = {10.1103/PhysRev.115.1342},
  url = {https://link.aps.org/doi/10.1103/PhysRev.115.1342}
}

@article{li2021nonequilibrium,
  title = {Nonequilibrium steady-state theory of photodoped Mott insulators},
  author = {Li, Jiajun and Eckstein, Martin},
  journal = {Phys. Rev. B},
  volume = {103},
  issue = {4},
  pages = {045133},
  numpages = {13},
  year = {2021},
  month = {Jan},
  publisher = {American Physical Society},
  doi = {10.1103/PhysRevB.103.045133},
  url = {https://link.aps.org/doi/10.1103/PhysRevB.103.045133}
}

@article{picano2021quantum,
  title = {Quantum Boltzmann equation for strongly correlated electrons},
  author = {Picano, Antonio and Li, Jiajun and Eckstein, Martin},
  journal = {Phys. Rev. B},
  volume = {104},
  issue = {8},
  pages = {085108},
  numpages = {11},
  year = {2021},
  month = {Aug},
  publisher = {American Physical Society},
  doi = {10.1103/PhysRevB.104.085108},
  url = {https://link.aps.org/doi/10.1103/PhysRevB.104.085108}

}

@article{schuler2020nessi,
  title={NESSi: The non-equilibrium systems simulation package},
  author={Sch{\"u}ler, Michael and Gole{\v{z}}, Denis and Murakami, Yuta and Bittner, Nikolaj and Herrmann, Andreas and Strand, Hugo UR and Werner, Philipp and Eckstein, Martin},
  journal={Comput. Phys. Commun.},
  volume={257},
  pages={107484},
  year={2020},
  url={https://doi.org/10.1016/j.cpc.2020.107484},
  publisher={Elsevier}
}

@article{baym1961conservation,
  title = {Conservation Laws and Correlation Functions},
  author = {Baym, Gordon and Kadanoff, Leo P.},
  journal = {Phys. Rev.},
  volume = {124},
  issue = {2},
  pages = {287--299},
  numpages = {0},
  year = {1961},
  month = {Oct},
  publisher = {American Physical Society},
  doi = {10.1103/PhysRev.124.287},
  url = {https://link.aps.org/doi/10.1103/PhysRev.124.287}
}

@article{barnes1976new,
  title={New method for the Anderson model},
  author={Barnes, SE},
  journal={J. Phys. F: Met. Phys.},
  volume={6},
  number={7},
  pages={1375},
  year={1976},
  url={https://iopscience.iop.org/article/10.1088/0305-4608/6/7/018},
  publisher={IOP Publishing}
}

@article{coleman1984new,
  title = {New approach to the mixed-valence problem},
  author = {Coleman, Piers},
  journal = {Phys. Rev. B},
  volume = {29},
  issue = {6},
  pages = {3035--3044},
  numpages = {0},
  year = {1984},
  month = {Mar},
  publisher = {American Physical Society},
  doi = {10.1103/PhysRevB.29.3035},
  url = {https://link.aps.org/doi/10.1103/PhysRevB.29.3035}
}

@article{aoki2014nonequilibrium,
  title = {Nonequilibrium dynamical mean-field theory and its applications},
  author = {Aoki, Hideo and Tsuji, Naoto and Eckstein, Martin and Kollar, Marcus and Oka, Takashi and Werner, Philipp},
  journal = {Rev. Mod. Phys.},
  volume = {86},
  issue = {2},
  pages = {779--837},
  numpages = {59},
  year = {2014},
  month = {Jun},
  publisher = {American Physical Society},
  doi = {10.1103/RevModPhys.86.779},
  url = {https://link.aps.org/doi/10.1103/RevModPhys.86.779}
}

@article{murakami2018high,
  title = {High-Harmonic Generation in Mott Insulators},
  author = {Murakami, Yuta and Eckstein, Martin and Werner, Philipp},
  journal = {Phys. Rev. Lett.},
  volume = {121},
  issue = {5},
  pages = {057405},
  numpages = {5},
  year = {2018},
  month = {Aug},
  publisher = {American Physical Society},
  doi = {10.1103/PhysRevLett.121.057405},
  url = {https://link.aps.org/doi/10.1103/PhysRevLett.121.057405}
}

@article{muhlbacher2008real,
  title = {Real-Time Path Integral Approach to Nonequilibrium Many-Body Quantum Systems},
  author = {M\"uhlbacher, Lothar and Rabani, Eran},
  journal = {Phys. Rev. Lett.},
  volume = {100},
  issue = {17},
  pages = {176403},
  numpages = {4},
  year = {2008},
  month = {May},
  publisher = {American Physical Society},
  doi = {10.1103/PhysRevLett.100.176403},
  url = {https://link.aps.org/doi/10.1103/PhysRevLett.100.176403}
}

@article{werner2009diagrammatic,
  title = {Diagrammatic Monte Carlo simulation of nonequilibrium systems},
  author = {Werner, Philipp and Oka, Takashi and Millis, Andrew J.},
  journal = {Phys. Rev. B},
  volume = {79},
  issue = {3},
  pages = {035320},
  numpages = {18},
  year = {2009},
  month = {Jan},
  publisher = {American Physical Society},
  doi = {10.1103/PhysRevB.79.035320},
  url = {https://link.aps.org/doi/10.1103/PhysRevB.79.035320}
}

@article{eckstein2009thermalization,
  title = {Thermalization after an Interaction Quench in the Hubbard Model},
  author = {Eckstein, Martin and Kollar, Marcus and Werner, Philipp},
  journal = {Phys. Rev. Lett.},
  volume = {103},
  issue = {5},
  pages = {056403},
  numpages = {4},
  year = {2009},
  month = {Jul},
  publisher = {American Physical Society},
  doi = {10.1103/PhysRevLett.103.056403},
  url = {https://link.aps.org/doi/10.1103/PhysRevLett.103.056403}
}

@article{tsuji2011dynamical,
  title = {Dynamical Band Flipping in Fermionic Lattice Systems: An ac-Field-Driven Change of the Interaction from Repulsive to Attractive},
  author = {Tsuji, Naoto and Oka, Takashi and Werner, Philipp and Aoki, Hideo},
  journal = {Phys. Rev. Lett.},
  volume = {106},
  issue = {23},
  pages = {236401},
  numpages = {4},
  year = {2011},
  month = {Jun},
  publisher = {American Physical Society},
  doi = {10.1103/PhysRevLett.106.236401},
  url = {https://link.aps.org/doi/10.1103/PhysRevLett.106.236401}
}

@article{werner2010weak,
  title = {Weak-coupling quantum Monte Carlo calculations on the Keldysh contour: Theory and application to the current-voltage characteristics of the Anderson model},
  author = {Werner, Philipp and Oka, Takashi and Eckstein, Martin and Millis, Andrew J.},
  journal = {Phys. Rev. B},
  volume = {81},
  issue = {3},
  pages = {035108},
  numpages = {11},
  year = {2010},
  month = {Jan},
  publisher = {American Physical Society},
  doi = {10.1103/PhysRevB.81.035108},
  url = {https://link.aps.org/doi/10.1103/PhysRevB.81.035108}
}

@article{tsuji2013nonequilibrium,
  title = {Nonequilibrium dynamical mean-field theory based on weak-coupling perturbation expansions: Application to dynamical symmetry breaking in the Hubbard model},
  author = {Tsuji, Naoto and Werner, Philipp},
  journal = {Phys. Rev. B},
  volume = {88},
  issue = {16},
  pages = {165115},
  numpages = {28},
  year = {2013},
  month = {Oct},
  publisher = {American Physical Society},
  doi = {10.1103/PhysRevB.88.165115},
  url = {https://link.aps.org/doi/10.1103/PhysRevB.88.165115}
}

@article{pruschke1989anderson,
  title={The Anderson model with finite Coulomb repulsion},
  author={Pruschke, Th and Grewe, N},
  journal={Z. Phys. B Condens. Matter},
  volume={74},
  number={4},
  pages={439--449},
  year={1989},
  publisher={Springer},
  url={https://link.springer.com/article/10.1007/BF01311391},
}

@article{keiter1971diagrammatic,
  title={Diagrammatic approach to the anderson model for dilute alloys},
  author={Keiter, H and Kimball, JC},
  journal={J. Appl. Phys.},
  volume={42},
  number={4},
  pages={1460--1461},
  year={1971},
  publisher={American Institute of Physics},
  url={https://doi.org/10.1063/1.1660293}
}

@article{kunzel2024numerically,
  title = {Numerically Exact Simulation of Photodoped Mott Insulators},
  author = {K\"unzel, Fabian and Erpenbeck, Andr\'e and Werner, Daniel and Arrigoni, Enrico and Gull, Emanuel and Cohen, Guy and Eckstein, Martin},
  journal = {Phys. Rev. Lett.},
  volume = {132},
  issue = {17},
  pages = {176501},
  numpages = {6},
  year = {2024},
  month = {Apr},
  publisher = {American Physical Society},
  doi = {10.1103/PhysRevLett.132.176501},
  url = {https://link.aps.org/doi/10.1103/PhysRevLett.132.176501},
}

@article{Kaye2024,
  title = {Decomposing Imaginary-Time Feynman Diagrams Using Separable Basis Functions: Anderson Impurity Model Strong-Coupling Expansion},
  author = {Kaye, Jason and Huang, Zhen and Strand, Hugo U. R. and Gole\ifmmode \check{z}\else \v{z}\fi{}, Denis},
  journal = {Phys. Rev. X},
  volume = {14},
  issue = {3},
  pages = {031034},
  numpages = {22},
  year = {2024},
  month = {Aug},
  publisher = {American Physical Society},
  doi = {10.1103/PhysRevX.14.031034},
  url = {https://link.aps.org/doi/10.1103/PhysRevX.14.031034}
}

@article{paprotzki2025high,
  title={High order strong-coupling expansion for X-ray absorption on a dynamically screened impurity},
  author={Paprotzki, Eva and Eckstein, Martin},
  journal={arXiv preprint arXiv:2501.05825},
  url={https://doi.org/10.48550/arXiv.2501.05825},
  doi={10.48550/arXiv.2501.05825},
  year={2025}
}

\end{document}